\documentclass[twocolumn,prl,amsmath,amssymb, aps,superscriptaddress, nofootinbib]{revtex4-2}
 
\usepackage[titletoc,toc,title]{appendix}
\usepackage{titletoc}
\usepackage{titlesec}
\usepackage[version=4]{mhchem} 
\usepackage[normalem]{ulem}
\usepackage{mhchem}
\usepackage{amsmath}
\usepackage{amssymb}
\usepackage{amsfonts}
\usepackage{graphicx}
\usepackage{xcolor}
\usepackage{xfrac}
\usepackage{footmisc}
\usepackage{comment}
\usepackage{pifont}
\usepackage{times}
\usepackage[hidelinks]{hyperref}
\usepackage{enumitem}
\usepackage{centernot}
\usepackage{cancel}
\usepackage{slashed}
\usepackage{relsize}
\usepackage{amsmath,amssymb,mathrsfs}
\usepackage{times}
\usepackage{epsfig}
\usepackage{verbatim}
\usepackage{bm}
\usepackage[utf8]{inputenc}
\usepackage{graphics}
\usepackage{graphicx,epsfig,amssymb,amsmath,color,cancel}
\usepackage{subfigure}
\usepackage[english]{babel}
\usepackage{adjustbox}
\usepackage{physics}

\usepackage{array}
\newcolumntype{P}[1]{>{\centering\arraybackslash}p{#1}}
\newcolumntype{M}[1]{>{\centering\arraybackslash}m{#1}}
\usepackage{tabularray}

\definecolor{zima_blue}{HTML}{1393C1}
\hypersetup{setpagesize=false,bookmarksnumbered=true,bookmarksopen=true,%
colorlinks=true,linkcolor=zima_blue,urlcolor=zima_blue,citecolor=zima_blue,linktocpage=false}

\newcommand{\units}[1]{\textrm{\ #1}}

\newcommand\blfootnote[1]{%
  \begingroup
  \renewcommand\thefootnote{}\footnote{#1}%
  \addtocounter{footnote}{-1}%
  \endgroup
}

\begin{document}

\title{Primordial Black Holes from Supercooled Phase Transitions}

\author{Yann Gouttenoire}
\email{yann.gouttenoire@gmail.com}
\affiliation{School of Physics and Astronomy, Tel-Aviv University, Tel-Aviv 69978, Israel}
\author{Tomer Volansky}
\email{tomer.volansky@gmail.com}
\affiliation{School of Physics and Astronomy, Tel-Aviv University, Tel-Aviv 69978, Israel}

\begin{abstract}

Cosmological first-order phase transitions (1stOPTs) are said to be strongly supercooled when the nucleation temperature is much smaller than the critical temperature. These are 
often encountered 
in theories that admit a nearly scale-invariant potential, 
for which the bounce action decreases only logarithmically with 
temperature. 
During supercooled 1stOPTs the equation of state of the universe undergoes a 
rapid and 
drastic change, 
transitioning 
from vacuum-domination to radiation-domination. 
The statistical variations in 
bubble nucleation histories imply that 
distinct causal patches percolate at slightly different times. Patches which percolate the latest undergo the longest vacuum-domination stage
and as a consequence 
develop large over-densities triggering their 
collapse into primordial black holes (PBHs). 
We derive an analytical 
approximation 
for the probability of a patch to collapse into a PBH as a function of the 1stOPT duration, $\beta^{-1}$, and deduce the expected PBH abundance. We 
find that 1stOPTs which take more than $15\%$ of a Hubble time to complete ($\beta/H \lesssim 7$) produce observable PBHs.
Their abundance 
is independent of the duration of the supercooling phase,
in agreement with the de Sitter no hair conjecture. 
\end{abstract}

\maketitle

\section{INTRODUCTION}

Primordial black holes (PBHs) have been the
object of intense research activities since the detection of gravitational waves from mergers of solar-mass black holes in 2015 \cite{LIGOScientific:2018mvr}. The detection of black holes with sub-solar masses would be 
considered
evidence for the gravitational collapse of large overdensities 
which pre-existed
in the primordial plasma \cite{Carr:1974nx}. A variety of mechanisms have been proposed for generating such inhomogeneities, e.g. inflaton ultra slow-roll~\cite{Carr:1993aq,Ivanov:1994pa}, collapse of cosmic strings~\cite{Hawking:1987bn,Caldwell:1995fu,Jenkins:2020ctp,Blanco-Pillado:2021klh}, of domain walls~\cite{Vachaspati:2017hjw,Ferrer:2018uiu,Gelmini:2022nim,Gelmini:2023ngs}, of scalar condensates~\cite{Dolgov:1992pu,Kasai:2022vhq,Cotner:2016cvr,Martin:2019nuw} or in a dissipative 
dark
sector~\cite{Chang:2018bgx,Flores:2020drq,Domenech:2023afs,Chakraborty:2022mwu}. 
Overdensities and the formation of PBHs can also be associated with 
cosmological first-order phase transitions (1stOPTs) where by and large, four mechanisms have been identified: 
bubble collisions~\cite{Hawking:1982ga,Moss:1994iq,Ashoorioon:2020hln,Jung:2021mku}, matter squeezing by bubble walls~\cite{Crawford:1982yz,Gross:2021qgx,Baker:2021sno,Kawana:2021tde,Huang:2022him}, transitions to a metastable vacuum during inflation~\cite{Garriga:2015fdk,Deng:2016vzb,Deng:2017uwc,Kusenko:2020pcg} and the collapse of delayed false vacuum patches~\cite{Sato:1981bf,Maeda:1981gw,Sato:1981gv,Kodama:1981gu,Kodama:1982sf,Hsu:1990fg,Liu:2021svg,Hashino:2021qoq,He:2022amv,Kawana:2022olo,Lewicki:2023ioy,Gehrman:2023esa}. 

\begin{figure}[ht!]
\centering
\vspace{0.5cm}
\includegraphics[width=250pt]{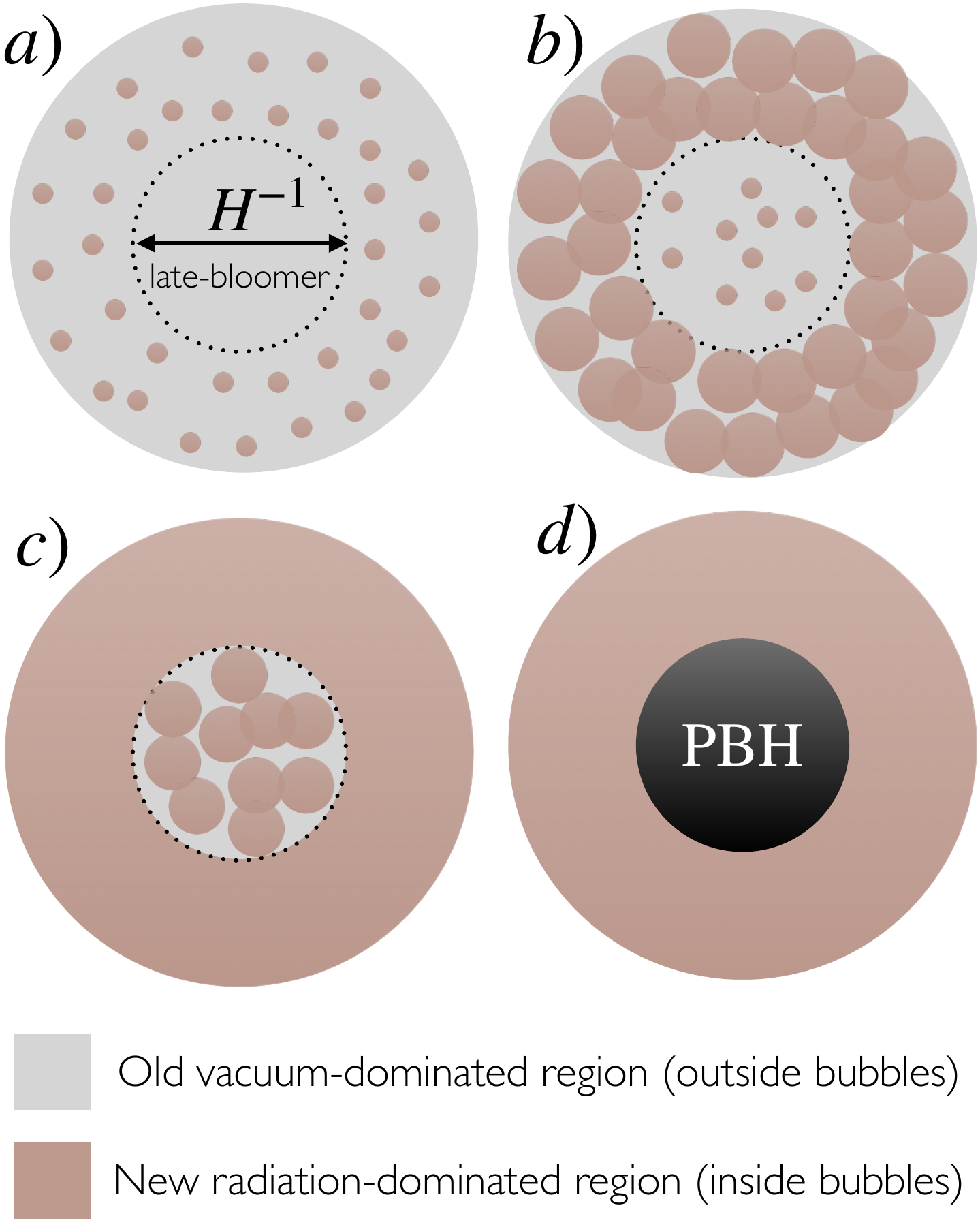}
\caption{\small  \label{fig:keynote}  \textbf{  {\normalsize The supercooled late-blooming mechanism}}:\\ \textbf{a)} The nucleation of bubbles through quantum or thermal tunneling is a random process. Within certain causal patches -- such as the one delimited with a black dotted circle and labeled \textbf{``late-bloomer''} --  bubble nucleation can start later than the background. \textbf{b) and c)} In the supercooled limit, false vacuum regions in \textbf{gray} are vacuum-dominated while true vacuum regions in \textbf{brown} are energetically dominated by components which redshift like radiation (see App.~\ref{app:PT_dynamics}).  As a result,  the background 
is rapidly redshifting while  
late-bloomers  
admit a nearly constant energy density. \textbf{d)} This inhomogeneity in the equation of state 
generates a Hubble-size over-density in the radiation fluid which, above a certain threshold, collapses into a PBH.}
\vspace{-0.7cm}
\end{figure}

In this work, we revisit the last of these mechanisms and show that PBHs can be abundantly produced in the supercooling regime, e.g. when the energy density of the universe is dominated by the latent heat of a phase transition. The latent heat acts as a cosmological constant which causes the universe to inflate until the transition completes and the energy is converted into radiation once bubbles nucleate and percolate.
As illustrated in Fig.~\ref{fig:keynote}, since bubble nucleation is a stochastic event, and since regions outside bubbles expand faster than those inside, a delayed nucleation within a  causal patch would develop high curvature and collapse into a PBH. 
We find that any 1stOPT whose ``duration'' $\beta^{-1}$ is longer than one tenth of Hubble time,
\begin{equation}
\label{eq:beta_def}
\beta ~\equiv~\frac{1}{\Gamma_{\mathsmaller{\rm V}}}\frac{d{\Gamma_{\mathsmaller{\rm V}}}}{dt}\lesssim 7H\,,
\end{equation}
produces PBHs with observational consequences. Here $\Gamma_{\mathsmaller{\rm V}}\equiv \Gamma/V$ is the bubble nucleation rate per unit of volume (and hence a dimension-4 parameter).  The mass of these PBHs is given by the mass inside the sound horizon, cf. Eq.~\eqref{eq:MPBHs}. 

\begin{figure*}[t!]
\centering
\raisebox{0cm}{\makebox{\includegraphics[width=0.65\textwidth]{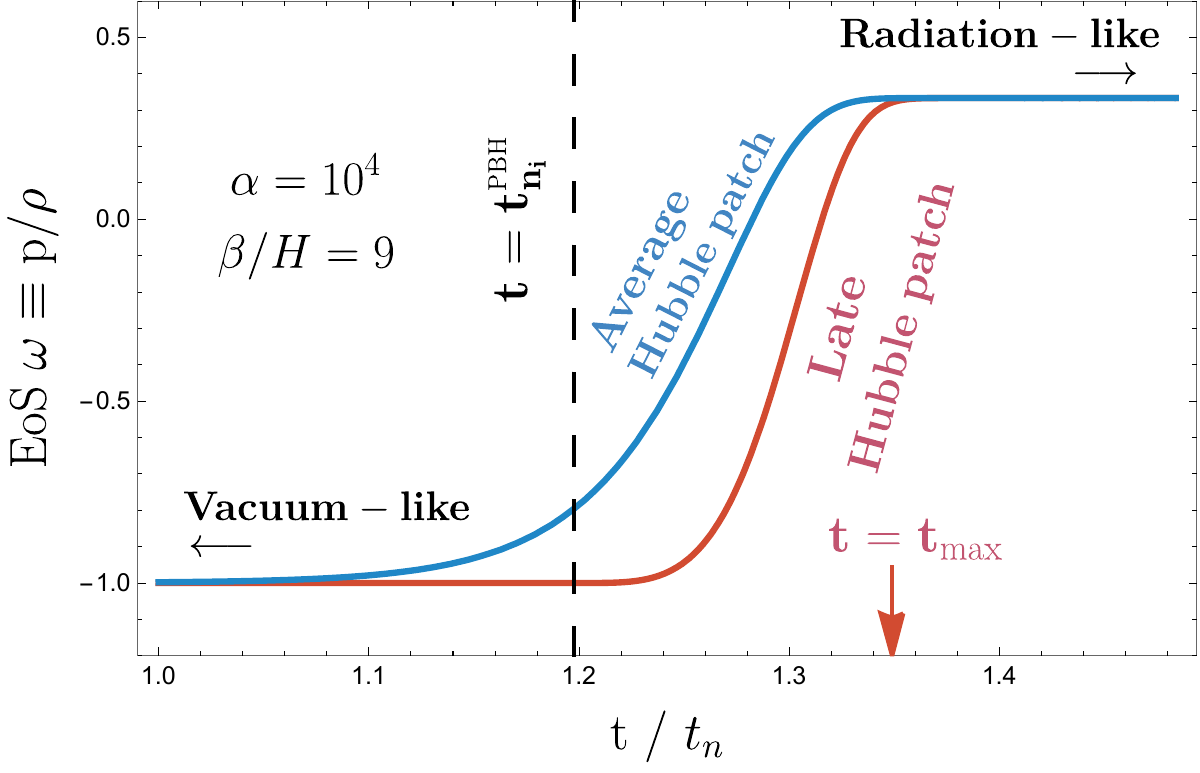}}}
\caption{\small  \label{fig:EOS} 
During a supercooled first-order PT, the universe changes  from a vacuum-like equation of state (EoS) with $\omega = -1$ to a radiation EoS with $\omega = 1/3$. Depending on their bubble nucleation history, distinct Hubble patches follow different EoSs. We compare the EoS of the background (\textbf{blue}) to the EoS of a late-blooming patch (\textbf{red}) inside which nucleation only starts after a critical time $t_{n_i}^{\rm \mathsmaller{PBH}}$.
As a result, the radiation-like cooling phase of such a patch begins late with respect to the 
background, resulting in an over-density (see Fig.~\ref{fig:rho_evolution}) which reaches a maximal density contrast at $t_{\rm max}$ which, if $\delta\rho/\rho > \delta_c$, collapses into a PBH.
 We use $\omega = -(\dot{a}^2+2a\ddot{a})/3\dot{a}^2$ \cite{Kolb:1990vq} with $a(t)$ the solution of Eq.~\eqref{eq:rho_R_text} for $t_{n_i}=t_c$ (\textbf{blue}) and $t_{n_i}=t_{n_i}^{\rm PBH}$ (\textbf{red}) so that it just passes the threshold $\delta\rho/\rho = \delta_c$ in Eq.~\eqref{eq:PBHs_threshold_0} at $t=t_{\rm max}$.}
\end{figure*}

\section{SUPERCOOLED PHASE TRANSITIONS}
In App.~\ref{app:PT_dynamics} we argue that during a 1stOPT the universe can be viewed as composed of a vacuum component $\rho_{\rm V}$ and a radiation component $\rho_{\rm R}$, with equation of state $p=-\rho$ and $p=\rho/3$, respectively
\begin{equation}
\rho_{\rm tot} = \rho_{\rm V} + \rho_{\rm R}\, .
\end{equation} 
Before being converted into radiation through percolation,
$\rho_{\rm V}$ is initially given by the difference $\Delta V$ between the false and true vacuum 
energy densities evaluated at zero-temperature.
Bubbles nucleate around the temperature $T_n$ given by the instantaneous criterion 
\begin{equation}
\label{eq:inst_approx}
    \Gamma_{\mathsmaller{\rm V}} (T_{n}) = H^4(T_{n})\,.
\end{equation}
We assume an exponential nucleation rate per unit of volume 
\begin{equation}
\label{eq:tunneling_rate_def_0}
\Gamma_{\mathsmaller{\rm V}}(t) = \Gamma_0 e^{\beta t}\,\qquad \text{with} \quad \Gamma_0 = H^4(T_{n}) e^{-\beta t_n}\,,
\end{equation}
where the expression for $\Gamma_0$ follows from Eq.~\eqref{eq:inst_approx} and where $t_n=-\int^{T_n}\frac{dT}{TH(T)}$ is the universe cosmic time when its temperature is $T_n$. We stress that  Eq.~\eqref{eq:tunneling_rate_def_0} should be viewed as a Taylor-expansion of the bounce action around $t_n$ at first order.  Any large higher-order corrections would change our results.
We introduce the temperature $T_{\rm eq}$ when the universe becomes vacuum-dominated
\begin{equation}
\label{eq:Teq_def}
 \frac{\pi^2}{30}g_* T_{\rm eq}^4 \equiv \Delta V\,,
\end{equation}
where $g_*$ is the number of relativistic degrees of freedom. 
The ``strength" of the 1stOPT is quantified by the latent heat parameter
\begin{equation}
\label{eq:alpha_def}
\alpha ~\equiv ~ \frac{\rho_{\rm V}}{\rho_{\rm R}}\Big|_{T=T_n}~ =~ \left(\frac{T_{\rm eq}}{T_n}\right)^4 \equiv e^{4N_e}\,, 
\end{equation}
and it is said to be supercooled if $\alpha >1$. 
When the temperature lies within the interval $ T_{\rm eq}> T > T_n $, the universe undergoes a short inflation period of $N_e$ e-foldings.\footnote{The inflation stage operating during the supercooled phase transition studied in this work is a priori unrelated to the primordial inflation which explains the homogeneous and anisotropic CMB.}
The inflationary stage ends by the nucleation of bubbles followed by the expansion of their walls assuming successful percolation. Then, under the assumption of instantaneous reheating,  the universe is heated back to $T_{\rm eq}$,
up to change in degrees of freedom. 

\section{PBH FORMATION}
 
Quantum or thermal tunneling at the origin of each bubble nucleation is a stochastic event. 
As a consequence, the time when bubbles percolate is a random variable whose value depends on the nucleation and expansion history of the $\simeq (\beta/H)^3/6$ bubbles in a causal patch.\footnote{The number density $n_b$ of bubbles grows as $dn_b/dt = F(t;t_{n_i})\Gamma_{\rm \mathsmaller{V}}(t)$, cf. Eq.~\eqref{eq:dNb_Gamma_Vfalse}, where $F$ is the false vacuum fraction defined in Eq.~\eqref{eq:F_vacuum_fraction_text}. Time integration between $-\infty$ to $+\infty$ neglecting Hubble expansion gives $n_b \simeq  \beta^3/8\pi$. We deduce the number of bubbles per Hubble patch $n_b\times 4\pi H^{-3}/3\simeq (\beta/H)^3/6$. }
As we discuss in App.~\ref{app:PT_dynamics}, during bubble growth, the vacuum energy $\rho_{\rm V}$ is converted into a mixture of relativistic expanding bubble walls, relativistic kinetic and thermal energy damped into the plasma and 
relativistic scalar waves associated with bubble collisions. 
These contributions all redshift like radiation $a^{-4}$ and so we denote them by $\rho_{\rm R}$. 

The time of percolation is the time when most of the vacuum energy $\rho_{\rm V}$ has been converted into radiation $\rho_{\rm R}$.
Any delay of the percolation time in a specific causal patch necessarily generates an overdensity of $\rho_{\rm R}$ with respect to the average background. This is because this late Hubble patch is still inflating while the radiation density in its neighborhood has already started redshifting away. 
\begin{figure*}[t!]
\centering
\raisebox{0cm}{\makebox{\includegraphics[ width=0.8\textwidth, scale=1]{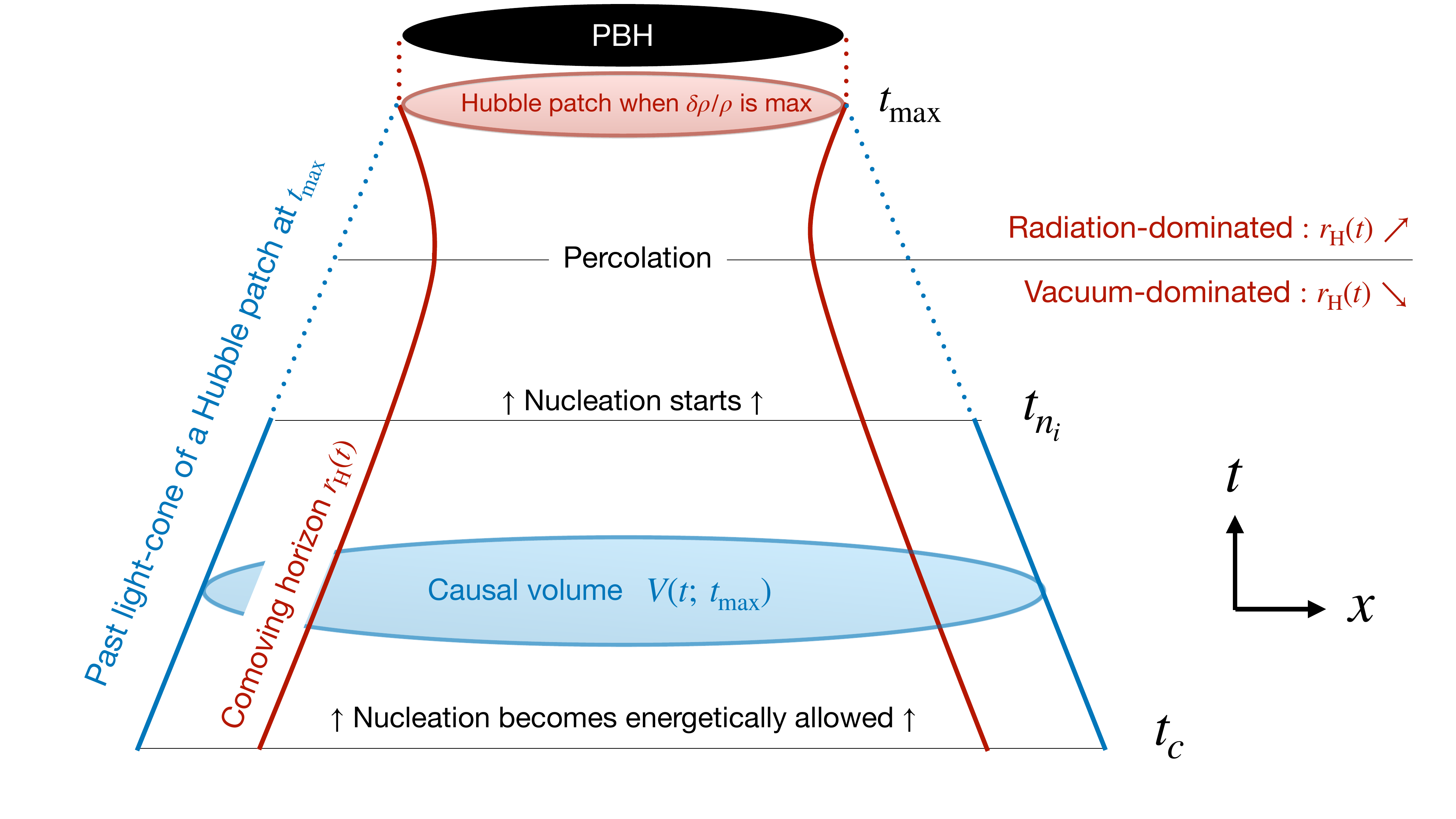}}}
\caption{  \label{fig:space-time_diag} 
  An illustration 
  depicting in chronological order the various steps leading to the eventual collapse of a Hubble patch into a PBH during a supercooled PT. The comoving Hubble horizon $r_{\rm H}(t)$ shown with \textbf{red lines}, shrinks when the universe is vacuum-dominated and grows when it becomes radiation-dominated. 
  With \textbf{blue lines} we show the trajectory of the outermost walls which would enter  the Hubble volume at $t_{\rm max}$ (see Eq.~\eqref{eq:wall_traj} in App.~\ref{sec:numerical_treatment}) and coincide with the past light-cone of the Hubble volume at $t_{\rm max}$. 
  A space-like slice bounded by this light-cone has the volume $V(t;t_{\rm max})$ and is shown in \textbf{shaded blue} (see Eq.~\eqref{eq:causal_volume_text}). The survival probability $\mathcal{P}_{\rm surv}\left(t_{n_i};t_{\rm max}\right)$ (see Eq.~\eqref{eq:proba_tni_PBHs_0}) is the probability of having no bubbles nucleated inside $V(t'; t_{\rm max})$ between $t_c< t'<t_{n_i}$. 
}
\end{figure*} 
The time evolution of the radiation component $\rho_{\rm R}$ follows from energy-momentum conservation in the expanding universe,
\begin{equation}
\label{eq:rho_R_text}
\dot{\rho}_{\rm R}(t;t_{n_i})  + 4H\rho_{\rm R}(t;t_{n_i}) = - \dot{\rho}_{V}(t;t_{n_i}) ,\quad H=\sqrt{\frac{\rho_{\rm V}+\rho_{\rm R}}{3M_{\rm pl}^2}}\,.
\end{equation}
Here $t_{n_i}$ is a free parameter which sets the time at which the first bubble is nucleated in a given causal patch \cite{Kodama:1982sf}.
During bubble growth, the vacuum component $\rho_{V}$ evolves as 
\begin{equation}
\label{eq:rho_V_text}
\rho_{V}(t;t_{n_i}) =  F(t;\, t_{n_i})\Delta V\,,
\end{equation}
where $F(t;\,t_{n_i})$ is the volume fraction of remaining false vacuum at time $t$ \cite{Guth:1979bh}. In  App.~\ref{app:vacuum_energy} 
we rederive this function:
\begin{equation}
\label{eq:F_vacuum_fraction_text}
F(t;\, t_{n_i}) ~= ~ \textrm{exp}\left[-\int_{t_{n_i}}^t dt{'}\, \Gamma_{\mathsmaller{\rm V}}(t^{'}) a(t{'})^3 \frac{4}{3}\pi \,r^3(t;t')\right]\,,
\end{equation}
where $r(t;t')$ is the comoving radius of a bubble which nucleated at time $t'$ and expanded at speed of light until $t$
\begin{equation}
\label{eq:r_t_tp}
  r(t;t')= \int_{t^{'}}^t\frac{d\tilde{t}}{a(\tilde{t})}\,.
\end{equation}
The above equation neglects the comoving bubble radius at nucleation.
In Figs.~\ref{fig:EOS} and \ref{fig:rho_evolution} we show, using Eq.~\eqref{eq:rho_R_text}, the evolution of the equation of state $\omega$ and energy densities $\rho_{\rm R}$, $\rho_{\rm V}$ of a Hubble patch in which bubbles start nucleating later than the background. 
The larger $t_{n_i}$ in Eq.~\eqref{eq:F_vacuum_fraction_text}, the later nucleation 
begins, and the denser with respect to the background the patch becomes after percolation.
If the over-density, $\delta(t;t_{n_i})$, of a lately-nucleated Hubble patch with respect to the background 
is larger than the critical threshold  \cite{Carr:1975qj}
\begin{equation}
\label{eq:PBHs_threshold_0}
\delta(t;t_{n_i})  \equiv \frac{\rho_{\rm tot}(t;t_{n_i})-\rho_{\rm tot}^{\rm bkg}(t)}{\rho_{\rm R}^{\rm bkg}(t)}~ > ~ \delta_{\rm c} \simeq 0.50\,, 
\end{equation}
then this late Hubble patch collapses into a PBH. 
Here $\rho_{\rm tot}(t;t_{n_i})$ denotes the total energy density at time $t$, in a late patch in which the first nucleation event occurs at a time $t_{n_i}$.  On the other hand, $\rho_{\rm tot}^{\rm bkg}(t)\equiv \rho_{\rm tot}(t;t_c)$ represents the background (average) energy density, where nucleation may initiate as early as the time $t_c$ when the phase transition becomes energetically permitted. 

During the radiation domination era, the collapse threshold has been found to range between $\delta_c \in [0.40,~0.67]$, depending on the profile shape \cite{Musco:2018rwt,Escriva:2019phb}, with $\delta_c = 0.40$ for the broadest profiles, and $\delta_c = 0.67$ for the sharpest ones. Past literature have widely used values around $\delta_c\simeq 0.45$ \cite{Jedamzik:1999am,Green:2004wb,Musco:2004ak,Musco:2012au,Harada:2015yda}.
Awaiting future studies on the density profile shape generated by 1stOPT, we assume $\delta_c = 0.50$.

\begin{figure*}[t!]
    \centering
\raisebox{0cm}{\makebox{\includegraphics[ width=0.7\textwidth, scale=1]{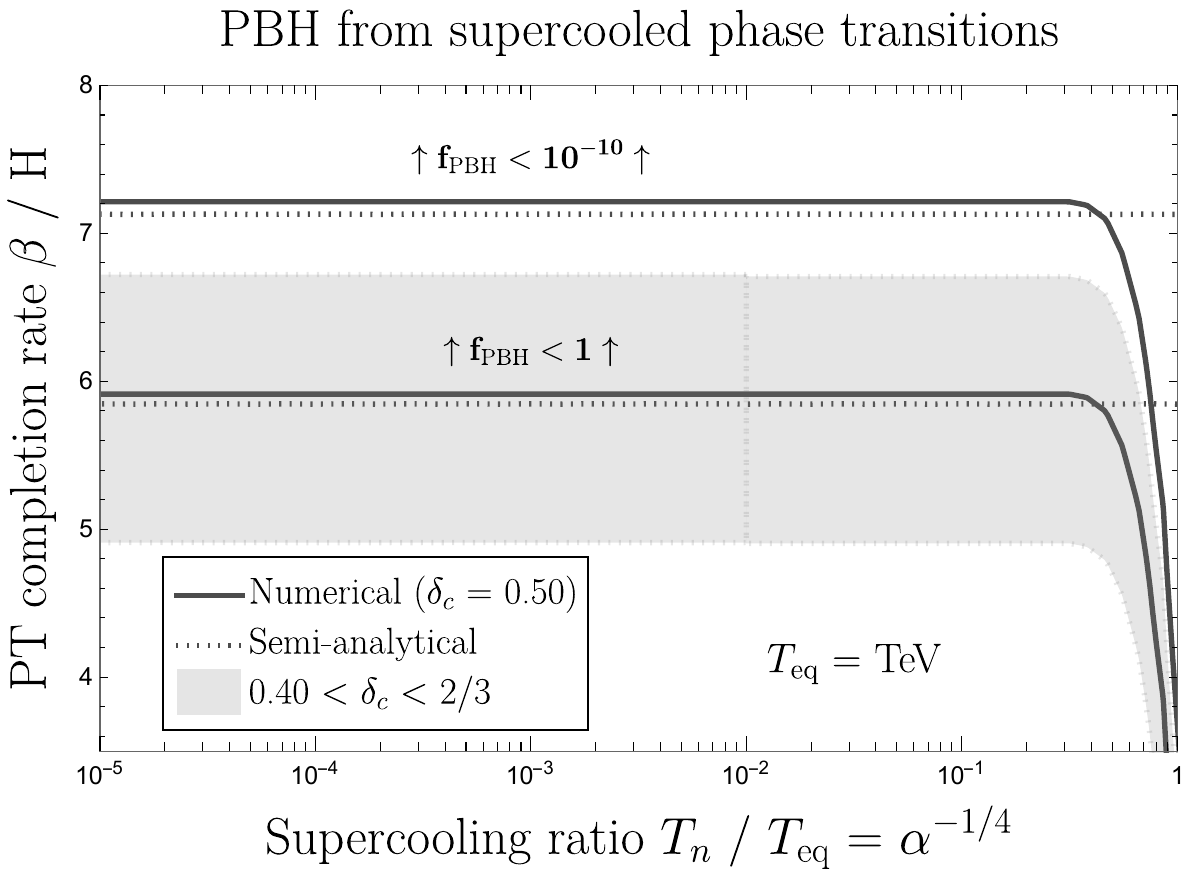}}}
\caption{\small  \label{fig:PBHs_DM_TnOTc_delta_c} Predictions for the PBH density, $f_{\rm PBH}\equiv \rho_{\rm PBH}/\rho_{\rm DM}$, in the $\alpha^{-1/4}$--$\beta/H$ plane.  As discussed in the text, $\alpha$ signifies the strength of the supercooled PT while $\beta^{-1}$  encodes its duration.  For sufficiently long ($\beta\lesssim 6H$) and strong ($\alpha \gtrsim 100$) supercooled PT, significant production of PBHs is expected due to the existence of regions in which bubble nucleation is delayed, resulting in large over-densities.    
The {\bf gray bands} show the dependence of the PBHs abundance on the critical threshold $\delta_{c}=\delta \rho/\rho$ above which gravitational collapse occurs. The {\bf solid} ({\bf dotted}) lines are the numerical (semi-analytical) predictions for $f_{\rm PBH}$ assuming $\delta_c=0.50$. The PBH mass is given by the mass within the sound horizon at the time of the collapse. The temperature $T_{\rm eq}$ marks the beginning of the inflationary phase, during which the universe super-cools until bubbles nucleate at temperature $T_n$. The visible plateau associated with the asymptotic independence of the PBHs abundance on the strength of the phase transition (and hence the duration during which the universe is inflating) 
is a manifestation of de Sitter no hair conjecture~\cite{Wald:1983ky}.}
\end{figure*}
Two comments are now in order.  First, it may seem that the density contrast, $\delta(t;t_{n_i})$, would 
remain constant once all patches have transitioned into a radiation-dominated state 
and should therefore redshift in a similar fashion.
Instead, it reaches a maximum value at a time $t_{\rm max}$ shortly after the percolation of the late-blooming patch, and then decreases to zero as $t$ approaches infinity, as shown in bottom panel of Fig.~\ref{fig:rho_evolution}. This behavior results from the late-blooming patch experiencing a faster expansion 
(and hence a faster redshift) 
due to its excess energy density. This behavior comes from the choice of defining the Hubble constant in Eq.~\eqref{eq:rho_R_text} as a local quantity, which takes different values for distinct causal patches. Another approach -- possibly more realistic -- would be to define a uniform Hubble constant identical for all causal patches and introduce a patch-dependent curvature component $K = a^2 (\rho / 3M_{\rm pl}^2 - H^2)$ in the Friedmann equation. We leave this approach for future work.
Second, one may wonder why we only consider overdensities produced on a full Hubble patch, and ignore those that may occur in smaller regions which go through the phase transitions late.  It is well known that the Schwarzschild radius $r_{\rm s} \equiv 2GM = 8\pi G\rho\,r_{\rm H}^3/3$ of a Hubble patch is equal to its Hubble horizon $r_{\rm H}=H^{-1}$. This explains why an overdensity of $\delta_c\simeq 50\%$ is sufficient to make it collapse into a PBH. In principle, the delay of nucleation in regions of sub-Hubble size $r<r_{\rm H}$ could also produce PBHs. However, since $r_s \propto \rho\, r^{3}$, the overdensity needs to be larger than the critical density $\delta_c$ in Eq.~\eqref{eq:PBHs_threshold_0} by a factor $(r_{\rm H}/r)^2$. The probability for it to happens is exponentially suppressed, as we now discuss. 

For a given Hubble patch of radius $H^{-1}(t=t_{\rm max})$ 
(the Hubble radius when the density contrast $\delta(t;t_{n_i}^{\rm \mathsmaller{PBH}})$ reaches its maximum value), 
we introduce the probability $\mathcal{P}_{\rm surv}(t_{n_i};t_{\rm max})$ that no nucleation occurs in the past light-cone of a Hubble patch at $t_{\rm max}$ before the time $t_{n_i}$ \cite{Kodama:1982sf} (see App.~\ref{sec:collapse_probability} for a derivation): 
\begin{equation}
\label{eq:proba_tni_PBHs_0}
\mathcal{P}_{\rm surv} \left(t_{n_i};t_{\rm max}\right) = \exp\hspace{-0.1cm}\left[ -\hspace{-0.1cm}\int_{t_c}^{t_{n_i}} \hspace{-0.2cm}dt{'}\, \Gamma_{\mathsmaller{\rm V}}(t{'}) a(t{'})^3V(t';t_{\rm max}) \right]\,.
\end{equation}
 $V(t'; t_{\rm max})$ is the space-like slice at time $t'$ bounded by the past light-cone of a Hubble patch at $t_{\rm max}$:
\begin{equation}
\label{eq:causal_volume_text}
    V(t';t_{\rm max})  = \frac{4\pi}{3}\left(r_{\rm H}(t_{\rm max}) + r(t_{\rm max};t{'}) \right)^3\,,
\end{equation}
with $r_{\rm H}\equiv (aH)^{-1}$ and $r(t;t')$ defined in Eq.~\eqref{eq:r_t_tp}. 
The probability that a Hubble patch collapses into a PBH is given by
\begin{equation}
    \mathcal{P}_{\rm coll} \equiv \mathcal{P}_{\rm surv}(t_{n_i}^{\rm \mathsmaller{PBH}};t_{\rm max}).
\end{equation}
where $t_{n_i}^{\rm \mathsmaller{PBH}}$ is the minimum delay of the onset of nucleation for a Hubble patch to reach the critical threshold $\delta_c$ in Eq.~\eqref{eq:PBHs_threshold_0}. We refer the reader to Fig.~\ref{fig:space-time_diag} as well as Fig.~\ref{fig:sequential_stages} in App.~\ref{sec:energies}  for summaries of the different stages leading to the formation of PBHs.

\begin{figure*}[t!]
    \centering
    \includegraphics[width=0.75\textwidth]{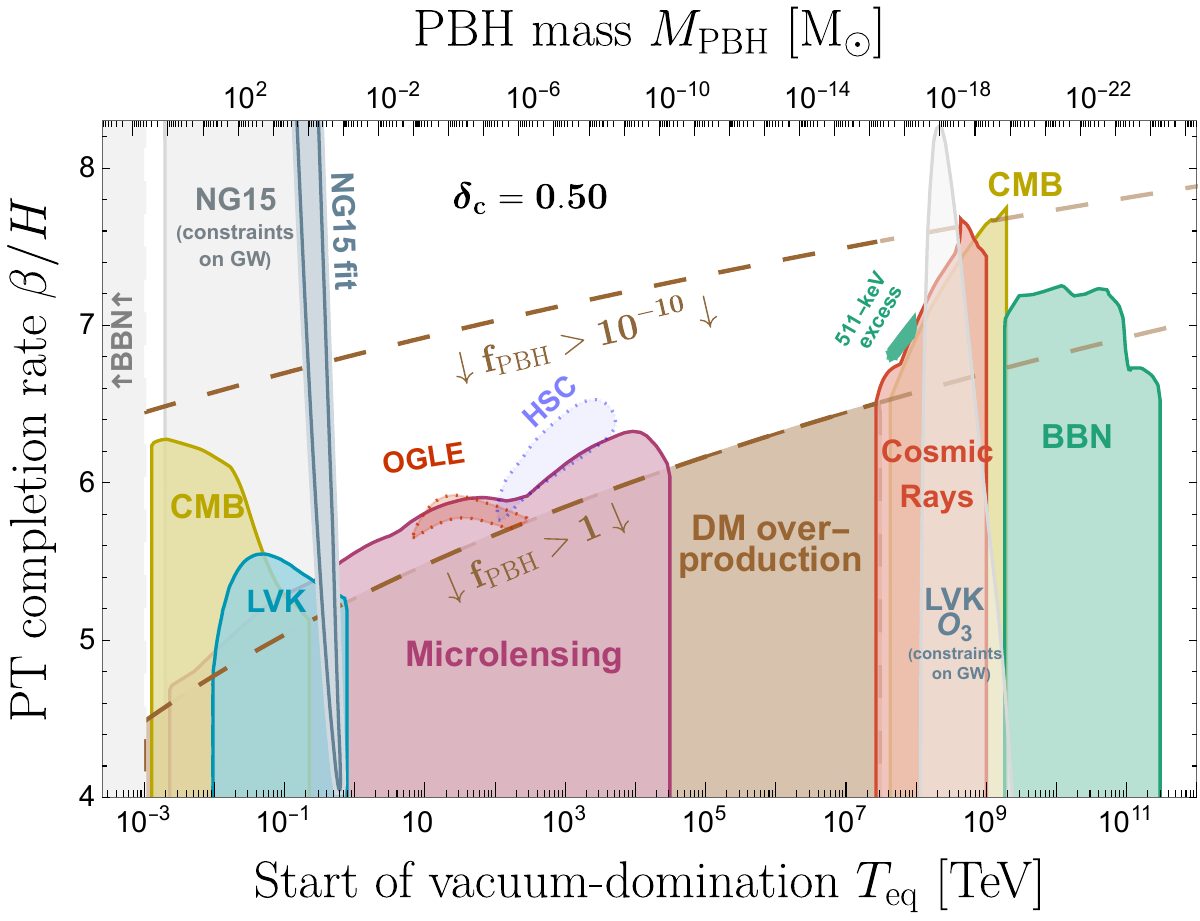}
    \caption{\label{fig:PBHconstraints} \small  
Constraints on strong ($\alpha>100$) supercooled PTs for which PBHs are expected to be produced, shown in the $T_{\rm eq}$--$\beta/H$ plane (with the PBH mass shown on the top x-axis).   As discussed in the text, $T_{\rm eq}$ is the temperature below which an intermediate period of inflation occurs while the universe undergoes super-cooling. It is equal to the maximal reheating temperature after the phase transition up to ratio of relativistic degrees of freedom. The quantity $\beta^{-1}$ encodes the duration of the phase transition.      
In {\bf yellow}, regions excluded by CMB either due to the expected  accretion of predicted large PBHs ({\bf left})~\cite{Ali-Haimoud:2016mbv,Poulin:2017bwe,Serpico:2020ehh}  or due to the evaporation of small PBHs  ({\bf right})~\cite{Poulin:2016anj,Stocker:2018avm,Poulter:2019ooo}, are shown.  The {\bf cyan} region is excluded due to expected merging events in LIGO-Virgo~\cite{DeLuca:2020qqa} while the {\bf purple} region is excluded from the MACHO~\cite{MACHO:2000qbb}, Eros~\cite{EROS-2:2006ryy}, OGLE~\cite{Niikura:2019kqi}, and HSC~\cite{Niikura:2017zjd} microlensing experiments.    The region in which DM over-closes the universe is  shown in {\bf brown}, exclusion from unobserved cosmic-rays fluxes~\cite{Carr:2009jm,Boudaud:2018hqb} is in {\bf red}, and evaporation during BBN~\cite{Carr:2009jm} is in {\bf green}. The \textbf{green spot} is the best fit region for contributing to the 511~keV excess \cite{DeRocco:2019fjq,Laha:2019ssq,Keith:2021guq}. The {\bf dotted dark blue} and {\bf dotted orange} regions are the best-fit regions which address the anomalous microlensing events reported in HSC and OGLE data respectively~\cite{Niikura:2019kqi,Niikura:2017zjd,Sugiyama:2021xqg}. Finally, in dashed and solid {\bf gray} we indicate where gravitational waves from bubble collision fall within the detectability of pulsar timing arrays~\cite{NANOGrav:2021flc,NANOGrav:2023gor,Chen:2021rqp,EPTA:2023fyk,Goncharov:2021oub,Reardon:2023gzh,Xu:2023wog,Antoniadis:2022pcn,InternationalPulsarTimingArray:2023mzf} and exclusion of LIGO-Virgo~\cite{Romero:2021kby}.  Additionally, in the region labeled "BBN" in gray, the reheating temperature is lower than the temperature of neutrino decoupling, $T_{\rm eq} \lesssim \text{MeV}$ \cite{Rubakov:2017xzr}, which is excluded.  The above limits are recasted from various existing constraints shown in Fig.~\ref{fig:PBHs_constraints}. We have fixed the PBHs threshold to $\delta_c=0.50$.} 
\end{figure*}

As we show in App.~\ref{sec:analytical_treatment}, in the limit $ \alpha \gtrsim 10^{2}$ one can approximate the collapsed fraction by the analytic formula, 
\begin{equation}
\label{eq:proba_coll_ana}
\mathcal{P}_{\rm coll}\simeq \exp\left[-a\left( \frac{\beta}{H_n} \right)^{\!b}\left(1+\delta_{\rm c}\right)^{c\frac{\beta}{H_n}} \right]\,,
\end{equation}
elucidating the exponential dependence on $\beta/H_n$ and $\delta_c$, where $a\simeq 1.024$, $b\simeq 0.6921$, $c\simeq 0.8831$ are fitting parameters. Eq.~\eqref{eq:proba_coll_ana} is valid for $\beta/H \in [3,8]$ and $\delta_c \in [0.4,2/3]$.\footnote{The formula in the first version of this work was only valid for $\delta_c\in [0.4,0.5]$.}
Remarkably it is independent of the energy scale of the phase transition $T_{\rm eq}$ and of the duration of the supercooling stage $N_e$ defined in Eq.~\eqref{eq:alpha_def}. For $N_e \gtrsim 1$, the universe enters an almost pure vacuum state with no trace of the initial radiation abundance~\cite{Wald:1983ky},
 which explains the absence of $\alpha$ in Eq.~\eqref{eq:proba_coll_ana}.
The fraction $f_{\rm \mathsmaller{PBH}}$ of DM in the form of PBHs today is given by (see App.~\ref{sec:PBH_abundance}):
\begin{equation}
\label{eq:f_PBH_main}
f_{\rm \mathsmaller{PBH}} \simeq \left(\frac{\mathcal{P}_{\rm coll}}{ 2.2 \times 10^{-8}}\right)\left( \frac{T_{\rm eq}}{140~\rm MeV} \right)\,.
\end{equation}
In order to have 1$\%$ of the DM density in the form of PBHs, only a fraction $P_{\rm \mathsmaller{PBH}}\sim 10^{-8}$ of Hubble patches need to collapse at the $\rm QCD$ epoch and only $10^{-12}$ at the $\rm TeV$ epoch. 
The mass of PBHs is given by the mass inside the sound horizon $c_s H^{-1}$ at the time of the collapse \cite{Escriva:2021pmf}
\begin{align}
\label{eq:MPBHs}
M_{\rm \mathsmaller{PBH}} \simeq M_{\rm sun}~\left(\frac{20}{g_*(T_{\rm eq})} \right)^{1/2}\left(\frac{140~\rm MeV}{T_{\rm eq}} \right)^2,
\end{align}
where $c_s =1/\sqrt{3}$ was used. 
The regions which predict sizable PBH abundance are shown on the $\alpha$-$\beta$ plane in Fig.~\ref{fig:PBHs_DM_TnOTc_delta_c}. 

The parameter $\beta^{-1}$ encapsulates the typical time scale associated with the variations of the nucleation times in different Hubble patches, and as such, a strong suppression of the PBH abundance is anticipated for large values of $\beta/H$.
Conversely, $\alpha$ encodes the 
amount 
of energy density stored in the form of vacuum energy during the phase transition and
hence the production of PBHs
becomes inefficient for small values of $\alpha$. These features are clearly visible in Fig.~\ref{fig:PBHs_DM_TnOTc_delta_c}. For further information on the mechanism of PBH formation, we refer the reader to App.~\ref{app:PBHformation}.

\section{ASTROPHYSICAL AND COSMOLOGICAL CONSTRAINTS}

\subsection{PBH DETECTABILITY}

The standard search strategy for supercooled phase transitions (PT) are gravitational waves (GWs) sourced by bubble collision and the associated plasma dynamics~\cite{Caprini:2015zlo,Caprini:2019egz,Gouttenoire:2021kjv,Gouttenoire:2022gwi}.  
The possible production of PBHs allows for a novel and complementary strategy  to identify the presence of a supercooled PT in the early universe and here we summarize existing constraints (for a review on the cosmological and astrophysical constraints on PBHs see~\cite{Carr:2020xqk}).

In Fig.~\ref{fig:PBHconstraints} we show the viable parameter space in the $\beta/H$-$T_{\rm eq}$ plane, for the supercooled 1stOPT scenario discussed in this letter. The Gray-shaded regions represent observables that are not related to PBHs. Firstly, for temperatures below $T_{\rm eq}\lesssim \rm MeV$ a supercooled PT would inject entropy after the onset of Big-Bang Nucleosynthesis (BBN), which is excluded. Secondly, the non-detection of a stochastic GW background at the LIGO/Virgo interferometers~\cite{Romero:2021kby} excludes supercooled PTs around $T_{\rm eq}\sim 10^8~\rm GeV$, while the region around $T_{\rm eq}\sim 100~\rm MeV$ is of interest~ \cite{Gouttenoire:2023bqy,Ellis:2023oxs,Lewicki:2024ghw} for the recent hint for a nano-Hertz GW signal found in pulsar timing array datasets 
NANOgrav~\cite{NANOGrav:2021flc,NANOGrav:2023gor}, EPTA~\cite{Chen:2021rqp,EPTA:2023fyk}, PPTA~\cite{Goncharov:2021oub,Reardon:2023gzh}, CPTA~\cite{Xu:2023wog} and IPTA~\cite{Antoniadis:2022pcn,InternationalPulsarTimingArray:2023mzf}. 
The GW signal sourced by bubble collision is derived using the bulk flow model~\cite{Jinno:2017fby,Konstandin:2017sat}.

On the other hand, the colored regions in Fig.~\ref{fig:PBHconstraints} depict the constraints and regions of interest connected with the presence of PBHs. Sufficiently small PBHs, with masses below $\lesssim 10^{15}\units{g}$, evaporate and do not survive to present day~\cite{Hawking:1974rv,Hawking:1975vcx}.  Such PBHs, which in the framework discussed here correspond to $T_{\rm eq}\gtrsim 10^7 ~{\rm GeV}$  (see Eq.~\ref{eq:MPBHs}), would radiate at early times and are therefore constrained.  The yellow region (on the right) in Fig.~\ref{fig:PBHconstraints} shows such limits from Cosmic Microwave Background (CMB) anisotropies~\cite{Poulin:2016anj,Stocker:2018avm,Poulter:2019ooo}.  The red region corresponds to constraints from measured fluxes of extra-galactic photons on Earth~\cite{Carr:2009jm} and from the $e^\pm$ flux measured by Voyager 1~\cite{Boudaud:2018hqb}. In the green spot, they could contribute to the $511~\rm keV$ excess \cite{DeRocco:2019fjq,Laha:2019ssq,Keith:2021guq}. Finally, constraints on energy injection  at BBN are shown in green~\cite{Carr:2009jm,Acharya:2020jbv,Keith:2020jww}.

Conversely, supercooled PTs happening at lower temperatures, $T_{\rm eq} \lesssim 10^7{~\rm GeV}$, produce PBHs which would contribute to the Dark Matter (DM) relic density today. For sufficiently small $\beta/H$ they would be over-produced (see Eqs.~\eqref{eq:proba_coll_ana} and~\eqref{eq:f_PBH_main}), thereby overclosing the universe, as shown in brown.
Microlensing constraints
displayed in purple are 
from MACHO~\cite{MACHO:2000qbb}, Eros~\cite{EROS-2:2006ryy}, OGLE~\cite{Niikura:2019kqi} and HSC~\cite{Niikura:2017zjd}, 
relevant for supercooled PTs occurring around the electroweak epoch $T_{\rm eq} \sim \rm 100~\rm GeV$~\cite{Gouttenoire:2023pxh}. Best-fit regions \cite{Sugiyama:2021xqg} for
the recently observed microlensing events in OGLE~\cite{Niikura:2019kqi} and HSC data~\cite{Niikura:2017zjd} are shown with dotted outlines in orange and dark blue.
The cyan-colored region shows constraints from the LIGO/Virgo interferometers \cite{LIGOScientific:2019kan,DeLuca:2020qqa}, relevant for a 
supercooled PT occuring around the QCD confinement temperature.
Finally, in yellow (on the left) we also show the excluded region due to CMB~\cite{Ali-Haimoud:2016mbv,Poulin:2017bwe,Serpico:2020ehh}, relevant for  supercooled PTs occurring just before the onset of BBN, thereby predicting $10^{1-4}$ solar-mass PBHs whose accretion dynamics are constrained.


\section{CONCLUSION}

Supercooled phase transitions take place when mass scales emerge from the soft breaking of scale invariance \cite{Coleman:1973jx,Witten:1980ez,Hempfling:1996ht,Bardeen:1995,Iso:2017uuu,vonHarling:2017yew}. They have remarkable cosmological consequences, including 
GWs from bubble collisions~\cite{Randall:2006py,Jinno:2016knw,Brdar:2018num,Brdar:2019qut,Marzo:2018,Ellis:2019oqb,Ellis:2020nnr,Baldes:2018emh,Prokopec:2018tnq,DelleRose:2019pgi,Kierkla:2022odc,Gouttenoire:2022gwi}, dilution of dangerous relics~\cite{Konstandin:2011dr,Hambye:2018qjv,Baldes:2020kam}, or particle production~\cite{Baldes:2020kam,Azatov:2021ifm,Baldes:2023fsp}, that can source the genesis of dark matter~\cite{Baldes:2021aph,Azatov:2021ifm,Baldes:2022oev,Wong:2023qon} or the baryonic asymmetry~\cite{Konstandin:2011ds,Servant:2014bla,Azatov:2021irb,Baldes:2021vyz}.

In this study, we investigated the formation of PBHs during supercooled PTs. A version of this mechanism was originally proposed in the 80s~\cite{Kodama:1982sf} using a simultaneous nucleation rate,
$\Gamma_{\mathsmaller{\rm V}}(t)\propto \delta(t-t_n)$, 
and was reconsidered more recently by~\cite{Liu:2021svg} using the exponential nucleation rate defined in Eq.~\eqref{eq:tunneling_rate_def_0}. Our study builds on these works, conducting a comprehensive analysis of the mechanism, including detailed computations of the energy budget, bubble wall equation of state, and collapse probability. We consistently take into account, for the first time, the nucleation history throughout the entire past light-cone of the collapsing Hubble patch. In addition to the numerical results, we provide a ready-to-use semi-analytical formula for the PBH abundance. Our results suggest that the PBH formation is strongly dependent 
on the tunneling probability growth rate 
$\beta$ (the inverse of which encodes the duration of the phase transition), 
and is  independent of the strength $\alpha$ of the phase transition, as long as it super-cools (inflates) during more than one Hubble time.

We point out that experiments constraining PBH populations also constrain the parameter space of supercooled phase transitions.
The constraints derived in this work, which imply $\beta/H \lesssim 5-7$ over a wide range of reheating temperatures $T_{\rm eq}$, are the strongest to date. This study also highlights novel cosmological implications, including the possibility studied in \cite{Gouttenoire:2023bqy,Ellis:2023oxs,Lewicki:2024ghw}, for supercooled PTs occurring around the QCD epoch $T_{\rm eq}\sim 150~\rm MeV$ to produce PBHs in the LIGO-Virgo range~\cite{Romero:2021kby} and at the same time GWs from bubble collisions in the PTA window~\cite{NANOGrav:2021flc,NANOGrav:2023gor,Chen:2021rqp,EPTA:2023fyk,Goncharov:2021oub,Reardon:2023gzh,Xu:2023wog,Antoniadis:2022pcn,InternationalPulsarTimingArray:2023mzf}.
Additionally, supercooled phase transitions occurring around the electroweak epoch $T_{\rm eq}\sim 100~\rm GeV$, as predicted in conformal Higgs models~\cite{Gouttenoire:2023pxh}, would result with  PBHs falling within the reach of microlensing experiments, and could potentially account for the recently observed unusual events recorded in OGLE and HSC data \cite{Niikura:2019kqi,Niikura:2017zjd,Sugiyama:2021xqg}.

{\bf Note added.}---%
We now comment on references that appeared after the first submission of this paper to the arxiv.
In the present work, we consider a monochromatic PBH mass distribution defined by the mass inside the sound horizon. Instead, Refs.~\cite{Baldes:2023rqv, Lewicki:2024ghw} calculate the extended mass distribution.
In the present work, we only consider the fluctuation of the nucleation time of the first bubble. Instead, Ref.~\cite{Lewicki:2024ghw} includes the fluctuation of the nucleation time of the first $j_c$ bubbles.
Refs.~\cite{Lewicki:2023ioy,Flores:2024lng} study PBHs formation in purely vacuum-dominated patches only, before any bubble is nucleated, with Ref.~\cite{Flores:2024lng} raising doubt that curvature perturbation are efficiently generated after nucleation has started. In the present work, we treat late-blooming patches as evolving independently from the background and do not consider the effects of the curvature $K$ that would result from their interaction.

{\bf Acknowledgements.}---%
The authors thank Yoann G\'{e}nolini, Daniel Lozano Jarque and Filippo Sala for useful discussions. Additionally, the authors express their gratitude to Iason Baldes and Anish Ghoshal for pointing out typographical errors in Eq.~\eqref{eq:N_patches} and Eq.~\eqref{eq:tni_PBHs_approx}, respectively, in the first version of this work.
YG is grateful to the Azrieli Foundation for the award of an Azrieli Fellowship.  TV is supported, in part, by the Israel Science Foundation (grant No. 1862/21), by the Binational Science Foundation (grant No. 2020220),   and by the European Research Council (ERC) under the EU Horizon 2020 Programme (ERC-CoG-2015 - Proposal n. 682676 LDMThExp).

\clearpage
\appendix
\onecolumngrid

\fontsize{11}{13}\selectfont






\renewcommand{\tocname}{\Large  Table of contents
\vspace{1 cm}}%

\titleformat{\section}
{\normalfont\fontsize{12}{14}\bfseries  \centering }{\thesection.}{1em}{}
\titleformat{\subsection}
{\normalfont\fontsize{12}{14}\bfseries \centering}{\thesubsection.}{1em}{}
\titleformat{\subsubsection}
{\normalfont\fontsize{12}{14}\bfseries \centering}{\thesubsubsection)}{1em}{}

\titleformat{\paragraph}
{\normalfont\fontsize{12}{14}\bfseries  }{\thesection:}{1em}{}

 {
 \hypersetup{linkcolor=black}
 \tableofcontents
 }
\newpage
\section {Quantities and notations}
\label{sec:energies}

\begin{figure}[ht!]
\centering
\raisebox{0cm}{\makebox{\includegraphics[ width=0.9\textwidth, scale=1]{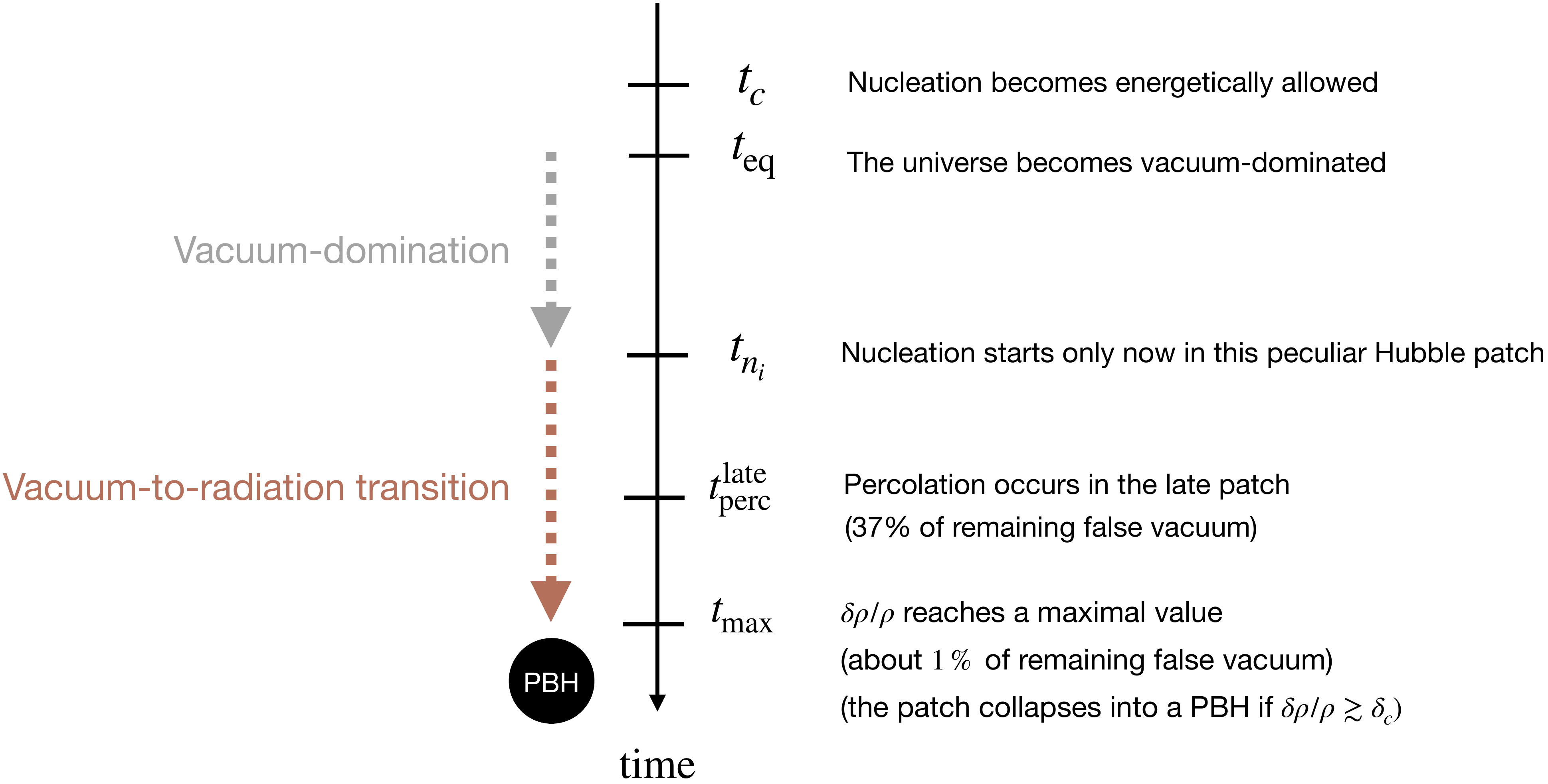}}}
\caption{  \label{fig:sequential_stages} 
A 
diagram that illustrates the sequential stages, from top to bottom, leading a given Hubble patch to collapse into a PBH. Bubble nucleation becomes energetically allowed 
 at $t_c$ but is postponed until $t_{n_i}$. Meanwhile the Hubble patch remains vacuum-dominated. This results in an overdensity peak at $t_{\rm max}$, shortly after the percolation time $t_{\rm perc}^{\rm late}$. The delayed patch collapses into a PBH if the density contrast surpasses the critical value $\delta_c$.
}
\end{figure}

In Fig.~\ref{fig:sequential_stages} and Table~\ref{tab:table1} we summarize the notations, scales and dynamics which lead to the formation of PBHs.  
\begin{table}[h!t]
  \begin{center}
    \begin{tblr}{|Q[c,2.5cm]|Q[c,2.5cm]|Q[c,10cm]|}
      \hline
      \textbf{Quantity} & \textbf{Equation} & \textbf{Definition}  \\
      \hline \hline
        $t_c$ & Eq.~\eqref{eq:proba_tni_PBHs_0} & Time when nucleation becomes energetically allowed \\
      \hline
      $t_{\rm eq}$ & Eq.~\eqref{eq:Teq_def} & Time when the universe becomes vacuum-dominated \\
      \hline
      $t_{n_i}$ & Eq.~\eqref{eq:F_vacuum_fraction_text} & Time when the first bubble is nucleated in a given causal patch \\
      \hline
      $t_{n_i}^{\rm PBH}$ & Eq.~\eqref{eq:PBHs_threshold_2} & Critical delay of nucleation beyond which a patch collapses into a PBH \\
      \hline
      $t_{n}$ & Eq.~\eqref{eq:inst_approx} & Instantaneous nucleation time, defined through $\Gamma(t_n)\equiv H^4(t_n)$ \\
      \hline
      $t_{\rm perc}^{\rm late}$ & Eq.~\eqref{eq:tstar_bkg_late} & Percolation time in a given late patch ($\sim 37\%$ of remaining false vacuum) \\
      \hline
      $t_{\rm perc}^{\rm bkg}$ & Eq.~\eqref{eq:tstar_bkg_late} & Average percolation time in the background ($\sim 37\%$ of remaining false vacuum) \\
    \hline
      $t_{\rm max}$ & Eq.~\eqref{eq:t_max} & Time when a late patch reaches a maximal density contrast $\delta\rho/\rho$ with respect to the background ($\sim 1\%$ of remaining false vacuum)\\
      \hline \hline
      $r(t;t')$ & Eq.~\eqref{eq:r_t_tp} &  Radius at time $t$ of a bubble which nucleated at time $t'$ \\
      \hline
      $V(t';t_{\rm max})$ & Eq.~\eqref{eq:causal_volume} & Space-like slice at time $t'$ bounded by the past light-cone of a Hubble patch at time $t_{\rm max}$. We have $V(t';t_{\rm max})=4\pi\, r_{\rm out}\left(t';t_{\rm max}\right)^3/3$ (see below)\\
      \hline
      $r_{\rm out}\left(t';t_{\rm max}\right)$ & Eq.~\eqref{eq:wall_traj} & Comoving distance between the center of a Hubble patch at $t_{\rm max}$ and the center of a bubble nucleating at $t'$ whose wall enter the horizon at $t_{\rm max}$. It represents the past light-cone of a Hubble patch at $t_{\rm max}$.\\
      \hline
    \end{tblr}
    \label{tab:table1}
    \caption{The formalism presented in this paper involves the use of various time and length quantities, whose definitions are summarized above.  }
  \end{center}
\end{table}

\section{Phase transition dynamics}
\label{app:PT_dynamics}
In this section we show that the energy density of the universe during a supercooled 1stOPT can be divided into vacuum energy ($\rho_V$), 
expanding bubble walls ($\rho_{\rm wall}$), energy damped into the plasma ($\rho_{\rm plasma}$), residue of bubble collisions ($\rm \rho_{\rm coll}$), a growing thermal component ($\rho_{\rm th}$), and the pre-existing super-cooled plasma ($\rho_{\rm cool}$): 
\begin{equation}
\rho_{\rm tot} = \rho_{\rm V} + \rho_{\rm wall}+ \rho_{\rm plasma}+ \rho_{\rm scalar}+\rho_{\rm th}+\rho_{\rm cool}\,.
\end{equation}
Below we briefly discuss each of these components, showing that all but the vacuum energy behaves like radiation thereby establishing the claim in the paper that effectively, during the 1stOPT, the energy budget is composed of vacuum energy and radiation.  

\subsection{Vacuum energy}
\label{app:vacuum_energy}
In the supercooled limit and before nucleation commences, finite-temperature corrections to the potential of the scalar field driving the phase transition are small \cite{Quiros:1999jp}, and the energy density, $\rho_{\rm V}$, of the scalar field is to a good approximation given by the energy difference,  $\Delta V$, between the two relevant minima in the potential at zero temperature. As nucleation proceeds and bubbles expand, it decreases as
\begin{equation}
\label{eq:rho_V}
\rho_{V}(t;t_{n_i}) =  F(t;\, t_{n_i})\Delta V\,,
\end{equation}
where 
\begin{equation}
\label{eq:def_F_V_ratio}
    F(t;\,t_{n_i})\equiv V_{\rm false}(t;\,t_{n_i})/V_{\rm tot}
\end{equation}
is the volume fraction of remaining false vacuum at time $t$ with $t_{n_i}$ the time at which the first bubble is nucleated. $V_{\rm false}(t;t_{n_i})$ is the comoving volume of false vacuum at $t$, assuming bubble nucleation starts at $t_{n_i}$, and $V_{\rm tot}$ is the total comoving volume.\footnote{The comoving volume is related to the physical volume by the scale factor $a(t)$ of the universe:
$V_{\textrm{comoving}} = V_{\textrm{physical}}/a(t)^3$.} The expression for $F(t;\,t_{n_i})$ was originally derived in 1979 by~\cite{Guth:1979bh} and here we reproduce it following~\cite{Hindmarsh:2019phv}.
The number of bubbles nucleated per unit of comoving volume between $t'$ and $t'+dt'$ is
\begin{equation}
\label{eq:dNb_Gamma_Vfalse}
    dN_b(t') = \Gamma_{\mathsmaller{\rm V}}(t^{'}) a(t{'})^3 V_{\rm false}(t';t_{n_i})dt'\,.
\end{equation}
The reduction of false vacuum volume $V_{\rm false}(t;t_{n_i})$ between time $t$ and $t+dt$ due to the growth of the $dN_b(t')$ bubbles nucleated between $t'$ and $t'+dt'$ is
\begin{equation}
\label{eq:d2V}
    d^2V_{\rm false}(t;t_{n_i}) = - \frac{V_{\rm false}(t;t_{n_i})}{V_{\rm false}(t';t_{n_i})}4\pi r(t;t')^2 dr \,dN_b(t')\,,
\end{equation}
where  $dr = v_w dt/a(t)$ and $r(t;t')$ is the comoving radius at $t$ of a bubble having wall velocity $v_w$ and being 
nucleated at $t'$: 
\begin{equation}
\label{eq:comoving_radius}
r(t;t')~= ~  \int_{t^{'}}^t d\tilde{t}~\frac{v_w(\tilde{t}) }{a(\tilde{t})}.
\end{equation}
The above equation neglects the comoving bubble radius $r_{\rm nuc}$ at nucleation.\footnote{Assuming $r_{\rm nuc}\simeq T_{\rm nuc}$, we expect corrections from the finite bubble size at nucleation to the false vacuum fraction in Eq.~\eqref{eq:F_vacuum_fraction} to be of order 
\begin{equation}
    r_{\rm nuc}\beta \simeq 1.3 \times 10^{-13} \left(\frac{T_{\rm eq}}{10^2T_n} \right)\left(\frac{T_{\rm eq}}{\rm TeV} \right) \frac{\beta/H}{10}.
\end{equation}} The ratio $V_{\rm false}(t;t_{n_i})/V_{\rm false}(t';t_{n_i})$ prevents over-counting by removing regions which have been converted into the true vacuum between $t'$ and $t$. 
Integrating $t'$ over all nucleation times between $t_{n_i}$ and $t$ gives 
\begin{equation}
    dV_{\rm false}(t;t_{n_i}) = -  V_{\rm false}(t;t_{n_i}) \frac{v_w dt}{a(t)} \int_{t_{n_i}}^t dt'\, \Gamma_{\rm \mathsmaller{V}}(t') a(t')^34\pi r(t;t')^2 .
\end{equation}
Dividing by the total comoving volume $V_{\rm tot}$ and using Eq.~\eqref{eq:def_F_V_ratio}, we get
\begin{equation}
    \frac{dF(t;\,t_{n_i})}{dt} = -F(t;\,t_{n_i})\frac{v_w}{a(t)}\int_{t_{n_i}}^t dt'\, \Gamma_{\rm \mathsmaller{V}}(t') a(t')^34\pi r(t;t')^2 .
\end{equation}
The solution to the last equation is
\begin{equation}
\label{eq:F_vacuum_fraction}
F(t;\,  t_{n_i}) ~\equiv ~ e^{-I(t;\,t_{n_i})},\qquad \textrm{with} \quad  I(t;\,t_{n_i})\equiv \int_{t_{n_i}}^t dt{'}\, \Gamma_{\mathsmaller{\rm V}}(t^{'}) a(t{'})^3 \frac{4\pi}{3}r(t;t')^3.
\end{equation}
$I(t;\,t_{n_i})$ can be interpreted as the average number of bubbles inside which, a given point is contained assuming that bubble walls can pass through each others. The ratio $V_{\rm false}(t;t_{n_i})/V_{\rm false}(t';t_{n_i})$ in Eq.~\eqref{eq:d2V}, which is introduced to avoid multiple counting due to overlapping bubbles, is responsible for the exponential behavior in Eq.~\eqref{eq:F_vacuum_fraction}. 

\subsection{Radiation energy}
{\bf Expanding bubble wall.}   
$\rho_{\rm wall}$ describes the energy density stored in the relativistic bubble walls. We denote by $\gamma_w(t;t_n)$ the Lorentz factor of a  wall at time $t$, nucleated at time $t_n$, and by $\sigma$ the wall energy per unit area. The energy density stored in the expanding walls (which are yet to be collide) is~\cite{Turner:1992tz}
\begin{equation}
\label{eq:rho_wall_def}
\rho_{\rm wall}(t)  = F(t;\,  t_{n_i}) \int_{t_{n_i}}^t dt{'}\, \Gamma_{\mathsmaller{\rm V}}(t^{'}) a(t{'})^3 4\pi r(t;t')^2 \gamma_w(t;t') \sigma  /a(t)\,.
\end{equation}
The factor $F(t;\,  t_{n_i})$ gets rid of regions which have already collided.
The scale factor $a(t)$ at the denominator comes from the ratio between the bubble wall energy $4\pi \gamma_w \sigma (ar)^2$ and the comoving volume $a^3$. Below in Eq.~\eqref{eq:rho_wall_def_2} we check that Eq.~\eqref{eq:rho_wall_def} satisfies energy-momentum conservation.
The equation of motion of a bubble wall in FRW background reads, e.g. \cite{Deng:2020mds}
\begin{equation}
\ddot{R} + 3H \dot{R} + \frac{2}{R}\left(1 - \dot{R}^2\right) = \frac{\Delta V-\mathcal{P}_{\rm fric}}{\sigma} \left( 1- \dot{R}^2\right)^{3/2}\,,
\end{equation}
where the bubble physical radius  $R$ is related to the comoving one in Eq.~\eqref{eq:comoving_radius} by 
\begin{equation}
  R= a r\,.  
\end{equation}
We have introduced the friction pressure $\mathcal{P}_{\rm fric}$ due to wall-plasma interactions \cite{Gouttenoire:2021kjv}.
Using $\dot{\gamma}_w =  \gamma_w^3 v_w \dot{v}_w$, we obtain
\begin{equation}
\label{eq:eom_gamma}
\frac{d\gamma_w}{dt}  + 3 H v_w^2 \gamma_w + \frac{2}{R} v_w \gamma_w = \frac{\Delta V-\mathcal{P}_{\rm fric}}{\sigma}v_w\,.
\end{equation}
Plugging Eq.~\eqref{eq:eom_gamma} into Eq.~\eqref{eq:rho_wall_def}, we get the time evolution of the bubble wall energy density
\begin{equation}
\label{eq:rho_wall_time_evolution}
\dot{\rho}_{\rm wall}  + (1+3v_w^2)H\rho_{\rm wall} = - \dot{\rho}_{V} +  \frac{\mathcal{P}_{\rm fric}}{\Delta V}\dot{\rho}_{V} - \dot{I}\rho_{\rm wall}\,,
\end{equation}
where $I\equiv I(t;t_{n_i})$ is defined in Eq.~\eqref{eq:F_vacuum_fraction}.
We obtain that relativistic bubble walls, $v_w=1$, redshift as radiation $\rho\propto a^{-4}$ while non-relativistic bubble walls, $v_w \ll 1$, have the usual equation of state of domain walls at rest $\rho\propto a^{-1}$ \cite{Vilenkin:2000jqa}. This is one of the main results of this appendix.

{\textbf{Reheated energy density.} The second term on the right hand side of Eq.~\eqref{eq:rho_wall_time_evolution} describes the energy transfer rate into the plasma, in the form of sound waves, turbulence and heat due to wall-particle interactions. We collectively denote by $\rho_{\rm plasma}$ the energy density of those plasma excitations.
The third term on the right hand side of Eq.~\eqref{eq:rho_wall_time_evolution} describes the conversion of energy into dissolved scalar configurations as a consequence of bubble wall collisions. 
Collision of relativistic bubble walls are expected to produce relativistic configurations of the scalar field driving the phase transition and of fields coupled to it, which have been called ``scalar waves'' \cite{Watkins:1991zt,Kolb:1996jr,Konstandin:2011ds,Falkowski:2012fb,Cutting:2018tjt,Cutting:2020nla}. We denote this energy density component as $\rho_{\rm scalar}$. After production, plasma excitation $\rho_{\rm plasma}$ and scalar waves $\rho_{\rm scalar}$ thermalize to form the reheated thermal bath \cite{Aarts:2000mg,Micha:2002ey,Arrizabalaga:2005tf} which we denote by $\rho_{\rm th}$.
Introducing $(\tau_{\rm plasma},\omega_{\rm plasma})$ and $(\tau_{\rm scalar},\omega_{\rm scalar})$ the lifetimes and equations of state of plasma excitations and scalar waves respectively, and starting from Eq.~\eqref{eq:rho_wall_time_evolution}, we can write the coupled continuity equations 
\begin{align}
\label{eq:rho_plasma}
&\dot{\rho}_{\rm plasma}  + 3(1+\omega_{\rm plasma})H\rho_{\rm plasma} = + \frac{\mathcal{P}_{\rm fric}}{\Delta V}\dot{\rho}_{V} -  \rho_{\rm plasma}/\tau_{\rm plasma}\,, \\
\label{eq:rho_scalar}
&\dot{\rho}_{\rm scalar}  + 3(1+\omega_{\rm scalar})H\rho_{\rm scalar} = +\dot{I}\rho_{\rm wall} -  \rho_{\rm scalar}/\tau_{\rm scalar}\,, \\
\label{eq:rho_th}
&\dot{\rho}_{\rm th}  + 4H\rho_{\rm th} = + \rho_{\rm plasma}/\tau_{\rm plasma}+ \rho_{\rm scalar}/\tau_{\rm scalar}\,.
\end{align}
The equation of state of plasma excitations, e.g. sound waves, turbulence and heat, is expected to be radiation-like\footnote{It is worth noting that our prediction contradicts the claims reported in Ref.~\cite{Niedermann:2020dwg}.}, $\omega_{\rm plasma}= 1/3$. 
For relativistic bubble walls, scalar waves are expected to be relativistic \cite{Watkins:1991zt,Kolb:1996jr,Konstandin:2011ds,Falkowski:2012fb,Cutting:2018tjt,Cutting:2020nla}, and therefore to also follow a radiation-like equation of state $\omega_{\rm scalar}=1/3$. A detailed study of the dynamics and its implications for the formation of  PBHs is postponed to future work. In particular, if the universe just after percolation is dominated by long-lived and non-relativistic scalars waves, with $\tau_{\rm scalar}H\gtrsim 1$ and $\omega_{\rm scalar}\simeq 0$, then the formation of PBHs would take place during  matter-domination where the collapse threshold is much lower than during radiation-domination \cite{Harada:2016mhb}.

Finally, we comment that the super-cooled plasma energy density $\rho_{\rm cool}$ which is already present in the universe before the 1stOPT begins, is also relativistic and is diluted by the vacuum-domination period.   Consequently, this component plays little role.

{\textbf{Total radiation energy density.}}
To summarize, the energy density components associated with relativistic bubble walls all behave like radiation and we can decompose the total energy density of the universe as the sum of a vacuum and radiation component
\begin{equation}
\label{eq:rho_tot}
    \rho_{\rm tot} = \rho_{\rm V} + \rho_{\rm R},
\end{equation}
with
\begin{equation}
\rho_{\rm R} \equiv \rho_{\rm wall}+ \rho_{\rm plasma}+ \rho_{\rm scalar}+\rho_{\rm th}+\rho_{\rm cool}\,.
\end{equation}
Summing Eqs.~\eqref{eq:rho_wall_time_evolution}, \eqref{eq:rho_plasma}, \eqref{eq:rho_scalar} and \eqref{eq:rho_th}, we obtain the evolution 
\begin{equation}
\label{eq:rho_R}
\dot{\rho}_{\rm R}(t;t_{n_i})  + 4H\rho_{\rm R}(t;t_{n_i}) = - \dot{\rho}_{V}(t;t_{n_i}) \,,
\end{equation}
with $\rho_{\rm V}(t;t_{n_i})$ given by Eq.~\eqref{eq:rho_V} and the scale factor given by
\begin{equation}
\label{eq:Friedman_eq}
H(t;t_{n_i}) = \frac{\dot{a}}{a}= \sqrt{\frac{\rho_{\rm V}(t;t_{n_i})+\rho_{\rm R}(t;t_{n_i})}{3M_{\rm Pl}^2}}\,,
\end{equation}
with $M_{\rm pl} = 2.44\times 10^{18}\rm ~GeV$. 
Note that before nucleation begins, the vacuum energy is constant $\rho_{\rm V} = \Delta V$ and the Friedmann equation in Eq.~\eqref{eq:Friedman_eq} has an exact solution:
\begin{equation}
\label{eq:scale_fac_sinh}
    a(t) = a_{\rm eq}\sqrt{\sinh{\left(\sqrt{2} H_{\rm eq} t \right)}}\,,\qquad \text{with}\quad H_{\rm eq} = \sqrt{\frac{2\Delta V}{3M_{\rm pl}^2}}\,,
\end{equation}
which for $t\gg H_{\rm eq}^{-1}$, evolves as 
\begin{equation}
     \label{eq:scale_factor_exp_0}
    a(t) \propto \exp( H t), \qquad \text{with}\quad H = \sqrt{\frac{\Delta V}{3M_{\rm pl}^2}}.
\end{equation}
Assuming that $\rho_{\rm V}\simeq \Delta V$ is still valid at the instantaneous nucleation time $t_n$ in Eq.~\eqref{eq:definition_Hn}, we can plug Eq.~\eqref{eq:scale_fac_sinh} into Eq.~\eqref{eq:alpha_def} and obtain
\begin{equation}
\label{eq:definition_tn}
    t_n = \sqrt{\frac{3M_{\rm pl}^2}{4\Delta V}}\times\textrm{arcsh}(\sqrt{\alpha})\,.
\end{equation}

{\textbf{Latent heat fractions.}} 
If we neglect Hubble expansion $HR \ll 1$ and thermal friction $\mathcal{P}_{\rm fric} \ll \Delta V$, the solution of Eq.~\eqref{eq:eom_gamma} is
\begin{equation}
\gamma_w(R) ~=~\frac{\Delta V R}{3\sigma} + \frac{c}{R^2}\,,
\end{equation}
where $c$ depends on the initial condition. Neglecting the second term and plugging it into Eq.~\eqref{eq:rho_wall_def}, we get 
\begin{equation}
\label{eq:rho_wall_def_2}
\frac{\rho_{\rm V}}{\Delta V}  =  F, \qquad \frac{\rho_{\rm wall}}{\Delta V}  =  F_{\rm wall},\qquad \frac{\rho_{\rm coll}}{\Delta V}  = F_{\rm coll},
\end{equation}
with 
\begin{equation}
F \equiv e^{-I},\qquad F_{\rm wall} \simeq  Ie^{-I},\qquad  F_{\rm coll} \simeq \left(1-e^{-I} \right)\left(1-I e^{-I} \right)= 1-(1+I)e^{-I}.
\end{equation}
The quantities $F$, $F_{\rm wall}$ and $F_{\rm coll}$ are the fractions of latent heat $\Delta V$  stored in false vacuum, expanding bubble walls, and collided components, respectively. Their sum is unity due to energy conservation once the Hubble expansion is neglected. This confirms that Eq.~\eqref{eq:rho_wall_def} is correct.}  Since the expansion is crucial for  generating the inhomogeneities required for the formation of PBHs, Eqs.~\eqref{eq:rho_wall_def_2} can not replace the numerical integration of Eq.~\eqref{eq:rho_R}.

\begin{figure}[t!]
\centering
\raisebox{0cm}{\makebox{\includegraphics[ width=0.7\textwidth, scale=1]{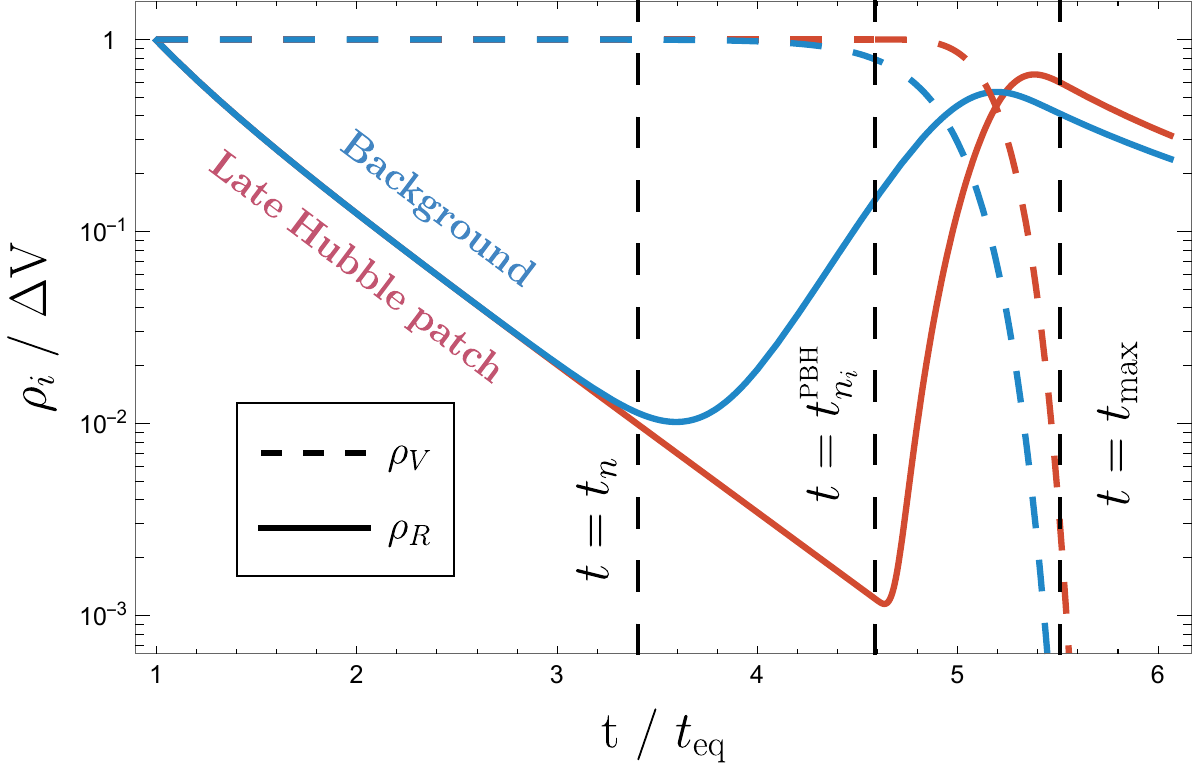}}}
\raisebox{0cm}{\makebox{\includegraphics[ width=0.7\textwidth, scale=1]{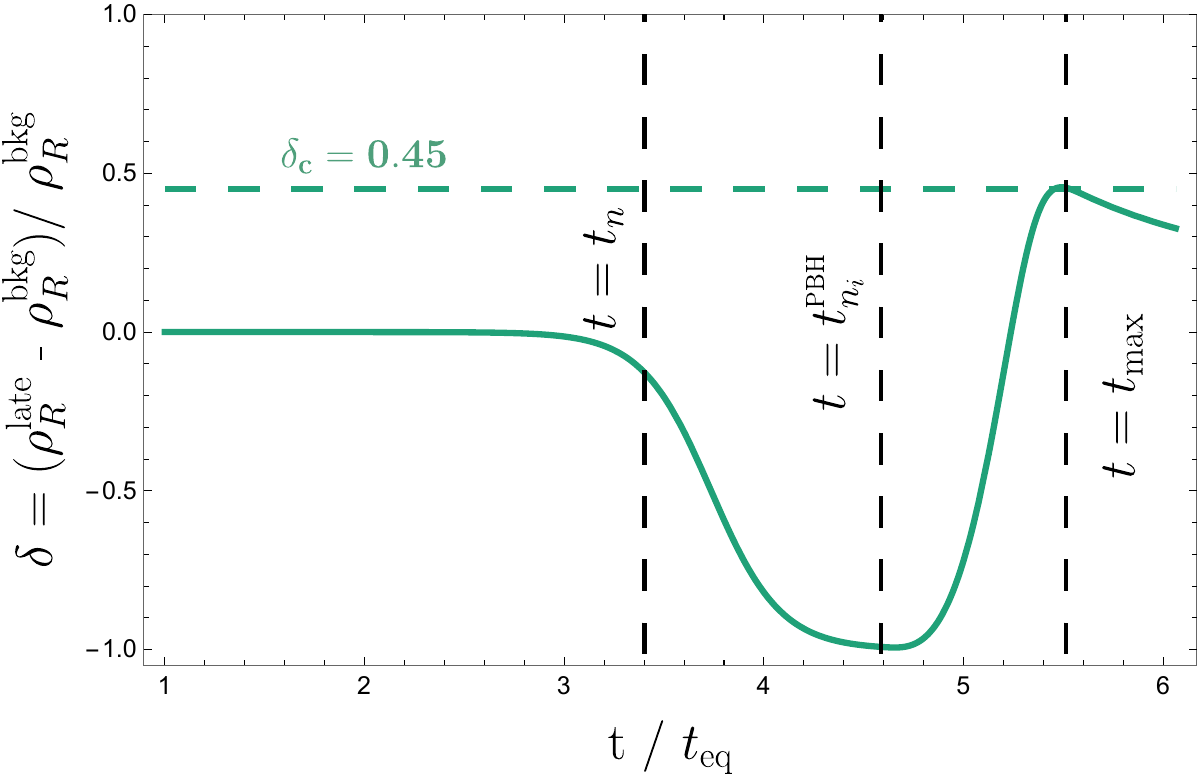}}}
\caption{  \label{fig:rho_evolution} Time evolution of the vacuum (dotted) and radiation (solid) energy density during a supercooled 1stOPT. {\bf Top:} we compare the evolution of the background (\textbf{blue}) for which nucleation starts around $t_n$, to the evolution of a late causal patch (\textbf{red}) for which nucleation is delayed until $t_{n_i}^{\rm \mathsmaller{PBH}}$, causing it to spend more time inflating and to reach the critical overdensity $\delta \simeq 0.45$.  {\bf Bottom:} the resulting density contract for the evolution is shown. The delayed patch collapses into a PBH close to  the time $t_{\rm max}$ when the density contrast $\delta$ reaches its peak value.  We have set $\beta/H=9$, $\alpha = 10^2$ and $T_{\rm eq}=0.01~\rm TeV$, even though we found the plots to be invariant under change of $T_{\rm eq}$.}
\end{figure}

\section{PBH formation dynamics}
\label{app:PBHformation}

\subsection{Supercooled late-blooming mechanism }
\label{sec:numerical_treatment}
  
Initially, the universe enters a stage of vacuum-domination at time $t_{\rm eq}$. After a time $t_c$ which may be shorter or longer than $t_{\rm eq}$, bubble nucleation becomes energetically possible but the universe continues to supercool until the time $t_n$ defined by the instantaneous approximation 
\begin{equation}
\label{eq:definition_Hn}
  \Gamma_{\mathsmaller{\rm V}}(t_n) \simeq H_n^4\qquad \textrm{where}\quad  H_n~ \equiv~ H(T_n)~=\sqrt{\frac{(1+\alpha^{-1})\Delta V}{3 M_{\rm pl}^2}},
\end{equation} 
with $\alpha$ being defined in Eq.~\eqref{eq:alpha_def}.
This is when most of the bubbles are nucleated in the universe.  Around the time $t_{\rm perc}^{\rm bkg}$ defined when $37\%$ of the universe has transitioned into the true vacuum, 
 the universe percolates.  However, some Hubble patches remain false-vacuum-dominated until a time $t_{n_i}$. These patches percolate at a time $t_{\rm perc}^{\rm late}>t_{\rm perc}^{\rm bkg}$. As a result of the difference in equation of state, the density contrast $\delta(t;t_{n_i})$ in late-blooming patches increases and reaches a maximum value at a time 
\begin{equation}
\label{eq:t_max}
   t_{\rm max}:\quad \quad \delta(t_{\rm max};t_{n_i}^{\rm PBH}) \equiv \underset{t}{\textrm{Max}}~\delta(t;t_{n_i}^{\rm PBH}).
\end{equation}
The time $t_{\rm max}$ is slightly larger than $t_{\rm perc}^{\rm late}$ as a result of the residual 37$\%$ of vacuum energy continuing to enhance the density contrast until only a small fraction (approximately 1$\%$) of the false vacuum remains.

{\textbf{PBH threshold.}}
The density contrast $\delta(t;t_{n_i})$ reads:
\begin{equation}
\label{eq:PBHs_threshold}
 \delta(t;t_{n_i}) \equiv \frac{\rho_{\rm tot}(t;t_{n_i})-\rho_{\rm tot}^{\rm bkg}(t)}{\rho_{\rm tot}^{\rm bkg}(t)},\qquad \textrm{with} \quad \rho_{\rm tot}^{\rm bkg}\equiv \rho_{\rm tot}(t;t_c)\,, 
\end{equation}
where $\rho_{\rm tot}(t;t_{n_i})$, defined to be the solution of Eq.~\eqref{eq:rho_R}, is the total energy density at time $t$ in a Hubble patch which remained $100\%$ false-vacuum-dominated until time $t_{n_i}$.
A late-blooming Hubble patch collapses into a PBH if
the contrast density $\delta(t;t_{n_i})$ becomes larger than the critical threshold $ \delta_{c} \simeq 0.50$.
We define $t_{n_i}^{\rm PBH}$ the minimal nucleation delay for this to happen. It is such that the peak value of $\delta(t;t_{n_i}^{\rm PBH}) $ at $t_{\rm max}$ saturates the critical threshold
\begin{equation}
\label{eq:PBHs_threshold_2}
 t_{n_i}^{\rm PBH}:\quad\quad \delta(t_{\rm max};t_{n_i}^{\rm PBH}) \equiv \delta_{c}.
\end{equation}
In Fig.~\ref{fig:rho_evolution} we numerically integrate Eq.~\eqref{eq:rho_R} and show in  the top figure the evolution of $\rho_R(t;t_c)$ and $\rho_V(t;t_c)$ in an average Hubble patch (blue) and a late-blooming Hubble patch (red) for which nucleation starts later than average and saturates the threshold defined in Eq.~\eqref{eq:PBHs_threshold_2}.   The bottom figure shows the evolution of the corresponding contrast density, with the horizontal dashed line indicating the critical value.

{\textbf{Collapse probability.}}  
\label{sec:collapse_probability}
For a PBH to form, it is necessary that the past light-cone of a Hubble-sized region remains in the false vacuum until the critical time $t_{n_i}^{\rm PBH}$. The probability of this occurrence, referred to as the collapse probability, can be expressed as follows:
\begin{equation}
\label{eq:P_coll_def_Psurv}
    \mathcal{P}_{\rm coll} \equiv \mathcal{P}_{\rm surv}(t_{n_i}^{\rm\mathsmaller{PBH}}; t_{\rm max})\,,
\end{equation}
with $t_{n_i}^{\rm PBH}$ defined in Eq.~\eqref{eq:PBHs_threshold_2}. Here, $\mathcal{P}_{\rm surv}(t_{n_i};t_{\rm max})$ represents the probability that the past light-cone of a Hubble patch at time $t_{\rm max}$ remains false-vacuum-dominated until time $t_{n_i}$. The Hubble patch must be evaluated at the time of its collapse into a PBH. Since a detailed understanding of the collapse dynamics is beyond the scope of this paper, we approximate it as the Hubble patch at time $t_{\rm max}$ when the contrast density $\delta(t;t_{n_i})$ reaches its peak value, cf. Eq.~\eqref{eq:t_max}. 
 To derive the expression for $\mathcal{P}_{\rm surv}(t_{n_i};t_{\rm max})$, we introduce the volume $V(t';t_{\rm max})$ which is the space-like slice at time $t'$ in the past light-cone of the Hubble patch at $t_{\rm max}$, (see Fig.~\ref{fig:space-time_diag}) 
\begin{equation}
\label{eq:causal_volume}
    V(t';t_{\rm max}) = \frac{4\pi}{3}\left(r_{\rm H}(t_{\rm max}) + r(t_{\rm max};t{'}) \right)^3\,,\qquad \textrm{with}\quad  r_{\rm H}(t) \equiv \frac{1}{a(t)H(t)}\,.
\end{equation}
Here $r(t_{\rm max};t')$, defined in Eq.~\eqref{eq:comoving_radius}, is the comoving distance traveled by a bubble wall between $t'$ and $t_{\rm max}$. 
The probability that a bubble nucleates in this volume in between the times $t'$ and $t'+dt'$ is
\begin{equation}
    dP_{\rm nuc}(t';t_{\rm max}) = dt'\, \Gamma_{\mathsmaller{\rm V}}(t{'}) a(t{'})^3 V(t';t_{\rm max}) \,.
\end{equation}
The probability $\mathcal{P}_{\rm surv}(t_{n_i};t_{\rm max})$ that no nucleation takes place in the past light-cone of the Hubble patch at $t_{\rm max}$ in the finite interval $(t_c;t_{n_i})$ is the $N\to \infty$ limit of the product $\prod_{k=1}^N\left[1-dP_{\rm nuc}(t_k;t_{\rm max})\right]$ where $1-dP_{\rm nuc}(t_k;t_{\rm max})$ is the probability of remaining in the false vacuum between $t_k=t_c+(t_{n_i}-t_c)k/N$ and $t_{k+1}$. We obtain the survival probability of the past light-cone of $H^{-1}(t_{\rm max})$ until time $t_{n_i}$
\begin{equation}
\label{eq:proba_surv_t_ni}
\mathcal{P}_{\rm surv}\left(t_{n_i}; t_{\rm max}\right) = \exp\left[ -\int_{t_c}^{t_{n_i}} dt'\, \Gamma_{\mathsmaller{\rm V}}(t{'}) a(t{'})^3V(t';t_{\rm max})  \right]\,.
\end{equation}
\begin{figure}[t!]
\centering
\raisebox{0cm}{\makebox{\includegraphics[ width=0.49\textwidth, scale=1]{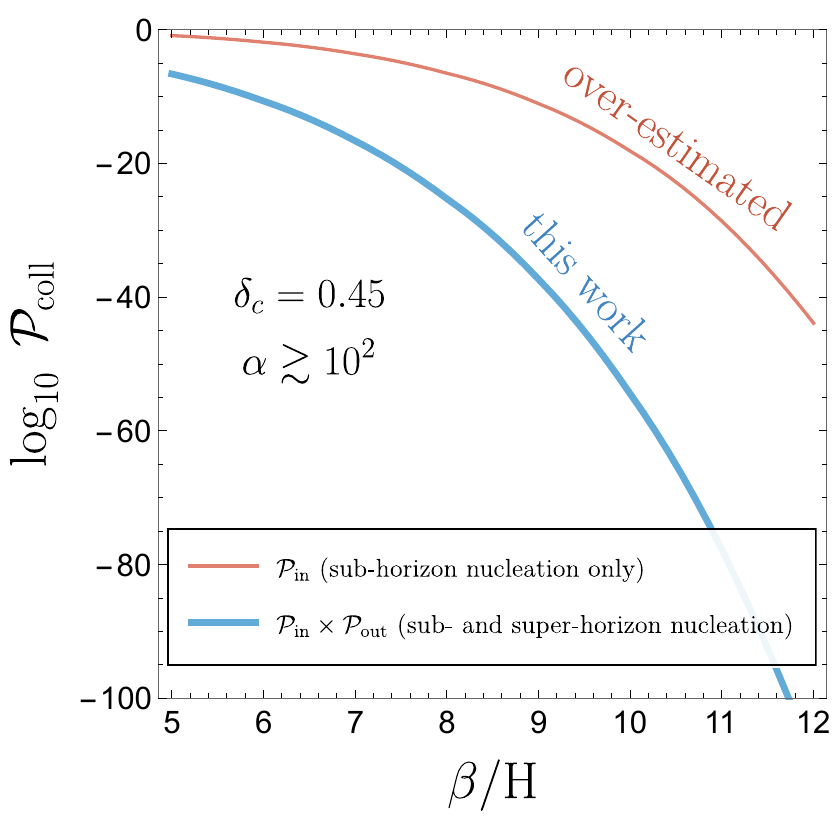}}}
\raisebox{0cm}{\makebox{\includegraphics[ width=0.49\textwidth, scale=1]{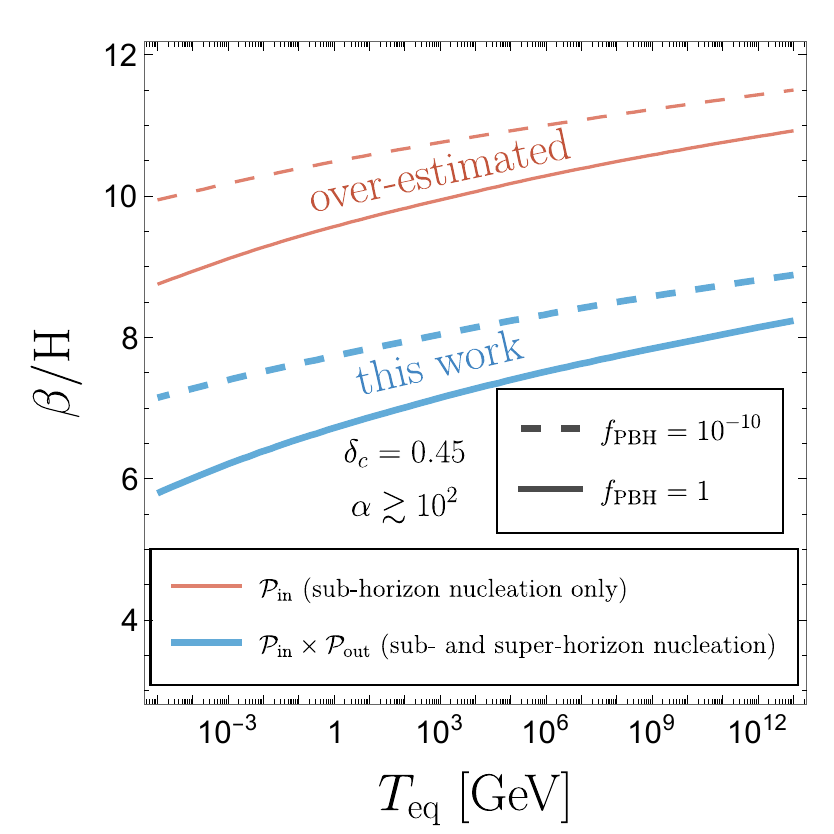}}}
\caption{  \label{fig:Psurv_betaOH_cone} {\bf Left:} the probability $\mathcal{P}_{\rm coll}$ in Eq.~\eqref{eq:P_coll_def_Psurv} for a Hubble patch to collapse into a PBH.  In {\bf blue} we show this probability  calculated by requiring  no nucleation to occur in its past-light cone until the critical delayed time $t_{n_i}^{\rm PBH}$.  In {\bf red} we show the naive expectation achieved by requiring no nucleation to occur within the Hubble patch only, i.e. within $V(t';t_{\rm max}) = 4\pi r^3_{\rm H}(t_{\rm max})/3$ instead of Eq.~\eqref{eq:causal_volume} (or equivalently assuming $\mathcal{P}_{\rm surv} = \mathcal{P}_{\rm in}$ instead of Eq.~\eqref{eq:proba_tni_PBHs}). This overestimate of the probability neglects bubbles nucleating outside the horizon but penetrating the horizon before the time of collapse around $t_{\rm max}$.   {\bf Right:} the required (inverse) duration $\beta/H$ of the phase transition as a function of the temperature $T_{\rm eq}$ at which the universe becomes vacuum-dominated, in order to achieve a PBH dark matter density of $1$ ({\bf solid lines}) and $10^{-10}$ ({\bf dashed lines}).  The colors indicate the same calculations for the collapse probability used in the left figure.  
}
\end{figure} 
In order to improve our understanding we can break-down the formula we just derived as follows.
First using that $dr(t_{\rm out};t_{n_i})=-v_w(t_{\rm out})dt_{\rm out}/a(t_{\rm out})$, cf. Eq.~\eqref{eq:comoving_radius}, and second performing an integration by part, we obtain
\begin{align}
    \textrm{log}~\mathcal{P}_{\rm surv}\left(t_{n_i}; t_{\rm max}\right) &= 4\pi\int_{t_c}^{t_{n_i}} dt'\, \Gamma_{\mathsmaller{\rm V}}(t{'}) a(t{'})^3\int_{t_c}^{t'} dt_{\rm out} \frac{v_w(t_{\rm out})}{a(t_{\rm out})}\left(r_{\rm H}(t_{\rm max}) + r(t_{\rm max};t_{\rm out}) \right)^2 \\
    &= 4\pi\int_{t_c}^{t_{n_i}} dt'\, \Gamma_{\mathsmaller{\rm V}}(t{'}) a(t{'})^3\int_{t_c}^{t_{n_i}} dt_{\rm out} \frac{v_w(t_{\rm out})}{a(t_{\rm out})}\left(r_{\rm H}(t_{\rm max}) + r(t_{\rm max};t_{\rm out}) \right)^2 \notag \\
    &\quad - 4\pi \int_{t_c}^{t_{n_i}} dt_{\rm out}\,\frac{v_w(t_{\rm out})}{a(t_{\rm out})}\left(r_{\rm H}(t_{\rm max}) + r(t_{\rm max};t_{\rm out}) \right)^2 \int_{t_c}^{t_{\rm out}} dt'\, \Gamma_{\mathsmaller{\rm V}}(t{'}) a(t{'})^3 \notag \\
       &= \int_{t_c}^{t_{n_i}} dt'\, \Gamma_{\mathsmaller{\rm V}}(t{'}) a(t{'})^3 \frac{4\pi}{3} r_{\rm H}(t_{\rm max})^3 + 4\pi \int^{r_{\rm out}(t_c;t_{\rm max})}_{r_{\rm H}(t_{n_i})} dr_{\rm out} \,r_{\rm out}^2 \int_{t_c}^{t_{\rm out}(r_{\rm out})} dt'\, \Gamma_{\mathsmaller{\rm V}}(t{'}) a(t{'})^3\, ,
\label{eq:decomposition_log_P_surv}
\end{align}
where in the last line we have performed the change of variables
\begin{equation}
\label{eq:wall_traj}
    t_{\rm out} \quad \to \quad r_{\rm out}\left(t_{\rm out};t_{\rm max}\right) = r_{\rm H}(t_{\rm max}) + r(t_{\rm max};t_{\rm out})\,.
\end{equation}
$r_{\rm out}\left(t_{\rm out};t_{\rm max}\right)$ is the distance from the center of a Hubble patch at $t_{\rm max}$ a bubble needs to nucleate at time $t_{\rm out}$ in order to penetrate inside the Hubble patch at $t_{\rm max}$. It defines the past light-cone of the Hubble patch at $t_{\rm max}$ shown in blue in Fig.~\ref{fig:space-time_diag}. $r_{\rm out}(t_c;t_{\rm max})$ is the furthest of those distances, $t_c$ being the time when nucleation becomes energetically allowed.  We use the label ``out'' to stress that the associated bubbles nucleate outside the Hubble patch at $t_{\rm max}$. 
Eq.~\eqref{eq:decomposition_log_P_surv} shows that the survival probability $\mathcal{P}_{\rm surv}\left(t_{n_i}; t_{\rm max}\right)$ can be decomposed as the product of two probabilities
\begin{equation}
\label{eq:proba_tni_PBHs}
\mathcal{P}_{\rm surv}\left(t_{n_i}; t_{\rm max}\right) = \mathcal{P}_{\rm in}\left(t_{n_i};  t_{\rm max}\right)  \times \mathcal{P}_{\rm out}\left(t_{n_i};  t_{\rm max}\right) \,,
\end{equation}
according to whether nucleation takes place inside (``in'') or outside (``out'') the Hubble patch at $t_{\rm max}$. The first factor
\begin{align}
\label{eq:P_hori}
\mathcal{P}_{\rm in}\left(t_{n_i};  t_{\rm max}\right) = \exp\left[ -\int_{t_c}^{t_{n_i}} dt{'}\, \Gamma_{\mathsmaller{\rm V}}(t{'}) a(t{'})^3V_{\rm H}^{\rm max} \right]\,,
\end{align}
is the probability of having no nucleation before time $t_{n_i}$ inside the Hubble patch at $t_{\rm max}$.
The second factor $\mathcal{P}_{\rm out}$ in Eq.~\eqref{eq:proba_tni_PBHs} is the probability of having no nucleation before time $t_{n_i}$ in regions of the past-like cone of the Hubble patch at $t_{\rm max}$, defined in Eq.~\eqref{eq:causal_volume}, which are outside the Hubble patch at $t_{\rm max}$
\begin{align}
\label{eq:P_cone}
\mathcal{P}_{\rm out}\left(t_{n_i};t_{\rm max}\right) = \exp\left[ -4\pi\int^{r_{\rm out}(t_c;t_{\rm max})}_{r_{\rm H}(t_{n_i})} dr_{\rm out} \,r_{\rm out}^2 \int_{t_c}^{t_{\rm out}\left(r_{\rm out};t_{\rm max}\right)}dt{'}\, \Gamma_{\mathsmaller{\rm V}}(t{'}) a(t{'})^3\right]\,.
\end{align}
It accounts for bubbles nucleating outside the Hubble horizon before $t_{n_i}$ but whose walls propagate inside the Hubble horizon before $t_{\rm max}$.
Here, $t_{\rm out}\left(r_{\rm out};t_{\rm max}\right)$ is the time at which a bubble needs to nucleate in order to enter the Hubble patch at $t_{\rm max}$, assuming it nucleates at a distance $r_{\rm out}>r_{\rm H}(t_{\rm max})$ from the center of the patch. The time $t_{\rm out}\left(r_{\rm out};t_{\rm max}\right)$ can be found from inverting the world line of the bubble wall $r_{\rm out}\left(t_{\rm out};t_{\rm max}\right)$ in Eq.~\eqref{eq:wall_traj}.  

To the best of our knowledge, the formula in Eq.~\eqref{eq:proba_surv_t_ni} was initially proposed by \cite{Kodama:1982sf} in 1982. However, the original version was simpler since it considered $t_{\rm max}$ and $t_{n_i}$ to be identical. This last point has been addressed more recently in \cite{Liu:2021svg} which however misses the bubble wall piece in Eq.~\eqref{eq:causal_volume}, or equivalently the factor $\mathcal{P}_{\rm out}$ in Eq.~\eqref{eq:proba_tni_PBHs}.\footnote{While this paper was in writing, a distinct derivation for the survival probability decomposition, Eq.~\eqref{eq:proba_tni_PBHs}, appeared in~\cite{Kawana:2022olo}. However, the formula put  }
\blfootnote{forward in~\cite{Kawana:2022olo} does not distinguish the time $t_{\rm max}$ of collapse and the time $t_{n_i}$ when nucleation starts, which implies that only patches which are}
\blfootnote{completely vacuum-dominated are allowed to collapse. Refs.~\cite{Lewicki:2023ioy,Flores:2024lng}, which appeared after the submission of the first version of this work on the}
\blfootnote{arxiv, also limit the gravitational collapse to areas that are entirely in a vacuum state, with $F=1$ in Eq.~\eqref{eq:F_vacuum_fraction}.}
\blfootnote{In contrast, the approach we present in this paper allows for gravitational collapse in regions with varying fraction of vacuum, ranging from $0\leq F \leq 1$. } In Fig.~\ref{fig:Psurv_betaOH_cone}, we show that the inclusion of the nucleation outside the Hubble patch considerably suppresses the collapse probability and moves the region of interest by $2$ units in $\beta/H$.  

\begin{figure}[t!]
\centering
\raisebox{0cm}{\makebox{\includegraphics[ width=0.7\textwidth, scale=1]{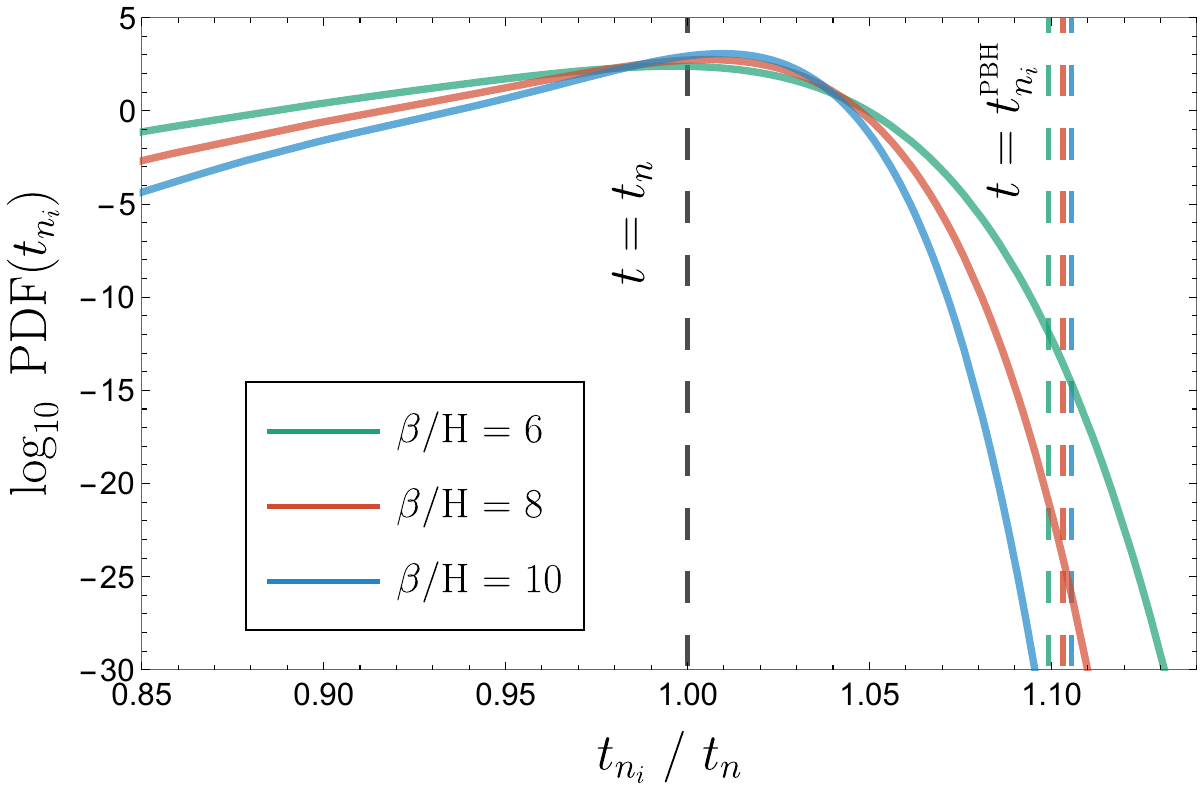}}}
\caption{  \label{fig:P_tni_PBHs} Probability Density Function (PDF) as a function of $t_{n_i}$ at which the first bubble nucleates, in units of the instantaneous nucleation time, $t_n$ defined through Eq.~\eqref{eq:definition_Hn}. Three different values of $\beta/H$ are shown, assuming a latent heat parameter $\alpha=10^4$ (see Eqs.~\eqref{eq:beta_def} and~\eqref{eq:alpha_def}).   In order to collapse into a PBH, the past light-cone of a given Hubble patch must remain false-vacuum-dominated until the second vertical dashed line $t_{n_i} = t_{n_i}^{\rm \mathsmaller{PBH}}$.  The survival probability defined in Eq.~\eqref{eq:P_coll_def_Psurv} is the PDF value  at $t_{n_i}^{\rm \mathsmaller{PBH}}$. 
}
\end{figure}

Finally, we use the survival probability $\mathcal{P}_{\rm surv}\left(t_{n_i}, t_{\rm max}\right)$ to compute the Probability Density Function (PDF) for the nucleation delay $t_{n_i}$ with a Monte-Carlo procedure.\footnote{We draw values $t_{n_i}$ from numerically solving the root equation $\mathcal{P}_{\rm surv}\left(t_{n_i}, t_{\rm max}\right)=[0,1]$ where $[0,1]$ is a random number between $0$ and $1$. \blfootnote{It has later been shown that the PDF for $t_{n_i}$ is a Poisson distribution \cite{Lewicki:2024ghw}.}}  In Fig.~\ref{fig:P_tni_PBHs}, we show the resulting PDFs as a function of $t_{n_i}/t$, for three values of $\beta/H$ and taking $\alpha=10^4$.  One can see that the PDFs peak around the ``instantaneous'' nucleation time $t_n$ given by Eq.~\eqref{eq:definition_tn}, and that further delay is exponentially suppressed in probability.

{\textbf{PBHs abundance.}}  
\label{sec:PBH_abundance}
The fraction $f_{\rm \mathsmaller{PBH}}=\rho_{\rm \mathsmaller{PBH},0}/\rho_{\rm \mathsmaller{DM},0}$ of Dark Matter (DM) in the form of PBHs today is given by
\begin{equation}
f_{\rm \mathsmaller{PBH}}  = \mathcal{P}_{\rm coll}\frac{M_{\rm \mathsmaller{PBH}}\mathcal{N}_{\rm patches}}{\rho_{\rm \mathsmaller{DM},0}\frac{4\pi}{3}H_0^{-3}} \simeq \left(\frac{\mathcal{P}_{\rm coll}}{ 6.0 \times 10^{-12}}\right)\left( \frac{T_{\rm eq}}{500~\rm GeV} \right)\,.
\end{equation}
where the collapsed fraction $\mathcal{P}_{\rm coll}$ is defined in Eq.~\eqref{eq:P_coll_def_Psurv}.
We introduced the current DM density $\rho_{\rm \mathsmaller{DM},0} \simeq 0.26 \times 3M_{\rm pl}^2H_0^2$ and the number $\mathcal{N}_{\rm patches}$ of Hubble patches, at the time when the universe temperature was $T_{\rm eq}$, in our past light-cone
\begin{align}
\label{eq:N_patches}
\mathcal{N}_{\rm patches}  
= \left( \frac{a_{\rm eq} H_{\rm eq}}{a_0 H_{0}} \right)^3\simeq 5.3 \times 10^{40} \left(\frac{g_*(T_{\rm eq})}{100}\right)^{1/2}\left( \frac{T_{\rm eq}}{500~\rm GeV} \right)^{3}.
\end{align}
The scale factor ratio $a_{\rm eq}/a_0$ is related to the temperature ratio assuming adiabatic universe expansion $a_{\rm eq}/a_{0}=(g_{*}(T_{0})/g_{*s}(T_{\rm eq}))^{1/3}(T_0/T_{\rm eq})$ with $g_{*}(T_{0})\simeq 3.94$.
The fraction of collapsed PBHs $\mathcal{P}_{\rm coll}$ and the associated DM fraction $f_{\rm \mathsmaller{PBH}}$ are shown in Fig.~\ref{fig:Psurv_betaOH_cone}. The resulting probability depends very sensitively on the duration $\beta^{-1}$ of the phase transition.
The PBHs abundance is shown in Fig.~\ref{fig:PBHs_DM_TnOTc_delta_c}.

\subsection{Analytical approximation}
\label{sec:analytical_treatment}
Using the approximation of the scale factor given by $a\simeq \exp\left(H_n t \right)$, the fraction of false vacuum in Eq.~\eqref{eq:F_vacuum_fraction} can be calculated as follows:
\begin{align}
\ln{F(t;t_{n_i})} &= - \int_{t_{n_i}}^t d\tilde{t}\, \Gamma_{\mathsmaller{\rm V}}(\tilde{t}) \,\frac{4\pi}{3H_n^3}\left(e^{-H_n\tilde{t}} - e^{-H_nt} \right)^3 \notag \\
&=-\frac{4\pi}{3}\frac{H_n}{\beta} e^{\beta(t_{n_i}-t_n)}\left(-1 + \frac{3e^{H_n(t_{n_i}-t)}}{1+H_n/\beta} - \frac{3e^{2H_n(t_{n_i}-t)}}{1+2H_n/\beta} + \frac{6 e^{\beta(t-t_{n_i})}}{(1+H_n/\beta)(2+H_n/\beta)(3+H_n/\beta)} \right) \notag \\
&\simeq -\frac{8\pi H_n^4}{\beta^4}\left(e^{\beta (t-t_n)}-e^{\beta (t_{n_i} - t_n)}\right), \label{eq:F_vs_t}
\end{align}
where $t_{n_i}$ represents the time when the first bubble is nucleated in the past light-cone and $t_n$ denotes the nucleation time in the instantaneous approximation defined in Eq.~\eqref{eq:definition_Hn}. In the last line we took the limits $\beta^{-1}, \, (t_{n_i}-t) \ll H_n^{-1}$. By inverting Eq.~\eqref{eq:F_vs_t}, we can obtain the time $t(F; t_{n_i})$ when only a fraction $F$ of the false vacuum remains:
\begin{equation}
\label{eq:tstar_def}
    t(F;t_{n_i}) \simeq \beta^{-1} \ln\left[e^{\beta t_{n_i}} + \frac{\beta^4}{8\pi H_n^4}e^{\beta t_n}\ln{(F^{-1})} \right].
\end{equation} 
Percolation occurs when the false vacuum fraction becomes smaller than some critical value $F_{\rm perc}$ (e.g. $F_{\rm perc}\simeq 37\%$)
\begin{equation}
    t_{\rm perc}(t_{n_i}) \equiv t(F_{\rm perc};t_{n_i}).
\end{equation}
The energy density in radiation, solution of Eq.~\eqref{eq:rho_R}, reads
\begin{equation}
\label{eq:rho_R_solution_0}
    \rho_{\rm R}(t;t_{n_i}) =  -\Delta V \int_{t_{n_i}}^t d\tilde{t}\, \left(\frac{a(\tilde{t})}{a(t)}\right)^{\!4} \frac{dF(\tilde{t}; t_{n_i})}{d\tilde{t}}.
\end{equation}
Upon approximating the false vacuum fraction $F(t;t_{n_i})$ by the Heaviside function
\begin{equation}
    F(t; t_{n_i}) \simeq \Theta\left(  t_{\rm perc}(t_{n_i})-t\right),
\end{equation}
we can derive an analytical estimate for Eq.~\eqref{eq:rho_R_solution_0}:
\begin{equation}
\label{eq:rho_bkg_evol}
    \rho_{\rm R}(t; t_{n_i}) \simeq  \Delta V  \left(\frac{a(t_{\rm perc}(t_{n_i}))}{a(t)}\right)^{\!4}  \Theta\left(t- t_{\rm perc}(t_{n_i}))\right) .
\end{equation}
The percolation time $t_{\rm perc}^{\rm bkg}$ in an average Hubble patch (``bkg'') where nucleation starts in principle as soon as it becomes energetically allowed $t_{n_i}\simeq t_c$, and the percolation time $t_{\rm perc}^{\rm late}$ in a late patch (``late'') where $t_{n_i}\gtrsim t_n$, read
\begin{equation}
\label{eq:tstar_bkg_late}
t_{\rm perc}^{\rm bkg} \equiv t(F_{\rm perc};t_c) \,,\qquad t_{\rm perc}^{\rm late} \equiv t(F_{\rm perc};t_{n_i})\, . 
\end{equation}
For $t \gtrsim t_{\rm perc}^{\rm bkg}$ in spite of being filled with newly formed radiation with equation of state $\omega(t) \simeq 1/3$, we numerically find that the scale factor of the universe still grows exponentially for a while longer:
 \begin{equation}
 \label{eq:scale_factor_exp}
 a(t) ~\simeq ~ \exp( H_nt)\,.
 \end{equation}
This is due to the inertia of the expansion rate encoded in the Friedmann equation $2a(t)\ddot{a}(t)=-(1+3\omega(t))\dot{a}^2(t)$. We approximate the Hubble factor in Eq.~\eqref{eq:scale_factor_exp} to be constant.
Meanwhile, late patches continue to be vacuum-dominated  until they are converted into radiation around the time $t_{\rm perc}^{\rm late}$ in Eq.~\eqref{eq:tstar_bkg_late} which we approximate to be equal to the time $t_{\rm max}$ in Eq.~\eqref{eq:t_max} when the density contrast reaches its maximal value\footnote{The accuracy of this approximation can be evaluated from the comment below Eq.~\eqref{eq:t_max}.}
 \begin{equation}
 \label{eq:t_perc_late_max}
     t_{\rm perc}^{\rm late} \simeq t_{\rm max}.
 \end{equation}
Using the exponential scale factor in Eq.~\eqref{eq:scale_factor_exp} along with the radiation energy density in Eq.~\eqref{eq:rho_bkg_evol}, we get the contrast density at $t_{\rm max}$ of a delayed patch with respect to the background, both characterized by Eq.~\eqref{eq:tstar_bkg_late}
\begin{equation}
\label{eq:contrast_density_ana}
\frac{\rho_{\rm R}^{\rm late}-\rho_{\rm R}^{\rm bkg}}{\rho_{\rm R}^{\rm bkg}}\Big|_{t=t_{\rm max}} \simeq \frac{1-e^{4H_n\left(t_{\rm perc}^{\rm bkg} - t_{\rm max}\right)}}{e^{4H_n\left(t_{\rm perc}^{\rm bkg} - t_{\rm max}\right)}}\,.
\end{equation}
The last calculation assumes the approximation specified in Eq.~\eqref{eq:t_perc_late_max} which neglects the redshift of radiation inside delayed patches before $t_{\rm max}$.
A delayed patch collapses if Eq.~\eqref{eq:contrast_density_ana} is above the critical threshold $\delta_c$.
Using Eqs.~\eqref{eq:tstar_def} and \eqref{eq:tstar_bkg_late} to express $t_{\rm perc}^{\rm bkg}$ and $t_{\rm max}$ as a function of the corresponding false vacuum fractions $F_{\rm perc}$ and $F_{\rm max}$, we compute the minimal delayed time $t_{n_i}^{\rm \mathsmaller{PBH}}$ of the onset of nucleation for a late patch to collapse into a PBH
\begin{equation}
\label{eq:tni_PBHs_approx}
\end{equation}
\begin{figure}[t!]
\centering
\raisebox{0cm}{\makebox{\includegraphics[ width=0.49\textwidth, scale=1]{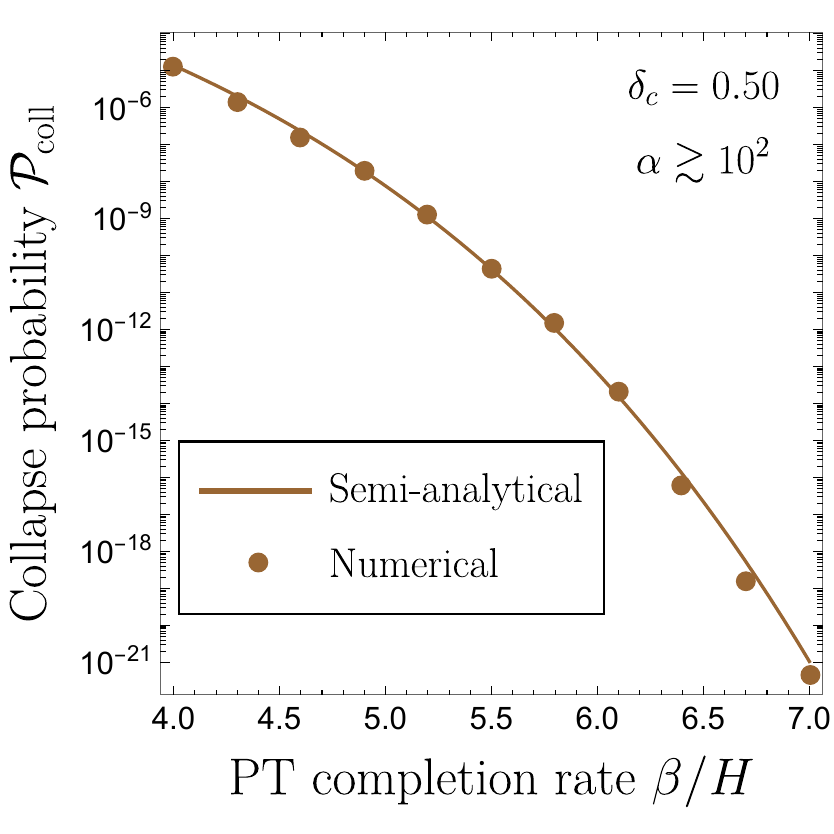}}}
\raisebox{0cm}{\makebox{\includegraphics[ width=0.49\textwidth, scale=1]{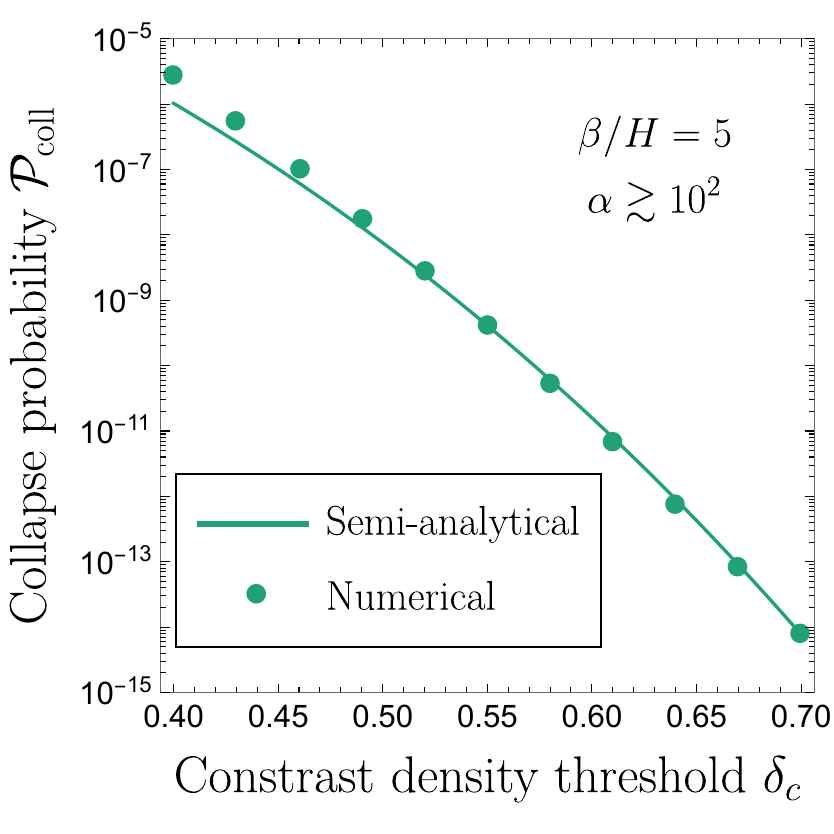}}}
\caption{  \label{fig:logP_vs_betaOH} Fraction of causal patches which collapse into a PBH $\mathcal{P}_{\rm coll} \equiv \mathcal{P}_{\rm surv}(t_{n_i}^{\rm PBH}; t_{\rm max})$ at the end of a supercooled phase transition. We assume that the supercooling period lasts at least $N_e = 0.25 \log(\alpha) \gtrsim 1$ e-folding so that the PBH abundance is independent of $N_e$. We show the dependence on the phase transition rate of completion $\beta/H$ (\textbf{left}) and on the density contrast threshold $\delta_c$ (\textbf{right}). We compare the numerical calculation discussed in Sec.~\ref{sec:numerical_treatment} to the semi-analytical formula, Eq.~\eqref{eq:proba_coll_ana_app}. }
\end{figure}
We now calculate the collapse probability using the probability $\mathcal{P}_{\rm coll} = \mathcal{P}_{\rm surv}(t_{n_i}^{\rm \mathsmaller{PBH}};t_{\rm max})$ defined in Eq.~\eqref{eq:proba_surv_t_ni} for no nucleation occurring within the past light-cone of a Hubble patch at $t_{\rm max}$ before the critical delayed time $t_{n_i}^{\rm \mathsmaller{PBH}}$. By using the exponential scale factor we can simplify the causal volume in Eq.~\eqref{eq:causal_volume} to  \begin{equation}
\label{eq:causal_volume_2}
    V(t';t_{\rm max}) \simeq \frac{4\pi}{3H_n^3}\left(e^{-{H_n t'}}-e^{-{H_n t_{\rm max}}}\left(1 -\frac{H_n}{H_{\rm max}}\right) \right)^3,
\end{equation}
where $H_n=H(t_n)$ and $H_{\rm max}=H(t_{\rm max})$.
To derive an analytical expression for the survival probability, we approximate $H_{\rm max}\simeq H_n$, leading to a simplified expression for the causal volume \begin{equation}
\label{eq:causal_volume_3}
    V(t';t_{\rm max}) \simeq \frac{4\pi}{3H_n^3}e^{-{3H_n t'}} .
\end{equation}
Substituting the exponential causal volume in Eq.~\eqref{eq:causal_volume_3} and the exponential tunneling rate in Eq.~\eqref{eq:tunneling_rate_def_0} into Eq.~\eqref{eq:proba_surv_t_ni}, we arrive at the survival probability
\begin{equation}
\label{eq:proba_tni_PBHs_approx}
 \mathcal{P}_{\rm surv}(t_{n_i}; t_{\rm max}) \simeq \exp\left[- \frac{4\pi}{3}\frac{H_n}{\beta} e^{\beta(t_{n_i}-t_n)}\right].
\end{equation}
Note that we have neglected a $t_c$-dependent term in Eq.~\eqref{eq:proba_tni_PBHs_approx}, which is valid for $t_{n_i}^{\rm \mathsmaller{PBH}} -t_c \gg \beta^{-1}$. Also note that because $H_{\rm max}< H_n$, the expression in Eq.~\eqref{eq:causal_volume_3} underestimates the causal volume, and the expression in Eq.~\eqref{eq:proba_tni_PBHs_approx} overestimates the survival probability.
Plugging Eq.~\eqref{eq:tni_PBHs_approx} into Eq.~\eqref{eq:proba_tni_PBHs_approx} gives
\begin{equation}
\label{eq:proba_tni_PBHs_approx_2}
\mathcal{P}_{\rm coll}~\simeq~ \exp\left[-\frac{1}{6}\left( \frac{\beta}{H_n} \right)^3\left((1+\delta_c)^{\frac{\beta}{4H_n}}\ln{(F_{\rm perc}^{-1})}-\ln{(F_{\rm max}^{-1})} \right)\right]\,,
\end{equation}
where $\mathcal{P}_{\rm coll}\equiv \mathcal{P}_{\rm surv}(t_{n_i}^{\rm \mathsmaller{PBH}};t_{\rm max})$.
This motivates the semi-analytical expression in the main text, cf.  Eq.~\eqref{eq:proba_coll_ana}, which we report here
\begin{equation}
\label{eq:proba_coll_ana_app}
\mathcal{P}_{\rm coll}\simeq \exp\left[-a\left( \frac{\beta}{H_n} \right)^{b}\left(1+\delta_{\rm c}\right)^{c\frac{\beta}{H_n}} \right]\,,
\end{equation}
where $a\simeq 1.024$, $b\simeq 0.6921$, $c\simeq 0.8831$ are parameters fitted against numerical calculations. We numerically find the value $b\simeq  0.6921$ to be preferred over the value $b\simeq  3$ suggested by Eq.~\eqref{eq:proba_tni_PBHs_approx_2}.
The observed discrepancy between Eqs.~\eqref{eq:proba_tni_PBHs_approx_2} and \eqref{eq:proba_coll_ana_app} is in line with the multiple levels of approximations employed in the calculation. A more thorough analytical investigation is warranted to gain a deeper understanding of this behavior, which we leave for future studies. We show the semi-analytical survival probability together with the numerical results in Figs.~\ref{fig:PBHs_DM_TnOTc_delta_c} and \ref{fig:logP_vs_betaOH}.

\newpage

\section{Constraints on PBHs}

\begin{figure}[ht!]
\centering
\includegraphics[width=0.75\textwidth]{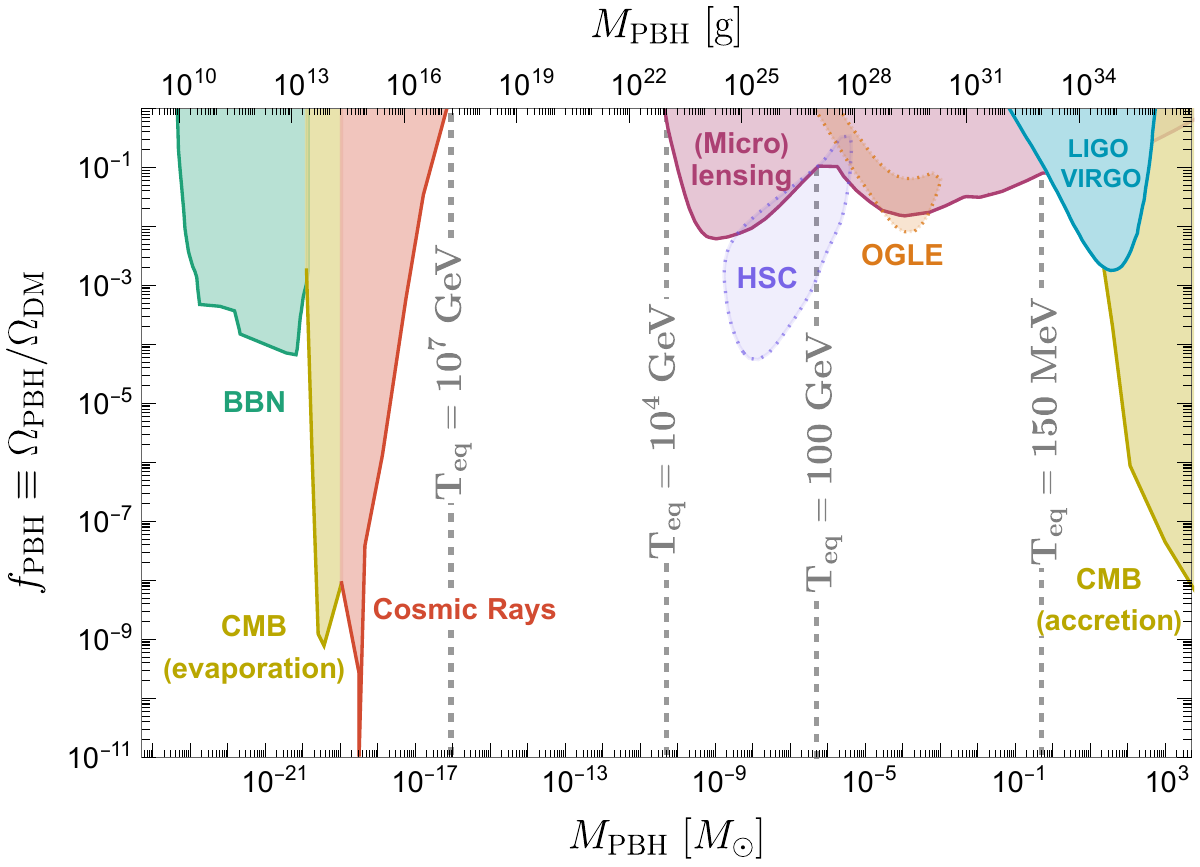}
\caption{  \label{fig:PBHs_constraints} Cosmological and astrophysical constraints on the abundance of PBHs. We indicate with vertical dashed lines the temperature $T_{\rm eq}$ when the universe starts supercooling during a phase transition which would form PBHs at the corresponding mass $M_{\rm \mathsmaller{PBH}}$. From right to left, we have accretion during CMB \cite{Ali-Haimoud:2016mbv,Poulin:2017bwe,Serpico:2020ehh}, merging events in LIGO-Virgo \cite{DeLuca:2020qqa}, microlensing experiments \cite{MACHO:2000qbb,EROS-2:2006ryy,Niikura:2019kqi,Niikura:2017zjd}, cosmic-ray fluxes \cite{Carr:2009jm,Boudaud:2018hqb,Laha:2019ssq,Dasgupta:2019cae,Laha:2020ivk,Ray:2021mxu}, evaporation during CMB \cite{Poulin:2016anj,Stocker:2018avm,Poulter:2019ooo}, and BBN \cite{Carr:2009jm}. The dotted regions are the posterior distributions inferred from anomalous microlensing events reported in HSC and OGLE data \cite{Niikura:2019kqi,Niikura:2017zjd,Sugiyama:2021xqg}.}
\end{figure}

 The abundance of primordial black holes (PBHs) below $10^{17}~\rm g$ is constrained through the process of Hawking radiation, which is the loss of mass and eventual evaporation of PBHs. If there are too many PBHs, they could have negative impacts on Big-Bang Nucleosynthesis (BBN), the Cosmic Microwave Background (CMB), and cosmic-rays. Instead, PBHs with asteroid mass $10^{17}~\rm g \lesssim M_{\rm PBH} \lesssim 10^{23}~\rm g$ could compose 100$\%$ of Dark Matter. At larger masses, PBH are constrained by microlensing, $\rm kHz$ GW interferometers and accretion in the CMB. We summarize the constraints together with the associated references in Fig.~\ref{fig:PBHs_constraints}.

\newpage

\bibliography{biblio}

\begin{thebibliography}{150}%
\makeatletter
\providecommand \@ifxundefined [1]{%
 \@ifx{#1\undefined}
}%
\providecommand \@ifnum [1]{%
 \ifnum #1\expandafter \@firstoftwo
 \else \expandafter \@secondoftwo
 \fi
}%
\providecommand \@ifx [1]{%
 \ifx #1\expandafter \@firstoftwo
 \else \expandafter \@secondoftwo
 \fi
}%
\providecommand \natexlab [1]{#1}%
\providecommand \enquote  [1]{``#1''}%
\providecommand \bibnamefont  [1]{#1}%
\providecommand \bibfnamefont [1]{#1}%
\providecommand \citenamefont [1]{#1}%
\providecommand \href@noop [0]{\@secondoftwo}%
\providecommand \href [0]{\begingroup \@sanitize@url \@href}%
\providecommand \@href[1]{\@@startlink{#1}\@@href}%
\providecommand \@@href[1]{\endgroup#1\@@endlink}%
\providecommand \@sanitize@url [0]{\catcode `\\12\catcode `\$12\catcode
  `\&12\catcode `\#12\catcode `\^12\catcode `\_12\catcode `\%12\relax}%
\providecommand \@@startlink[1]{}%
\providecommand \@@endlink[0]{}%
\providecommand \url  [0]{\begingroup\@sanitize@url \@url }%
\providecommand \@url [1]{\endgroup\@href {#1}{\urlprefix }}%
\providecommand \urlprefix  [0]{URL }%
\providecommand \Eprint [0]{\href }%
\providecommand \doibase [0]{https://doi.org/}%
\providecommand \selectlanguage [0]{\@gobble}%
\providecommand \bibinfo  [0]{\@secondoftwo}%
\providecommand \bibfield  [0]{\@secondoftwo}%
\providecommand \translation [1]{[#1]}%
\providecommand \BibitemOpen [0]{}%
\providecommand \bibitemStop [0]{}%
\providecommand \bibitemNoStop [0]{.\EOS\space}%
\providecommand \EOS [0]{\spacefactor3000\relax}%
\providecommand \BibitemShut  [1]{\csname bibitem#1\endcsname}%
\let\auto@bib@innerbib\@empty
\bibitem [{\citenamefont {Abbott}\ \emph
  {et~al.}(2019{\natexlab{a}})\citenamefont {Abbott} \emph
  {et~al.}}]{LIGOScientific:2018mvr}%
  \BibitemOpen
  \bibfield  {author} {\bibinfo {author} {\bibfnamefont {B.~P.}\ \bibnamefont
  {Abbott}} \emph {et~al.} (\bibinfo {collaboration} {LIGO Scientific,
  Virgo}),\ }\bibfield  {title} {\bibinfo {title} {{Gwtc-1: a
  Gravitational-Wave Transient Catalog of Compact Binary Mergers Observed by
  Ligo and Virgo during the First and Second Observing Runs}},\ }\href
  {https://doi.org/10.1103/PhysRevX.9.031040} {\bibfield  {journal} {\bibinfo
  {journal} {Phys. Rev. X}\ }\textbf {\bibinfo {volume} {9}},\ \bibinfo {pages}
  {031040} (\bibinfo {year} {2019}{\natexlab{a}})},\ \Eprint
  {https://arxiv.org/abs/1811.12907} {arXiv:1811.12907 [astro-ph.HE]}
  \BibitemShut {NoStop}%
\bibitem [{\citenamefont {Carr}\ and\ \citenamefont
  {Hawking}(1974)}]{Carr:1974nx}%
  \BibitemOpen
  \bibfield  {author} {\bibinfo {author} {\bibfnamefont {B.~J.}\ \bibnamefont
  {Carr}}\ and\ \bibinfo {author} {\bibfnamefont {S.~W.}\ \bibnamefont
  {Hawking}},\ }\bibfield  {title} {\bibinfo {title} {{Black holes in the early
  Universe}},\ }\href {https://doi.org/10.1093/mnras/168.2.399} {\bibfield
  {journal} {\bibinfo  {journal} {Mon. Not. Roy. Astron. Soc.}\ }\textbf
  {\bibinfo {volume} {168}},\ \bibinfo {pages} {399} (\bibinfo {year}
  {1974})}\BibitemShut {NoStop}%
\bibitem [{\citenamefont {Carr}\ and\ \citenamefont
  {Lidsey}(1993)}]{Carr:1993aq}%
  \BibitemOpen
  \bibfield  {author} {\bibinfo {author} {\bibfnamefont {B.~J.}\ \bibnamefont
  {Carr}}\ and\ \bibinfo {author} {\bibfnamefont {J.~E.}\ \bibnamefont
  {Lidsey}},\ }\bibfield  {title} {\bibinfo {title} {{Primordial Black Holes
  and Generalized Constraints on Chaotic Inflation}},\ }\href
  {https://doi.org/10.1103/PhysRevD.48.543} {\bibfield  {journal} {\bibinfo
  {journal} {Phys. Rev. D}\ }\textbf {\bibinfo {volume} {48}},\ \bibinfo
  {pages} {543} (\bibinfo {year} {1993})}\BibitemShut {NoStop}%
\bibitem [{\citenamefont {Ivanov}\ \emph {et~al.}(1994)\citenamefont {Ivanov},
  \citenamefont {Naselsky},\ and\ \citenamefont {Novikov}}]{Ivanov:1994pa}%
  \BibitemOpen
  \bibfield  {author} {\bibinfo {author} {\bibfnamefont {P.}~\bibnamefont
  {Ivanov}}, \bibinfo {author} {\bibfnamefont {P.}~\bibnamefont {Naselsky}},\
  and\ \bibinfo {author} {\bibfnamefont {I.}~\bibnamefont {Novikov}},\
  }\bibfield  {title} {\bibinfo {title} {{Inflation and Primordial Black Holes
  as Dark Matter}},\ }\href {https://doi.org/10.1103/PhysRevD.50.7173}
  {\bibfield  {journal} {\bibinfo  {journal} {Phys. Rev. D}\ }\textbf {\bibinfo
  {volume} {50}},\ \bibinfo {pages} {7173} (\bibinfo {year}
  {1994})}\BibitemShut {NoStop}%
\bibitem [{\citenamefont {Hawking}(1989)}]{Hawking:1987bn}%
  \BibitemOpen
  \bibfield  {author} {\bibinfo {author} {\bibfnamefont {S.~W.}\ \bibnamefont
  {Hawking}},\ }\bibfield  {title} {\bibinfo {title} {{Black Holes from Cosmic
  Strings}},\ }\href {https://doi.org/10.1016/0370-2693(89)90206-2} {\bibfield
  {journal} {\bibinfo  {journal} {Phys. Lett. B}\ }\textbf {\bibinfo {volume}
  {231}},\ \bibinfo {pages} {237} (\bibinfo {year} {1989})}\BibitemShut
  {NoStop}%
\bibitem [{\citenamefont {Caldwell}\ and\ \citenamefont
  {Casper}(1996)}]{Caldwell:1995fu}%
  \BibitemOpen
  \bibfield  {author} {\bibinfo {author} {\bibfnamefont {R.~R.}\ \bibnamefont
  {Caldwell}}\ and\ \bibinfo {author} {\bibfnamefont {P.}~\bibnamefont
  {Casper}},\ }\bibfield  {title} {\bibinfo {title} {{Formation of black holes
  from collapsed cosmic string loops}},\ }\href
  {https://doi.org/10.1103/PhysRevD.53.3002} {\bibfield  {journal} {\bibinfo
  {journal} {Phys. Rev. D}\ }\textbf {\bibinfo {volume} {53}},\ \bibinfo
  {pages} {3002} (\bibinfo {year} {1996})},\ \Eprint
  {https://arxiv.org/abs/gr-qc/9509012} {arXiv:gr-qc/9509012} \BibitemShut
  {NoStop}%
\bibitem [{\citenamefont {Jenkins}\ and\ \citenamefont
  {Sakellariadou}(2020)}]{Jenkins:2020ctp}%
  \BibitemOpen
  \bibfield  {author} {\bibinfo {author} {\bibfnamefont {A.~C.}\ \bibnamefont
  {Jenkins}}\ and\ \bibinfo {author} {\bibfnamefont {M.}~\bibnamefont
  {Sakellariadou}},\ }\bibfield  {title} {\bibinfo {title} {{Primordial black
  holes from cusp collapse on cosmic strings}},\ }\href@noop {} {\  (\bibinfo
  {year} {2020})},\ \Eprint {https://arxiv.org/abs/2006.16249}
  {arXiv:2006.16249 [astro-ph.CO]} \BibitemShut {NoStop}%
\bibitem [{\citenamefont {Blanco-Pillado}\ \emph {et~al.}(2021)\citenamefont
  {Blanco-Pillado}, \citenamefont {Olum},\ and\ \citenamefont
  {Vilenkin}}]{Blanco-Pillado:2021klh}%
  \BibitemOpen
  \bibfield  {author} {\bibinfo {author} {\bibfnamefont {J.~J.}\ \bibnamefont
  {Blanco-Pillado}}, \bibinfo {author} {\bibfnamefont {K.~D.}\ \bibnamefont
  {Olum}},\ and\ \bibinfo {author} {\bibfnamefont {A.}~\bibnamefont
  {Vilenkin}},\ }\bibfield  {title} {\bibinfo {title} {{No black holes from
  cosmic string cusps}},\ }\href@noop {} {\  (\bibinfo {year} {2021})},\
  \Eprint {https://arxiv.org/abs/2101.05040} {arXiv:2101.05040 [astro-ph.CO]}
  \BibitemShut {NoStop}%
\bibitem [{\citenamefont {Vachaspati}(2017)}]{Vachaspati:2017hjw}%
  \BibitemOpen
  \bibfield  {author} {\bibinfo {author} {\bibfnamefont {T.}~\bibnamefont
  {Vachaspati}},\ }\bibfield  {title} {\bibinfo {title} {{Lunar Mass Black
  Holes from QCD Axion Cosmology}},\ }\href@noop {} {\  (\bibinfo {year}
  {2017})},\ \Eprint {https://arxiv.org/abs/1706.03868} {arXiv:1706.03868
  [hep-th]} \BibitemShut {NoStop}%
\bibitem [{\citenamefont {Ferrer}\ \emph {et~al.}(2019)\citenamefont {Ferrer},
  \citenamefont {Masso}, \citenamefont {Panico}, \citenamefont {Pujolas},\ and\
  \citenamefont {Rompineve}}]{Ferrer:2018uiu}%
  \BibitemOpen
  \bibfield  {author} {\bibinfo {author} {\bibfnamefont {F.}~\bibnamefont
  {Ferrer}}, \bibinfo {author} {\bibfnamefont {E.}~\bibnamefont {Masso}},
  \bibinfo {author} {\bibfnamefont {G.}~\bibnamefont {Panico}}, \bibinfo
  {author} {\bibfnamefont {O.}~\bibnamefont {Pujolas}},\ and\ \bibinfo {author}
  {\bibfnamefont {F.}~\bibnamefont {Rompineve}},\ }\bibfield  {title} {\bibinfo
  {title} {{Primordial Black Holes from the QCD Axion}},\ }\href
  {https://doi.org/10.1103/PhysRevLett.122.101301} {\bibfield  {journal}
  {\bibinfo  {journal} {Phys. Rev. Lett.}\ }\textbf {\bibinfo {volume} {122}},\
  \bibinfo {pages} {101301} (\bibinfo {year} {2019})},\ \Eprint
  {https://arxiv.org/abs/1807.01707} {arXiv:1807.01707 [hep-ph]} \BibitemShut
  {NoStop}%
\bibitem [{\citenamefont {Gelmini}\ \emph
  {et~al.}(2023{\natexlab{a}})\citenamefont {Gelmini}, \citenamefont
  {Simpson},\ and\ \citenamefont {Vitagliano}}]{Gelmini:2022nim}%
  \BibitemOpen
  \bibfield  {author} {\bibinfo {author} {\bibfnamefont {G.~B.}\ \bibnamefont
  {Gelmini}}, \bibinfo {author} {\bibfnamefont {A.}~\bibnamefont {Simpson}},\
  and\ \bibinfo {author} {\bibfnamefont {E.}~\bibnamefont {Vitagliano}},\
  }\bibfield  {title} {\bibinfo {title} {{Catastrogenesis: DM, GWs, and PBHs
  from ALP string-wall networks}},\ }\href
  {https://doi.org/10.1088/1475-7516/2023/02/031} {\bibfield  {journal}
  {\bibinfo  {journal} {JCAP}\ }\textbf {\bibinfo {volume} {02}},\ \bibinfo
  {pages} {031}},\ \Eprint {https://arxiv.org/abs/2207.07126} {arXiv:2207.07126
  [hep-ph]} \BibitemShut {NoStop}%
\bibitem [{\citenamefont {Gelmini}\ \emph
  {et~al.}(2023{\natexlab{b}})\citenamefont {Gelmini}, \citenamefont {Hyman},
  \citenamefont {Simpson},\ and\ \citenamefont {Vitagliano}}]{Gelmini:2023ngs}%
  \BibitemOpen
  \bibfield  {author} {\bibinfo {author} {\bibfnamefont {G.~B.}\ \bibnamefont
  {Gelmini}}, \bibinfo {author} {\bibfnamefont {J.}~\bibnamefont {Hyman}},
  \bibinfo {author} {\bibfnamefont {A.}~\bibnamefont {Simpson}},\ and\ \bibinfo
  {author} {\bibfnamefont {E.}~\bibnamefont {Vitagliano}},\ }\bibfield  {title}
  {\bibinfo {title} {{Primordial black hole dark matter from catastrogenesis
  with unstable pseudo-Goldstone bosons}},\ }\href@noop {} {\  (\bibinfo {year}
  {2023}{\natexlab{b}})},\ \Eprint {https://arxiv.org/abs/2303.14107}
  {arXiv:2303.14107 [hep-ph]} \BibitemShut {NoStop}%
\bibitem [{\citenamefont {Dolgov}\ and\ \citenamefont
  {Silk}(1993)}]{Dolgov:1992pu}%
  \BibitemOpen
  \bibfield  {author} {\bibinfo {author} {\bibfnamefont {A.}~\bibnamefont
  {Dolgov}}\ and\ \bibinfo {author} {\bibfnamefont {J.}~\bibnamefont {Silk}},\
  }\bibfield  {title} {\bibinfo {title} {{Baryon Isocurvature Fluctuations at
  Small Scales and Baryonic Dark Matter}},\ }\href
  {https://doi.org/10.1103/PhysRevD.47.4244} {\bibfield  {journal} {\bibinfo
  {journal} {Phys. Rev. D}\ }\textbf {\bibinfo {volume} {47}},\ \bibinfo
  {pages} {4244} (\bibinfo {year} {1993})}\BibitemShut {NoStop}%
\bibitem [{\citenamefont {Kasai}\ \emph {et~al.}(2022)\citenamefont {Kasai},
  \citenamefont {Kawasaki},\ and\ \citenamefont {Murai}}]{Kasai:2022vhq}%
  \BibitemOpen
  \bibfield  {author} {\bibinfo {author} {\bibfnamefont {K.}~\bibnamefont
  {Kasai}}, \bibinfo {author} {\bibfnamefont {M.}~\bibnamefont {Kawasaki}},\
  and\ \bibinfo {author} {\bibfnamefont {K.}~\bibnamefont {Murai}},\ }\bibfield
   {title} {\bibinfo {title} {{Revisiting the Affleck-Dine Mechanism for
  Primordial Black Hole Formation}},\ }\href
  {https://doi.org/10.1088/1475-7516/2022/10/048} {\bibfield  {journal}
  {\bibinfo  {journal} {JCAP}\ }\textbf {\bibinfo {volume} {10}},\ \bibinfo
  {pages} {048}},\ \Eprint {https://arxiv.org/abs/2205.10148} {arXiv:2205.10148
  [astro-ph.CO]} \BibitemShut {NoStop}%
\bibitem [{\citenamefont {Cotner}\ and\ \citenamefont
  {Kusenko}(2017)}]{Cotner:2016cvr}%
  \BibitemOpen
  \bibfield  {author} {\bibinfo {author} {\bibfnamefont {E.}~\bibnamefont
  {Cotner}}\ and\ \bibinfo {author} {\bibfnamefont {A.}~\bibnamefont
  {Kusenko}},\ }\bibfield  {title} {\bibinfo {title} {{Primordial black holes
  from supersymmetry in the early universe}},\ }\href
  {https://doi.org/10.1103/PhysRevLett.119.031103} {\bibfield  {journal}
  {\bibinfo  {journal} {Phys. Rev. Lett.}\ }\textbf {\bibinfo {volume} {119}},\
  \bibinfo {pages} {031103} (\bibinfo {year} {2017})},\ \Eprint
  {https://arxiv.org/abs/1612.02529} {arXiv:1612.02529 [astro-ph.CO]}
  \BibitemShut {NoStop}%
\bibitem [{\citenamefont {Martin}\ \emph {et~al.}(2020)\citenamefont {Martin},
  \citenamefont {Papanikolaou},\ and\ \citenamefont {Vennin}}]{Martin:2019nuw}%
  \BibitemOpen
  \bibfield  {author} {\bibinfo {author} {\bibfnamefont {J.}~\bibnamefont
  {Martin}}, \bibinfo {author} {\bibfnamefont {T.}~\bibnamefont
  {Papanikolaou}},\ and\ \bibinfo {author} {\bibfnamefont {V.}~\bibnamefont
  {Vennin}},\ }\bibfield  {title} {\bibinfo {title} {{Primordial black holes
  from the preheating instability in single-field inflation}},\ }\href
  {https://doi.org/10.1088/1475-7516/2020/01/024} {\bibfield  {journal}
  {\bibinfo  {journal} {JCAP}\ }\textbf {\bibinfo {volume} {01}},\ \bibinfo
  {pages} {024}},\ \Eprint {https://arxiv.org/abs/1907.04236} {arXiv:1907.04236
  [astro-ph.CO]} \BibitemShut {NoStop}%
\bibitem [{\citenamefont {Chang}\ \emph {et~al.}(2019)\citenamefont {Chang},
  \citenamefont {Egana-Ugrinovic}, \citenamefont {Essig},\ and\ \citenamefont
  {Kouvaris}}]{Chang:2018bgx}%
  \BibitemOpen
  \bibfield  {author} {\bibinfo {author} {\bibfnamefont {J.~H.}\ \bibnamefont
  {Chang}}, \bibinfo {author} {\bibfnamefont {D.}~\bibnamefont
  {Egana-Ugrinovic}}, \bibinfo {author} {\bibfnamefont {R.}~\bibnamefont
  {Essig}},\ and\ \bibinfo {author} {\bibfnamefont {C.}~\bibnamefont
  {Kouvaris}},\ }\bibfield  {title} {\bibinfo {title} {{Structure Formation and
  Exotic Compact Objects in a Dissipative Dark Sector}},\ }\href
  {https://doi.org/10.1088/1475-7516/2019/03/036} {\bibfield  {journal}
  {\bibinfo  {journal} {JCAP}\ }\textbf {\bibinfo {volume} {03}},\ \bibinfo
  {pages} {036}},\ \Eprint {https://arxiv.org/abs/1812.07000} {arXiv:1812.07000
  [hep-ph]} \BibitemShut {NoStop}%
\bibitem [{\citenamefont {Flores}\ and\ \citenamefont
  {Kusenko}(2021)}]{Flores:2020drq}%
  \BibitemOpen
  \bibfield  {author} {\bibinfo {author} {\bibfnamefont {M.~M.}\ \bibnamefont
  {Flores}}\ and\ \bibinfo {author} {\bibfnamefont {A.}~\bibnamefont
  {Kusenko}},\ }\bibfield  {title} {\bibinfo {title} {{Primordial Black Holes
  from Long-Range Scalar Forces and Scalar Radiative Cooling}},\ }\href
  {https://doi.org/10.1103/PhysRevLett.126.041101} {\bibfield  {journal}
  {\bibinfo  {journal} {Phys. Rev. Lett.}\ }\textbf {\bibinfo {volume} {126}},\
  \bibinfo {pages} {041101} (\bibinfo {year} {2021})},\ \Eprint
  {https://arxiv.org/abs/2008.12456} {arXiv:2008.12456 [astro-ph.CO]}
  \BibitemShut {NoStop}%
\bibitem [{\citenamefont {Dom\`enech}\ \emph {et~al.}(2023)\citenamefont
  {Dom\`enech}, \citenamefont {Inman}, \citenamefont {Kusenko},\ and\
  \citenamefont {Sasaki}}]{Domenech:2023afs}%
  \BibitemOpen
  \bibfield  {author} {\bibinfo {author} {\bibfnamefont {G.}~\bibnamefont
  {Dom\`enech}}, \bibinfo {author} {\bibfnamefont {D.}~\bibnamefont {Inman}},
  \bibinfo {author} {\bibfnamefont {A.}~\bibnamefont {Kusenko}},\ and\ \bibinfo
  {author} {\bibfnamefont {M.}~\bibnamefont {Sasaki}},\ }\bibfield  {title}
  {\bibinfo {title} {{Halo Formation from Yukawa Forces in the Very Early
  Universe}},\ }\href@noop {} {\  (\bibinfo {year} {2023})},\ \Eprint
  {https://arxiv.org/abs/2304.13053} {arXiv:2304.13053 [astro-ph.CO]}
  \BibitemShut {NoStop}%
\bibitem [{\citenamefont {Chakraborty}\ \emph {et~al.}(2022)\citenamefont
  {Chakraborty}, \citenamefont {Chanda}, \citenamefont {Pandey},\ and\
  \citenamefont {Das}}]{Chakraborty:2022mwu}%
  \BibitemOpen
  \bibfield  {author} {\bibinfo {author} {\bibfnamefont {A.}~\bibnamefont
  {Chakraborty}}, \bibinfo {author} {\bibfnamefont {P.~K.}\ \bibnamefont
  {Chanda}}, \bibinfo {author} {\bibfnamefont {K.~L.}\ \bibnamefont {Pandey}},\
  and\ \bibinfo {author} {\bibfnamefont {S.}~\bibnamefont {Das}},\ }\bibfield
  {title} {\bibinfo {title} {{Formation and Abundance of Late-forming
  Primordial Black Holes as Dark Matter}},\ }\href
  {https://doi.org/10.3847/1538-4357/ac6ddd} {\bibfield  {journal} {\bibinfo
  {journal} {Astrophys. J.}\ }\textbf {\bibinfo {volume} {932}},\ \bibinfo
  {pages} {119} (\bibinfo {year} {2022})},\ \Eprint
  {https://arxiv.org/abs/2204.09628} {arXiv:2204.09628 [astro-ph.CO]}
  \BibitemShut {NoStop}%
\bibitem [{\citenamefont {Hawking}\ \emph {et~al.}(1982)\citenamefont
  {Hawking}, \citenamefont {Moss},\ and\ \citenamefont
  {Stewart}}]{Hawking:1982ga}%
  \BibitemOpen
  \bibfield  {author} {\bibinfo {author} {\bibfnamefont {S.~W.}\ \bibnamefont
  {Hawking}}, \bibinfo {author} {\bibfnamefont {I.~G.}\ \bibnamefont {Moss}},\
  and\ \bibinfo {author} {\bibfnamefont {J.~M.}\ \bibnamefont {Stewart}},\
  }\bibfield  {title} {\bibinfo {title} {{Bubble Collisions in the Very Early
  Universe}},\ }\href {https://doi.org/10.1103/PhysRevD.26.2681} {\bibfield
  {journal} {\bibinfo  {journal} {Phys. Rev. D}\ }\textbf {\bibinfo {volume}
  {26}},\ \bibinfo {pages} {2681} (\bibinfo {year} {1982})}\BibitemShut
  {NoStop}%
\bibitem [{\citenamefont {Moss}(1994)}]{Moss:1994iq}%
  \BibitemOpen
  \bibfield  {author} {\bibinfo {author} {\bibfnamefont {I.~G.}\ \bibnamefont
  {Moss}},\ }\bibfield  {title} {\bibinfo {title} {{Singularity Formation from
  Colliding Bubbles}},\ }\href {https://doi.org/10.1103/PhysRevD.50.676}
  {\bibfield  {journal} {\bibinfo  {journal} {Phys. Rev. D}\ }\textbf {\bibinfo
  {volume} {50}},\ \bibinfo {pages} {676} (\bibinfo {year} {1994})}\BibitemShut
  {NoStop}%
\bibitem [{\citenamefont {Ashoorioon}\ \emph {et~al.}(2021)\citenamefont
  {Ashoorioon}, \citenamefont {Rostami},\ and\ \citenamefont
  {Firouzjaee}}]{Ashoorioon:2020hln}%
  \BibitemOpen
  \bibfield  {author} {\bibinfo {author} {\bibfnamefont {A.}~\bibnamefont
  {Ashoorioon}}, \bibinfo {author} {\bibfnamefont {A.}~\bibnamefont
  {Rostami}},\ and\ \bibinfo {author} {\bibfnamefont {J.~T.}\ \bibnamefont
  {Firouzjaee}},\ }\bibfield  {title} {\bibinfo {title} {{Examining the end of
  inflation with primordial black holes mass distribution and gravitational
  waves}},\ }\href {https://doi.org/10.1103/PhysRevD.103.123512} {\bibfield
  {journal} {\bibinfo  {journal} {Phys. Rev. D}\ }\textbf {\bibinfo {volume}
  {103}},\ \bibinfo {pages} {123512} (\bibinfo {year} {2021})},\ \Eprint
  {https://arxiv.org/abs/2012.02817} {arXiv:2012.02817 [astro-ph.CO]}
  \BibitemShut {NoStop}%
\bibitem [{\citenamefont {Jung}\ and\ \citenamefont
  {Okui}(2021)}]{Jung:2021mku}%
  \BibitemOpen
  \bibfield  {author} {\bibinfo {author} {\bibfnamefont {T.~H.}\ \bibnamefont
  {Jung}}\ and\ \bibinfo {author} {\bibfnamefont {T.}~\bibnamefont {Okui}},\
  }\bibfield  {title} {\bibinfo {title} {{Primordial black holes from bubble
  collisions during a first-order phase transition}},\ }\href@noop {} {\
  (\bibinfo {year} {2021})},\ \Eprint {https://arxiv.org/abs/2110.04271}
  {arXiv:2110.04271 [hep-ph]} \BibitemShut {NoStop}%
\bibitem [{\citenamefont {Crawford}\ and\ \citenamefont
  {Schramm}(1982)}]{Crawford:1982yz}%
  \BibitemOpen
  \bibfield  {author} {\bibinfo {author} {\bibfnamefont {M.}~\bibnamefont
  {Crawford}}\ and\ \bibinfo {author} {\bibfnamefont {D.~N.}\ \bibnamefont
  {Schramm}},\ }\bibfield  {title} {\bibinfo {title} {{Spontaneous Generation
  of Density Perturbations in the Early Universe}},\ }\href
  {https://doi.org/10.1038/298538a0} {\bibfield  {journal} {\bibinfo  {journal}
  {Nature}\ }\textbf {\bibinfo {volume} {298}},\ \bibinfo {pages} {538}
  (\bibinfo {year} {1982})}\BibitemShut {NoStop}%
\bibitem [{\citenamefont {Gross}\ \emph {et~al.}(2021)\citenamefont {Gross},
  \citenamefont {Landini}, \citenamefont {Strumia},\ and\ \citenamefont
  {Teresi}}]{Gross:2021qgx}%
  \BibitemOpen
  \bibfield  {author} {\bibinfo {author} {\bibfnamefont {C.}~\bibnamefont
  {Gross}}, \bibinfo {author} {\bibfnamefont {G.}~\bibnamefont {Landini}},
  \bibinfo {author} {\bibfnamefont {A.}~\bibnamefont {Strumia}},\ and\ \bibinfo
  {author} {\bibfnamefont {D.}~\bibnamefont {Teresi}},\ }\bibfield  {title}
  {\bibinfo {title} {{Dark Matter as Dark Dwarfs and Other Macroscopic Objects:
  Multiverse Relics?}},\ }\href {https://doi.org/10.1007/JHEP09(2021)033}
  {\bibfield  {journal} {\bibinfo  {journal} {JHEP}\ }\textbf {\bibinfo
  {volume} {09}},\ \bibinfo {pages} {033}},\ \Eprint
  {https://arxiv.org/abs/2105.02840} {arXiv:2105.02840 [hep-ph]} \BibitemShut
  {NoStop}%
\bibitem [{\citenamefont {Baker}\ \emph {et~al.}(2021)\citenamefont {Baker},
  \citenamefont {Breitbach}, \citenamefont {Kopp},\ and\ \citenamefont
  {Mittnacht}}]{Baker:2021sno}%
  \BibitemOpen
  \bibfield  {author} {\bibinfo {author} {\bibfnamefont {M.~J.}\ \bibnamefont
  {Baker}}, \bibinfo {author} {\bibfnamefont {M.}~\bibnamefont {Breitbach}},
  \bibinfo {author} {\bibfnamefont {J.}~\bibnamefont {Kopp}},\ and\ \bibinfo
  {author} {\bibfnamefont {L.}~\bibnamefont {Mittnacht}},\ }\bibfield  {title}
  {\bibinfo {title} {{Detailed Calculation of Primordial Black Hole Formation
  during First-Order Cosmological Phase Transitions}},\ }\href@noop {} {\
  (\bibinfo {year} {2021})},\ \Eprint {https://arxiv.org/abs/2110.00005}
  {arXiv:2110.00005 [astro-ph.CO]} \BibitemShut {NoStop}%
\bibitem [{\citenamefont {Kawana}\ and\ \citenamefont
  {Xie}(2022)}]{Kawana:2021tde}%
  \BibitemOpen
  \bibfield  {author} {\bibinfo {author} {\bibfnamefont {K.}~\bibnamefont
  {Kawana}}\ and\ \bibinfo {author} {\bibfnamefont {K.-P.}\ \bibnamefont
  {Xie}},\ }\bibfield  {title} {\bibinfo {title} {{Primordial Black Holes from
  a Cosmic Phase Transition: the Collapse of Fermi-Balls}},\ }\href
  {https://doi.org/10.1016/j.physletb.2021.136791} {\bibfield  {journal}
  {\bibinfo  {journal} {Phys. Lett. B}\ }\textbf {\bibinfo {volume} {824}},\
  \bibinfo {pages} {136791} (\bibinfo {year} {2022})},\ \Eprint
  {https://arxiv.org/abs/2106.00111} {arXiv:2106.00111 [astro-ph.CO]}
  \BibitemShut {NoStop}%
\bibitem [{\citenamefont {Huang}\ and\ \citenamefont
  {Xie}(2022)}]{Huang:2022him}%
  \BibitemOpen
  \bibfield  {author} {\bibinfo {author} {\bibfnamefont {P.}~\bibnamefont
  {Huang}}\ and\ \bibinfo {author} {\bibfnamefont {K.-P.}\ \bibnamefont
  {Xie}},\ }\bibfield  {title} {\bibinfo {title} {{Primordial black holes from
  an electroweak phase transition}},\ }\href
  {https://doi.org/10.1103/PhysRevD.105.115033} {\bibfield  {journal} {\bibinfo
   {journal} {Phys. Rev. D}\ }\textbf {\bibinfo {volume} {105}},\ \bibinfo
  {pages} {115033} (\bibinfo {year} {2022})},\ \Eprint
  {https://arxiv.org/abs/2201.07243} {arXiv:2201.07243 [hep-ph]} \BibitemShut
  {NoStop}%
\bibitem [{\citenamefont {Garriga}\ \emph {et~al.}(2016)\citenamefont
  {Garriga}, \citenamefont {Vilenkin},\ and\ \citenamefont
  {Zhang}}]{Garriga:2015fdk}%
  \BibitemOpen
  \bibfield  {author} {\bibinfo {author} {\bibfnamefont {J.}~\bibnamefont
  {Garriga}}, \bibinfo {author} {\bibfnamefont {A.}~\bibnamefont {Vilenkin}},\
  and\ \bibinfo {author} {\bibfnamefont {J.}~\bibnamefont {Zhang}},\ }\bibfield
   {title} {\bibinfo {title} {{Black Holes and the Multiverse}},\ }\href
  {https://doi.org/10.1088/1475-7516/2016/02/064} {\bibfield  {journal}
  {\bibinfo  {journal} {JCAP}\ }\textbf {\bibinfo {volume} {02}},\ \bibinfo
  {pages} {064}},\ \Eprint {https://arxiv.org/abs/1512.01819} {arXiv:1512.01819
  [hep-th]} \BibitemShut {NoStop}%
\bibitem [{\citenamefont {Deng}\ \emph {et~al.}(2017)\citenamefont {Deng},
  \citenamefont {Garriga},\ and\ \citenamefont {Vilenkin}}]{Deng:2016vzb}%
  \BibitemOpen
  \bibfield  {author} {\bibinfo {author} {\bibfnamefont {H.}~\bibnamefont
  {Deng}}, \bibinfo {author} {\bibfnamefont {J.}~\bibnamefont {Garriga}},\ and\
  \bibinfo {author} {\bibfnamefont {A.}~\bibnamefont {Vilenkin}},\ }\bibfield
  {title} {\bibinfo {title} {{Primordial Black Hole and Wormhole Formation by
  Domain Walls}},\ }\href {https://doi.org/10.1088/1475-7516/2017/04/050}
  {\bibfield  {journal} {\bibinfo  {journal} {JCAP}\ }\textbf {\bibinfo
  {volume} {04}},\ \bibinfo {pages} {050}},\ \Eprint
  {https://arxiv.org/abs/1612.03753} {arXiv:1612.03753 [gr-qc]} \BibitemShut
  {NoStop}%
\bibitem [{\citenamefont {Deng}\ and\ \citenamefont
  {Vilenkin}(2017)}]{Deng:2017uwc}%
  \BibitemOpen
  \bibfield  {author} {\bibinfo {author} {\bibfnamefont {H.}~\bibnamefont
  {Deng}}\ and\ \bibinfo {author} {\bibfnamefont {A.}~\bibnamefont
  {Vilenkin}},\ }\bibfield  {title} {\bibinfo {title} {{Primordial Black Hole
  Formation by Vacuum Bubbles}},\ }\href
  {https://doi.org/10.1088/1475-7516/2017/12/044} {\bibfield  {journal}
  {\bibinfo  {journal} {JCAP}\ }\textbf {\bibinfo {volume} {12}},\ \bibinfo
  {pages} {044}},\ \Eprint {https://arxiv.org/abs/1710.02865} {arXiv:1710.02865
  [gr-qc]} \BibitemShut {NoStop}%
\bibitem [{\citenamefont {Kusenko}\ \emph {et~al.}(2020)\citenamefont
  {Kusenko}, \citenamefont {Sasaki}, \citenamefont {Sugiyama}, \citenamefont
  {Takada}, \citenamefont {Takhistov},\ and\ \citenamefont
  {Vitagliano}}]{Kusenko:2020pcg}%
  \BibitemOpen
  \bibfield  {author} {\bibinfo {author} {\bibfnamefont {A.}~\bibnamefont
  {Kusenko}}, \bibinfo {author} {\bibfnamefont {M.}~\bibnamefont {Sasaki}},
  \bibinfo {author} {\bibfnamefont {S.}~\bibnamefont {Sugiyama}}, \bibinfo
  {author} {\bibfnamefont {M.}~\bibnamefont {Takada}}, \bibinfo {author}
  {\bibfnamefont {V.}~\bibnamefont {Takhistov}},\ and\ \bibinfo {author}
  {\bibfnamefont {E.}~\bibnamefont {Vitagliano}},\ }\bibfield  {title}
  {\bibinfo {title} {{Exploring Primordial Black Holes from the Multiverse with
  Optical Telescopes}},\ }\href
  {https://doi.org/10.1103/PhysRevLett.125.181304} {\bibfield  {journal}
  {\bibinfo  {journal} {Phys. Rev. Lett.}\ }\textbf {\bibinfo {volume} {125}},\
  \bibinfo {pages} {181304} (\bibinfo {year} {2020})},\ \Eprint
  {https://arxiv.org/abs/2001.09160} {arXiv:2001.09160 [astro-ph.CO]}
  \BibitemShut {NoStop}%
\bibitem [{\citenamefont {Sato}\ \emph {et~al.}(1981)\citenamefont {Sato},
  \citenamefont {Sasaki}, \citenamefont {Kodama},\ and\ \citenamefont
  {Maeda}}]{Sato:1981bf}%
  \BibitemOpen
  \bibfield  {author} {\bibinfo {author} {\bibfnamefont {K.}~\bibnamefont
  {Sato}}, \bibinfo {author} {\bibfnamefont {M.}~\bibnamefont {Sasaki}},
  \bibinfo {author} {\bibfnamefont {H.}~\bibnamefont {Kodama}},\ and\ \bibinfo
  {author} {\bibfnamefont {K.-i.}\ \bibnamefont {Maeda}},\ }\bibfield  {title}
  {\bibinfo {title} {{Creation of Wormholes by First Order Phase Transition of
  a Vacuum in the Early Universe}},\ }\href
  {https://doi.org/10.1143/PTP.65.1443} {\bibfield  {journal} {\bibinfo
  {journal} {Prog. Theor. Phys.}\ }\textbf {\bibinfo {volume} {65}},\ \bibinfo
  {pages} {1443} (\bibinfo {year} {1981})}\BibitemShut {NoStop}%
\bibitem [{\citenamefont {Maeda}\ \emph {et~al.}(1982)\citenamefont {Maeda},
  \citenamefont {Sato}, \citenamefont {Sasaki},\ and\ \citenamefont
  {Kodama}}]{Maeda:1981gw}%
  \BibitemOpen
  \bibfield  {author} {\bibinfo {author} {\bibfnamefont {K.-i.}\ \bibnamefont
  {Maeda}}, \bibinfo {author} {\bibfnamefont {K.}~\bibnamefont {Sato}},
  \bibinfo {author} {\bibfnamefont {M.}~\bibnamefont {Sasaki}},\ and\ \bibinfo
  {author} {\bibfnamefont {H.}~\bibnamefont {Kodama}},\ }\bibfield  {title}
  {\bibinfo {title} {{Creation of De Sitter-schwarzschild Wormholes by a
  Cosmological First Order Phase Transition}},\ }\href
  {https://doi.org/10.1016/0370-2693(82)91151-0} {\bibfield  {journal}
  {\bibinfo  {journal} {Phys. Lett. B}\ }\textbf {\bibinfo {volume} {108}},\
  \bibinfo {pages} {98} (\bibinfo {year} {1982})}\BibitemShut {NoStop}%
\bibitem [{\citenamefont {Sato}\ \emph {et~al.}(1982)\citenamefont {Sato},
  \citenamefont {Kodama}, \citenamefont {Sasaki},\ and\ \citenamefont
  {Maeda}}]{Sato:1981gv}%
  \BibitemOpen
  \bibfield  {author} {\bibinfo {author} {\bibfnamefont {K.}~\bibnamefont
  {Sato}}, \bibinfo {author} {\bibfnamefont {H.}~\bibnamefont {Kodama}},
  \bibinfo {author} {\bibfnamefont {M.}~\bibnamefont {Sasaki}},\ and\ \bibinfo
  {author} {\bibfnamefont {K.-i.}\ \bibnamefont {Maeda}},\ }\bibfield  {title}
  {\bibinfo {title} {{Multiproduction of Universes by First Order Phase
  Transition of a Vacuum}},\ }\href
  {https://doi.org/10.1016/0370-2693(82)91152-2} {\bibfield  {journal}
  {\bibinfo  {journal} {Phys. Lett. B}\ }\textbf {\bibinfo {volume} {108}},\
  \bibinfo {pages} {103} (\bibinfo {year} {1982})}\BibitemShut {NoStop}%
\bibitem [{\citenamefont {Kodama}\ \emph {et~al.}(1981)\citenamefont {Kodama},
  \citenamefont {Sasaki}, \citenamefont {Sato},\ and\ \citenamefont
  {Maeda}}]{Kodama:1981gu}%
  \BibitemOpen
  \bibfield  {author} {\bibinfo {author} {\bibfnamefont {H.}~\bibnamefont
  {Kodama}}, \bibinfo {author} {\bibfnamefont {M.}~\bibnamefont {Sasaki}},
  \bibinfo {author} {\bibfnamefont {K.}~\bibnamefont {Sato}},\ and\ \bibinfo
  {author} {\bibfnamefont {K.-i.}\ \bibnamefont {Maeda}},\ }\bibfield  {title}
  {\bibinfo {title} {{Fate of Wormholes Created by First Order Phase Transition
  in the Early Universe}},\ }\href {https://doi.org/10.1143/PTP.66.2052}
  {\bibfield  {journal} {\bibinfo  {journal} {Prog. Theor. Phys.}\ }\textbf
  {\bibinfo {volume} {66}},\ \bibinfo {pages} {2052} (\bibinfo {year}
  {1981})}\BibitemShut {NoStop}%
\bibitem [{\citenamefont {Kodama}\ \emph {et~al.}(1982)\citenamefont {Kodama},
  \citenamefont {Sasaki},\ and\ \citenamefont {Sato}}]{Kodama:1982sf}%
  \BibitemOpen
  \bibfield  {author} {\bibinfo {author} {\bibfnamefont {H.}~\bibnamefont
  {Kodama}}, \bibinfo {author} {\bibfnamefont {M.}~\bibnamefont {Sasaki}},\
  and\ \bibinfo {author} {\bibfnamefont {K.}~\bibnamefont {Sato}},\ }\bibfield
  {title} {\bibinfo {title} {{Abundance of Primordial Holes Produced by
  Cosmological First Order Phase Transition}},\ }\href
  {https://doi.org/10.1143/PTP.68.1979} {\bibfield  {journal} {\bibinfo
  {journal} {Prog. Theor. Phys.}\ }\textbf {\bibinfo {volume} {68}},\ \bibinfo
  {pages} {1979} (\bibinfo {year} {1982})}\BibitemShut {NoStop}%
\bibitem [{\citenamefont {Hsu}(1990)}]{Hsu:1990fg}%
  \BibitemOpen
  \bibfield  {author} {\bibinfo {author} {\bibfnamefont {S.~D.~H.}\
  \bibnamefont {Hsu}},\ }\bibfield  {title} {\bibinfo {title} {{Black Holes
  From Extended Inflation}},\ }\href
  {https://doi.org/10.1016/0370-2693(90)90717-K} {\bibfield  {journal}
  {\bibinfo  {journal} {Phys. Lett. B}\ }\textbf {\bibinfo {volume} {251}},\
  \bibinfo {pages} {343} (\bibinfo {year} {1990})}\BibitemShut {NoStop}%
\bibitem [{\citenamefont {Liu}\ \emph {et~al.}(2022)\citenamefont {Liu},
  \citenamefont {Bian}, \citenamefont {Cai}, \citenamefont {Guo},\ and\
  \citenamefont {Wang}}]{Liu:2021svg}%
  \BibitemOpen
  \bibfield  {author} {\bibinfo {author} {\bibfnamefont {J.}~\bibnamefont
  {Liu}}, \bibinfo {author} {\bibfnamefont {L.}~\bibnamefont {Bian}}, \bibinfo
  {author} {\bibfnamefont {R.-G.}\ \bibnamefont {Cai}}, \bibinfo {author}
  {\bibfnamefont {Z.-K.}\ \bibnamefont {Guo}},\ and\ \bibinfo {author}
  {\bibfnamefont {S.-J.}\ \bibnamefont {Wang}},\ }\bibfield  {title} {\bibinfo
  {title} {{Primordial Black Hole Production during First-Order Phase
  Transitions}},\ }\href {https://doi.org/10.1103/PhysRevD.105.L021303}
  {\bibfield  {journal} {\bibinfo  {journal} {Phys. Rev. D}\ }\textbf {\bibinfo
  {volume} {105}},\ \bibinfo {pages} {L021303} (\bibinfo {year} {2022})},\
  \Eprint {https://arxiv.org/abs/2106.05637} {arXiv:2106.05637 [astro-ph.CO]}
  \BibitemShut {NoStop}%
\bibitem [{\citenamefont {Hashino}\ \emph {et~al.}(2021)\citenamefont
  {Hashino}, \citenamefont {Kanemura},\ and\ \citenamefont
  {Takahashi}}]{Hashino:2021qoq}%
  \BibitemOpen
  \bibfield  {author} {\bibinfo {author} {\bibfnamefont {K.}~\bibnamefont
  {Hashino}}, \bibinfo {author} {\bibfnamefont {S.}~\bibnamefont {Kanemura}},\
  and\ \bibinfo {author} {\bibfnamefont {T.}~\bibnamefont {Takahashi}},\
  }\bibfield  {title} {\bibinfo {title} {{Primordial Black Holes as a Probe of
  Strongly First-Order Electroweak Phase Transition}},\ }\href@noop {} {\
  (\bibinfo {year} {2021})},\ \Eprint {https://arxiv.org/abs/2111.13099}
  {arXiv:2111.13099 [hep-ph]} \BibitemShut {NoStop}%
\bibitem [{\citenamefont {He}\ \emph {et~al.}(2022)\citenamefont {He},
  \citenamefont {Li}, \citenamefont {Li},\ and\ \citenamefont
  {Wang}}]{He:2022amv}%
  \BibitemOpen
  \bibfield  {author} {\bibinfo {author} {\bibfnamefont {S.}~\bibnamefont
  {He}}, \bibinfo {author} {\bibfnamefont {L.}~\bibnamefont {Li}}, \bibinfo
  {author} {\bibfnamefont {Z.}~\bibnamefont {Li}},\ and\ \bibinfo {author}
  {\bibfnamefont {S.-J.}\ \bibnamefont {Wang}},\ }\bibfield  {title} {\bibinfo
  {title} {{Gravitational Waves and Primordial Black Hole Productions from
  Gluodynamics}},\ }\href@noop {} {\  (\bibinfo {year} {2022})},\ \Eprint
  {https://arxiv.org/abs/2210.14094} {arXiv:2210.14094 [hep-ph]} \BibitemShut
  {NoStop}%
\bibitem [{\citenamefont {Kawana}\ \emph {et~al.}(2022)\citenamefont {Kawana},
  \citenamefont {Kim},\ and\ \citenamefont {Lu}}]{Kawana:2022olo}%
  \BibitemOpen
  \bibfield  {author} {\bibinfo {author} {\bibfnamefont {K.}~\bibnamefont
  {Kawana}}, \bibinfo {author} {\bibfnamefont {T.}~\bibnamefont {Kim}},\ and\
  \bibinfo {author} {\bibfnamefont {P.}~\bibnamefont {Lu}},\ }\bibfield
  {title} {\bibinfo {title} {{PBH Formation from Overdensities in Delayed
  Vacuum Transitions}},\ }\href@noop {} {\  (\bibinfo {year} {2022})},\ \Eprint
  {https://arxiv.org/abs/2212.14037} {arXiv:2212.14037 [astro-ph.CO]}
  \BibitemShut {NoStop}%
\bibitem [{\citenamefont {Lewicki}\ \emph {et~al.}(2023)\citenamefont
  {Lewicki}, \citenamefont {Toczek},\ and\ \citenamefont
  {Vaskonen}}]{Lewicki:2023ioy}%
  \BibitemOpen
  \bibfield  {author} {\bibinfo {author} {\bibfnamefont {M.}~\bibnamefont
  {Lewicki}}, \bibinfo {author} {\bibfnamefont {P.}~\bibnamefont {Toczek}},\
  and\ \bibinfo {author} {\bibfnamefont {V.}~\bibnamefont {Vaskonen}},\
  }\bibfield  {title} {\bibinfo {title} {{Primordial black holes from strong
  first-order phase transitions}},\ }\href@noop {} {\  (\bibinfo {year}
  {2023})},\ \Eprint {https://arxiv.org/abs/2305.04924} {arXiv:2305.04924
  [astro-ph.CO]} \BibitemShut {NoStop}%
\bibitem [{\citenamefont {Gehrman}\ \emph {et~al.}(2023)\citenamefont
  {Gehrman}, \citenamefont {Shams Es~Haghi}, \citenamefont {Sinha},\ and\
  \citenamefont {Xu}}]{Gehrman:2023esa}%
  \BibitemOpen
  \bibfield  {author} {\bibinfo {author} {\bibfnamefont {T.~C.}\ \bibnamefont
  {Gehrman}}, \bibinfo {author} {\bibfnamefont {B.}~\bibnamefont {Shams
  Es~Haghi}}, \bibinfo {author} {\bibfnamefont {K.}~\bibnamefont {Sinha}},\
  and\ \bibinfo {author} {\bibfnamefont {T.}~\bibnamefont {Xu}},\ }\bibfield
  {title} {\bibinfo {title} {{The Primordial Black Holes that Disappeared:
  Connections to Dark Matter and MHz-GHz Gravitational Waves}},\ }\href@noop {}
  {\  (\bibinfo {year} {2023})},\ \Eprint {https://arxiv.org/abs/2304.09194}
  {arXiv:2304.09194 [hep-ph]} \BibitemShut {NoStop}%
\bibitem [{\citenamefont {Kolb}\ and\ \citenamefont
  {Turner}(1990)}]{Kolb:1990vq}%
  \BibitemOpen
  \bibfield  {author} {\bibinfo {author} {\bibfnamefont {E.~W.}\ \bibnamefont
  {Kolb}}\ and\ \bibinfo {author} {\bibfnamefont {M.~S.}\ \bibnamefont
  {Turner}},\ }\href {https://doi.org/10.1201/9780429492860} {\emph {\bibinfo
  {title} {{The Early Universe}}}},\ Vol.~\bibinfo {volume} {69}\ (\bibinfo
  {year} {1990})\BibitemShut {NoStop}%
\bibitem [{\citenamefont {Guth}\ and\ \citenamefont {Tye}(1980)}]{Guth:1979bh}%
  \BibitemOpen
  \bibfield  {author} {\bibinfo {author} {\bibfnamefont {A.~H.}\ \bibnamefont
  {Guth}}\ and\ \bibinfo {author} {\bibfnamefont {S.~H.~H.}\ \bibnamefont
  {Tye}},\ }\bibfield  {title} {\bibinfo {title} {{Phase Transitions and
  Magnetic Monopole Production in the Very Early Universe}},\ }\href
  {https://doi.org/10.1103/PhysRevLett.44.631} {\bibfield  {journal} {\bibinfo
  {journal} {Phys. Rev. Lett.}\ }\textbf {\bibinfo {volume} {44}},\ \bibinfo
  {pages} {631} (\bibinfo {year} {1980})},\ \bibinfo {note} {[Erratum:
  Phys.Rev.Lett. 44, 963 (1980)]}\BibitemShut {NoStop}%
\bibitem [{\citenamefont {Carr}(1975)}]{Carr:1975qj}%
  \BibitemOpen
  \bibfield  {author} {\bibinfo {author} {\bibfnamefont {B.~J.}\ \bibnamefont
  {Carr}},\ }\bibfield  {title} {\bibinfo {title} {{The Primordial Black Hole
  Mass Spectrum}},\ }\href {https://doi.org/10.1086/153853} {\bibfield
  {journal} {\bibinfo  {journal} {Astrophys. J.}\ }\textbf {\bibinfo {volume}
  {201}},\ \bibinfo {pages} {1} (\bibinfo {year} {1975})}\BibitemShut {NoStop}%
\bibitem [{\citenamefont {Musco}(2019)}]{Musco:2018rwt}%
  \BibitemOpen
  \bibfield  {author} {\bibinfo {author} {\bibfnamefont {I.}~\bibnamefont
  {Musco}},\ }\bibfield  {title} {\bibinfo {title} {{Threshold for primordial
  black holes: Dependence on the shape of the cosmological perturbations}},\
  }\href {https://doi.org/10.1103/PhysRevD.100.123524} {\bibfield  {journal}
  {\bibinfo  {journal} {Phys. Rev. D}\ }\textbf {\bibinfo {volume} {100}},\
  \bibinfo {pages} {123524} (\bibinfo {year} {2019})},\ \Eprint
  {https://arxiv.org/abs/1809.02127} {arXiv:1809.02127 [gr-qc]} \BibitemShut
  {NoStop}%
\bibitem [{\citenamefont {Escriv\`a}\ \emph {et~al.}(2020)\citenamefont
  {Escriv\`a}, \citenamefont {Germani},\ and\ \citenamefont
  {Sheth}}]{Escriva:2019phb}%
  \BibitemOpen
  \bibfield  {author} {\bibinfo {author} {\bibfnamefont {A.}~\bibnamefont
  {Escriv\`a}}, \bibinfo {author} {\bibfnamefont {C.}~\bibnamefont {Germani}},\
  and\ \bibinfo {author} {\bibfnamefont {R.~K.}\ \bibnamefont {Sheth}},\
  }\bibfield  {title} {\bibinfo {title} {{Universal threshold for primordial
  black hole formation}},\ }\href {https://doi.org/10.1103/PhysRevD.101.044022}
  {\bibfield  {journal} {\bibinfo  {journal} {Phys. Rev. D}\ }\textbf {\bibinfo
  {volume} {101}},\ \bibinfo {pages} {044022} (\bibinfo {year} {2020})},\
  \Eprint {https://arxiv.org/abs/1907.13311} {arXiv:1907.13311 [gr-qc]}
  \BibitemShut {NoStop}%
\bibitem [{\citenamefont {Jedamzik}\ and\ \citenamefont
  {Niemeyer}(1999)}]{Jedamzik:1999am}%
  \BibitemOpen
  \bibfield  {author} {\bibinfo {author} {\bibfnamefont {K.}~\bibnamefont
  {Jedamzik}}\ and\ \bibinfo {author} {\bibfnamefont {J.~C.}\ \bibnamefont
  {Niemeyer}},\ }\bibfield  {title} {\bibinfo {title} {{Primordial black hole
  formation during first order phase transitions}},\ }\href
  {https://doi.org/10.1103/PhysRevD.59.124014} {\bibfield  {journal} {\bibinfo
  {journal} {Phys. Rev. D}\ }\textbf {\bibinfo {volume} {59}},\ \bibinfo
  {pages} {124014} (\bibinfo {year} {1999})},\ \Eprint
  {https://arxiv.org/abs/astro-ph/9901293} {arXiv:astro-ph/9901293}
  \BibitemShut {NoStop}%
\bibitem [{\citenamefont {Green}\ \emph {et~al.}(2004)\citenamefont {Green},
  \citenamefont {Liddle}, \citenamefont {Malik},\ and\ \citenamefont
  {Sasaki}}]{Green:2004wb}%
  \BibitemOpen
  \bibfield  {author} {\bibinfo {author} {\bibfnamefont {A.~M.}\ \bibnamefont
  {Green}}, \bibinfo {author} {\bibfnamefont {A.~R.}\ \bibnamefont {Liddle}},
  \bibinfo {author} {\bibfnamefont {K.~A.}\ \bibnamefont {Malik}},\ and\
  \bibinfo {author} {\bibfnamefont {M.}~\bibnamefont {Sasaki}},\ }\bibfield
  {title} {\bibinfo {title} {{A New calculation of the mass fraction of
  primordial black holes}},\ }\href
  {https://doi.org/10.1103/PhysRevD.70.041502} {\bibfield  {journal} {\bibinfo
  {journal} {Phys. Rev. D}\ }\textbf {\bibinfo {volume} {70}},\ \bibinfo
  {pages} {041502} (\bibinfo {year} {2004})},\ \Eprint
  {https://arxiv.org/abs/astro-ph/0403181} {arXiv:astro-ph/0403181}
  \BibitemShut {NoStop}%
\bibitem [{\citenamefont {Musco}\ \emph {et~al.}(2005)\citenamefont {Musco},
  \citenamefont {Miller},\ and\ \citenamefont {Rezzolla}}]{Musco:2004ak}%
  \BibitemOpen
  \bibfield  {author} {\bibinfo {author} {\bibfnamefont {I.}~\bibnamefont
  {Musco}}, \bibinfo {author} {\bibfnamefont {J.~C.}\ \bibnamefont {Miller}},\
  and\ \bibinfo {author} {\bibfnamefont {L.}~\bibnamefont {Rezzolla}},\
  }\bibfield  {title} {\bibinfo {title} {{Computations of Primordial Black Hole
  Formation}},\ }\href {https://doi.org/10.1088/0264-9381/22/7/013} {\bibfield
  {journal} {\bibinfo  {journal} {Class. Quant. Grav.}\ }\textbf {\bibinfo
  {volume} {22}},\ \bibinfo {pages} {1405} (\bibinfo {year} {2005})},\ \Eprint
  {https://arxiv.org/abs/gr-qc/0412063} {arXiv:gr-qc/0412063} \BibitemShut
  {NoStop}%
\bibitem [{\citenamefont {Musco}\ and\ \citenamefont
  {Miller}(2013)}]{Musco:2012au}%
  \BibitemOpen
  \bibfield  {author} {\bibinfo {author} {\bibfnamefont {I.}~\bibnamefont
  {Musco}}\ and\ \bibinfo {author} {\bibfnamefont {J.~C.}\ \bibnamefont
  {Miller}},\ }\bibfield  {title} {\bibinfo {title} {{Primordial black hole
  formation in the early universe: critical behaviour and self-similarity}},\
  }\href {https://doi.org/10.1088/0264-9381/30/14/145009} {\bibfield  {journal}
  {\bibinfo  {journal} {Class. Quant. Grav.}\ }\textbf {\bibinfo {volume}
  {30}},\ \bibinfo {pages} {145009} (\bibinfo {year} {2013})},\ \Eprint
  {https://arxiv.org/abs/1201.2379} {arXiv:1201.2379 [gr-qc]} \BibitemShut
  {NoStop}%
\bibitem [{\citenamefont {Harada}\ \emph {et~al.}(2015)\citenamefont {Harada},
  \citenamefont {Yoo}, \citenamefont {Nakama},\ and\ \citenamefont
  {Koga}}]{Harada:2015yda}%
  \BibitemOpen
  \bibfield  {author} {\bibinfo {author} {\bibfnamefont {T.}~\bibnamefont
  {Harada}}, \bibinfo {author} {\bibfnamefont {C.-M.}\ \bibnamefont {Yoo}},
  \bibinfo {author} {\bibfnamefont {T.}~\bibnamefont {Nakama}},\ and\ \bibinfo
  {author} {\bibfnamefont {Y.}~\bibnamefont {Koga}},\ }\bibfield  {title}
  {\bibinfo {title} {{Cosmological long-wavelength solutions and primordial
  black hole formation}},\ }\href {https://doi.org/10.1103/PhysRevD.91.084057}
  {\bibfield  {journal} {\bibinfo  {journal} {Phys. Rev. D}\ }\textbf {\bibinfo
  {volume} {91}},\ \bibinfo {pages} {084057} (\bibinfo {year} {2015})},\
  \Eprint {https://arxiv.org/abs/1503.03934} {arXiv:1503.03934 [gr-qc]}
  \BibitemShut {NoStop}%
\bibitem [{\citenamefont {Wald}(1983)}]{Wald:1983ky}%
  \BibitemOpen
  \bibfield  {author} {\bibinfo {author} {\bibfnamefont {R.~M.}\ \bibnamefont
  {Wald}},\ }\bibfield  {title} {\bibinfo {title} {{Asymptotic behavior of
  homogeneous cosmological models in the presence of a positive cosmological
  constant}},\ }\href {https://doi.org/10.1103/PhysRevD.28.2118} {\bibfield
  {journal} {\bibinfo  {journal} {Phys. Rev. D}\ }\textbf {\bibinfo {volume}
  {28}},\ \bibinfo {pages} {2118} (\bibinfo {year} {1983})}\BibitemShut
  {NoStop}%
\bibitem [{\citenamefont {Ali-Ha\"\i{}moud}\ and\ \citenamefont
  {Kamionkowski}(2017)}]{Ali-Haimoud:2016mbv}%
  \BibitemOpen
  \bibfield  {author} {\bibinfo {author} {\bibfnamefont {Y.}~\bibnamefont
  {Ali-Ha\"\i{}moud}}\ and\ \bibinfo {author} {\bibfnamefont {M.}~\bibnamefont
  {Kamionkowski}},\ }\bibfield  {title} {\bibinfo {title} {{Cosmic microwave
  background limits on accreting primordial black holes}},\ }\href
  {https://doi.org/10.1103/PhysRevD.95.043534} {\bibfield  {journal} {\bibinfo
  {journal} {Phys. Rev. D}\ }\textbf {\bibinfo {volume} {95}},\ \bibinfo
  {pages} {043534} (\bibinfo {year} {2017})},\ \Eprint
  {https://arxiv.org/abs/1612.05644} {arXiv:1612.05644 [astro-ph.CO]}
  \BibitemShut {NoStop}%
\bibitem [{\citenamefont {Poulin}\ \emph
  {et~al.}(2017{\natexlab{a}})\citenamefont {Poulin}, \citenamefont {Serpico},
  \citenamefont {Calore}, \citenamefont {Clesse},\ and\ \citenamefont
  {Kohri}}]{Poulin:2017bwe}%
  \BibitemOpen
  \bibfield  {author} {\bibinfo {author} {\bibfnamefont {V.}~\bibnamefont
  {Poulin}}, \bibinfo {author} {\bibfnamefont {P.~D.}\ \bibnamefont {Serpico}},
  \bibinfo {author} {\bibfnamefont {F.}~\bibnamefont {Calore}}, \bibinfo
  {author} {\bibfnamefont {S.}~\bibnamefont {Clesse}},\ and\ \bibinfo {author}
  {\bibfnamefont {K.}~\bibnamefont {Kohri}},\ }\bibfield  {title} {\bibinfo
  {title} {{Cmb Bounds on Disk-Accreting Massive Primordial Black Holes}},\
  }\href {https://doi.org/10.1103/PhysRevD.96.083524} {\bibfield  {journal}
  {\bibinfo  {journal} {Phys. Rev. D}\ }\textbf {\bibinfo {volume} {96}},\
  \bibinfo {pages} {083524} (\bibinfo {year} {2017}{\natexlab{a}})},\ \Eprint
  {https://arxiv.org/abs/1707.04206} {arXiv:1707.04206 [astro-ph.CO]}
  \BibitemShut {NoStop}%
\bibitem [{\citenamefont {Serpico}\ \emph {et~al.}(2020)\citenamefont
  {Serpico}, \citenamefont {Poulin}, \citenamefont {Inman},\ and\ \citenamefont
  {Kohri}}]{Serpico:2020ehh}%
  \BibitemOpen
  \bibfield  {author} {\bibinfo {author} {\bibfnamefont {P.~D.}\ \bibnamefont
  {Serpico}}, \bibinfo {author} {\bibfnamefont {V.}~\bibnamefont {Poulin}},
  \bibinfo {author} {\bibfnamefont {D.}~\bibnamefont {Inman}},\ and\ \bibinfo
  {author} {\bibfnamefont {K.}~\bibnamefont {Kohri}},\ }\bibfield  {title}
  {\bibinfo {title} {{Cosmic microwave background bounds on primordial black
  holes including dark matter halo accretion}},\ }\href
  {https://doi.org/10.1103/PhysRevResearch.2.023204} {\bibfield  {journal}
  {\bibinfo  {journal} {Phys. Rev. Res.}\ }\textbf {\bibinfo {volume} {2}},\
  \bibinfo {pages} {023204} (\bibinfo {year} {2020})},\ \Eprint
  {https://arxiv.org/abs/2002.10771} {arXiv:2002.10771 [astro-ph.CO]}
  \BibitemShut {NoStop}%
\bibitem [{\citenamefont {Poulin}\ \emph
  {et~al.}(2017{\natexlab{b}})\citenamefont {Poulin}, \citenamefont
  {Lesgourgues},\ and\ \citenamefont {Serpico}}]{Poulin:2016anj}%
  \BibitemOpen
  \bibfield  {author} {\bibinfo {author} {\bibfnamefont {V.}~\bibnamefont
  {Poulin}}, \bibinfo {author} {\bibfnamefont {J.}~\bibnamefont
  {Lesgourgues}},\ and\ \bibinfo {author} {\bibfnamefont {P.~D.}\ \bibnamefont
  {Serpico}},\ }\bibfield  {title} {\bibinfo {title} {{Cosmological Constraints
  on Exotic Injection of Electromagnetic Energy}},\ }\href
  {https://doi.org/10.1088/1475-7516/2017/03/043} {\bibfield  {journal}
  {\bibinfo  {journal} {JCAP}\ }\textbf {\bibinfo {volume} {03}},\ \bibinfo
  {pages} {043}},\ \Eprint {https://arxiv.org/abs/1610.10051} {arXiv:1610.10051
  [astro-ph.CO]} \BibitemShut {NoStop}%
\bibitem [{\citenamefont {St\"ocker}\ \emph {et~al.}(2018)\citenamefont
  {St\"ocker}, \citenamefont {Kr\"amer}, \citenamefont {Lesgourgues},\ and\
  \citenamefont {Poulin}}]{Stocker:2018avm}%
  \BibitemOpen
  \bibfield  {author} {\bibinfo {author} {\bibfnamefont {P.}~\bibnamefont
  {St\"ocker}}, \bibinfo {author} {\bibfnamefont {M.}~\bibnamefont {Kr\"amer}},
  \bibinfo {author} {\bibfnamefont {J.}~\bibnamefont {Lesgourgues}},\ and\
  \bibinfo {author} {\bibfnamefont {V.}~\bibnamefont {Poulin}},\ }\bibfield
  {title} {\bibinfo {title} {{Exotic energy injection with ExoCLASS:
  Application to the Higgs portal model and evaporating black holes}},\ }\href
  {https://doi.org/10.1088/1475-7516/2018/03/018} {\bibfield  {journal}
  {\bibinfo  {journal} {JCAP}\ }\textbf {\bibinfo {volume} {03}},\ \bibinfo
  {pages} {018}},\ \Eprint {https://arxiv.org/abs/1801.01871} {arXiv:1801.01871
  [astro-ph.CO]} \BibitemShut {NoStop}%
\bibitem [{\citenamefont {Poulter}\ \emph {et~al.}(2019)\citenamefont
  {Poulter}, \citenamefont {Ali-Ha\"\i{}moud}, \citenamefont {Hamann},
  \citenamefont {White},\ and\ \citenamefont {Williams}}]{Poulter:2019ooo}%
  \BibitemOpen
  \bibfield  {author} {\bibinfo {author} {\bibfnamefont {H.}~\bibnamefont
  {Poulter}}, \bibinfo {author} {\bibfnamefont {Y.}~\bibnamefont
  {Ali-Ha\"\i{}moud}}, \bibinfo {author} {\bibfnamefont {J.}~\bibnamefont
  {Hamann}}, \bibinfo {author} {\bibfnamefont {M.}~\bibnamefont {White}},\ and\
  \bibinfo {author} {\bibfnamefont {A.~G.}\ \bibnamefont {Williams}},\
  }\bibfield  {title} {\bibinfo {title} {{CMB constraints on ultra-light
  primordial black holes with extended mass distributions}},\ }\href@noop {} {\
   (\bibinfo {year} {2019})},\ \Eprint {https://arxiv.org/abs/1907.06485}
  {arXiv:1907.06485 [astro-ph.CO]} \BibitemShut {NoStop}%
\bibitem [{\citenamefont {De~Luca}\ \emph {et~al.}(2020)\citenamefont
  {De~Luca}, \citenamefont {Franciolini}, \citenamefont {Pani},\ and\
  \citenamefont {Riotto}}]{DeLuca:2020qqa}%
  \BibitemOpen
  \bibfield  {author} {\bibinfo {author} {\bibfnamefont {V.}~\bibnamefont
  {De~Luca}}, \bibinfo {author} {\bibfnamefont {G.}~\bibnamefont
  {Franciolini}}, \bibinfo {author} {\bibfnamefont {P.}~\bibnamefont {Pani}},\
  and\ \bibinfo {author} {\bibfnamefont {A.}~\bibnamefont {Riotto}},\
  }\bibfield  {title} {\bibinfo {title} {{Primordial Black Holes Confront
  LIGO/Virgo data: Current situation}},\ }\href
  {https://doi.org/10.1088/1475-7516/2020/06/044} {\bibfield  {journal}
  {\bibinfo  {journal} {JCAP}\ }\textbf {\bibinfo {volume} {06}},\ \bibinfo
  {pages} {044}},\ \Eprint {https://arxiv.org/abs/2005.05641} {arXiv:2005.05641
  [astro-ph.CO]} \BibitemShut {NoStop}%
\bibitem [{\citenamefont {Alcock}\ \emph {et~al.}(2000)\citenamefont {Alcock}
  \emph {et~al.}}]{MACHO:2000qbb}%
  \BibitemOpen
  \bibfield  {author} {\bibinfo {author} {\bibfnamefont {C.}~\bibnamefont
  {Alcock}} \emph {et~al.} (\bibinfo {collaboration} {MACHO}),\ }\bibfield
  {title} {\bibinfo {title} {{The MACHO project: Microlensing results from 5.7
  years of LMC observations}},\ }\href {https://doi.org/10.1086/309512}
  {\bibfield  {journal} {\bibinfo  {journal} {Astrophys. J.}\ }\textbf
  {\bibinfo {volume} {542}},\ \bibinfo {pages} {281} (\bibinfo {year}
  {2000})},\ \Eprint {https://arxiv.org/abs/astro-ph/0001272}
  {arXiv:astro-ph/0001272} \BibitemShut {NoStop}%
\bibitem [{\citenamefont {Tisserand}\ \emph {et~al.}(2007)\citenamefont
  {Tisserand} \emph {et~al.}}]{EROS-2:2006ryy}%
  \BibitemOpen
  \bibfield  {author} {\bibinfo {author} {\bibfnamefont {P.}~\bibnamefont
  {Tisserand}} \emph {et~al.} (\bibinfo {collaboration} {EROS-2}),\ }\bibfield
  {title} {\bibinfo {title} {{Limits on the Macho Content of the Galactic Halo
  from the EROS-2 Survey of the Magellanic Clouds}},\ }\href
  {https://doi.org/10.1051/0004-6361:20066017} {\bibfield  {journal} {\bibinfo
  {journal} {Astron. Astrophys.}\ }\textbf {\bibinfo {volume} {469}},\ \bibinfo
  {pages} {387} (\bibinfo {year} {2007})},\ \Eprint
  {https://arxiv.org/abs/astro-ph/0607207} {arXiv:astro-ph/0607207}
  \BibitemShut {NoStop}%
\bibitem [{\citenamefont {Niikura}\ \emph
  {et~al.}(2019{\natexlab{a}})\citenamefont {Niikura}, \citenamefont {Takada},
  \citenamefont {Yokoyama}, \citenamefont {Sumi},\ and\ \citenamefont
  {Masaki}}]{Niikura:2019kqi}%
  \BibitemOpen
  \bibfield  {author} {\bibinfo {author} {\bibfnamefont {H.}~\bibnamefont
  {Niikura}}, \bibinfo {author} {\bibfnamefont {M.}~\bibnamefont {Takada}},
  \bibinfo {author} {\bibfnamefont {S.}~\bibnamefont {Yokoyama}}, \bibinfo
  {author} {\bibfnamefont {T.}~\bibnamefont {Sumi}},\ and\ \bibinfo {author}
  {\bibfnamefont {S.}~\bibnamefont {Masaki}},\ }\bibfield  {title} {\bibinfo
  {title} {{Constraints on Earth-Mass Primordial Black Holes from Ogle 5-Year
  Microlensing Events}},\ }\href {https://doi.org/10.1103/PhysRevD.99.083503}
  {\bibfield  {journal} {\bibinfo  {journal} {Phys. Rev. D}\ }\textbf {\bibinfo
  {volume} {99}},\ \bibinfo {pages} {083503} (\bibinfo {year}
  {2019}{\natexlab{a}})},\ \Eprint {https://arxiv.org/abs/1901.07120}
  {arXiv:1901.07120 [astro-ph.CO]} \BibitemShut {NoStop}%
\bibitem [{\citenamefont {Niikura}\ \emph
  {et~al.}(2019{\natexlab{b}})\citenamefont {Niikura} \emph
  {et~al.}}]{Niikura:2017zjd}%
  \BibitemOpen
  \bibfield  {author} {\bibinfo {author} {\bibfnamefont {H.}~\bibnamefont
  {Niikura}} \emph {et~al.},\ }\bibfield  {title} {\bibinfo {title}
  {{Microlensing Constraints on Primordial Black Holes with Subaru/Hsc
  Andromeda Observations}},\ }\href {https://doi.org/10.1038/s41550-019-0723-1}
  {\bibfield  {journal} {\bibinfo  {journal} {Nature Astron.}\ }\textbf
  {\bibinfo {volume} {3}},\ \bibinfo {pages} {524} (\bibinfo {year}
  {2019}{\natexlab{b}})},\ \Eprint {https://arxiv.org/abs/1701.02151}
  {arXiv:1701.02151 [astro-ph.CO]} \BibitemShut {NoStop}%
\bibitem [{\citenamefont {Carr}\ \emph {et~al.}(2010)\citenamefont {Carr},
  \citenamefont {Kohri}, \citenamefont {Sendouda},\ and\ \citenamefont
  {Yokoyama}}]{Carr:2009jm}%
  \BibitemOpen
  \bibfield  {author} {\bibinfo {author} {\bibfnamefont {B.~J.}\ \bibnamefont
  {Carr}}, \bibinfo {author} {\bibfnamefont {K.}~\bibnamefont {Kohri}},
  \bibinfo {author} {\bibfnamefont {Y.}~\bibnamefont {Sendouda}},\ and\
  \bibinfo {author} {\bibfnamefont {J.}~\bibnamefont {Yokoyama}},\ }\bibfield
  {title} {\bibinfo {title} {{New Cosmological Constraints on Primordial Black
  Holes}},\ }\href {https://doi.org/10.1103/PhysRevD.81.104019} {\bibfield
  {journal} {\bibinfo  {journal} {Phys. Rev. D}\ }\textbf {\bibinfo {volume}
  {81}},\ \bibinfo {pages} {104019} (\bibinfo {year} {2010})},\ \Eprint
  {https://arxiv.org/abs/0912.5297} {arXiv:0912.5297 [astro-ph.CO]}
  \BibitemShut {NoStop}%
\bibitem [{\citenamefont {Boudaud}\ and\ \citenamefont
  {Cirelli}(2019)}]{Boudaud:2018hqb}%
  \BibitemOpen
  \bibfield  {author} {\bibinfo {author} {\bibfnamefont {M.}~\bibnamefont
  {Boudaud}}\ and\ \bibinfo {author} {\bibfnamefont {M.}~\bibnamefont
  {Cirelli}},\ }\bibfield  {title} {\bibinfo {title} {{Voyager 1 $e^\pm$
  Further Constrain Primordial Black Holes as Dark Matter}},\ }\href
  {https://doi.org/10.1103/PhysRevLett.122.041104} {\bibfield  {journal}
  {\bibinfo  {journal} {Phys. Rev. Lett.}\ }\textbf {\bibinfo {volume} {122}},\
  \bibinfo {pages} {041104} (\bibinfo {year} {2019})},\ \Eprint
  {https://arxiv.org/abs/1807.03075} {arXiv:1807.03075 [astro-ph.HE]}
  \BibitemShut {NoStop}%
\bibitem [{\citenamefont {DeRocco}\ and\ \citenamefont
  {Graham}(2019)}]{DeRocco:2019fjq}%
  \BibitemOpen
  \bibfield  {author} {\bibinfo {author} {\bibfnamefont {W.}~\bibnamefont
  {DeRocco}}\ and\ \bibinfo {author} {\bibfnamefont {P.~W.}\ \bibnamefont
  {Graham}},\ }\bibfield  {title} {\bibinfo {title} {{Constraining Primordial
  Black Hole Abundance with the Galactic 511 keV Line}},\ }\href
  {https://doi.org/10.1103/PhysRevLett.123.251102} {\bibfield  {journal}
  {\bibinfo  {journal} {Phys. Rev. Lett.}\ }\textbf {\bibinfo {volume} {123}},\
  \bibinfo {pages} {251102} (\bibinfo {year} {2019})},\ \Eprint
  {https://arxiv.org/abs/1906.07740} {arXiv:1906.07740 [astro-ph.CO]}
  \BibitemShut {NoStop}%
\bibitem [{\citenamefont {Laha}(2019)}]{Laha:2019ssq}%
  \BibitemOpen
  \bibfield  {author} {\bibinfo {author} {\bibfnamefont {R.}~\bibnamefont
  {Laha}},\ }\bibfield  {title} {\bibinfo {title} {{Primordial Black Holes as a
  Dark Matter Candidate Are Severely Constrained by the Galactic Center 511 keV
  $\gamma$ -Ray Line}},\ }\href
  {https://doi.org/10.1103/PhysRevLett.123.251101} {\bibfield  {journal}
  {\bibinfo  {journal} {Phys. Rev. Lett.}\ }\textbf {\bibinfo {volume} {123}},\
  \bibinfo {pages} {251101} (\bibinfo {year} {2019})},\ \Eprint
  {https://arxiv.org/abs/1906.09994} {arXiv:1906.09994 [astro-ph.HE]}
  \BibitemShut {NoStop}%
\bibitem [{\citenamefont {Keith}\ and\ \citenamefont
  {Hooper}(2021)}]{Keith:2021guq}%
  \BibitemOpen
  \bibfield  {author} {\bibinfo {author} {\bibfnamefont {C.}~\bibnamefont
  {Keith}}\ and\ \bibinfo {author} {\bibfnamefont {D.}~\bibnamefont {Hooper}},\
  }\bibfield  {title} {\bibinfo {title} {{511~keV excess and primordial black
  holes}},\ }\href {https://doi.org/10.1103/PhysRevD.104.063033} {\bibfield
  {journal} {\bibinfo  {journal} {Phys. Rev. D}\ }\textbf {\bibinfo {volume}
  {104}},\ \bibinfo {pages} {063033} (\bibinfo {year} {2021})},\ \Eprint
  {https://arxiv.org/abs/2103.08611} {arXiv:2103.08611 [astro-ph.CO]}
  \BibitemShut {NoStop}%
\bibitem [{\citenamefont {Sugiyama}\ \emph {et~al.}(2021)\citenamefont
  {Sugiyama}, \citenamefont {Takada},\ and\ \citenamefont
  {Kusenko}}]{Sugiyama:2021xqg}%
  \BibitemOpen
  \bibfield  {author} {\bibinfo {author} {\bibfnamefont {S.}~\bibnamefont
  {Sugiyama}}, \bibinfo {author} {\bibfnamefont {M.}~\bibnamefont {Takada}},\
  and\ \bibinfo {author} {\bibfnamefont {A.}~\bibnamefont {Kusenko}},\
  }\bibfield  {title} {\bibinfo {title} {{Possible Evidence of QCD Axion Stars
  in Hsc and Ogle Microlensing Events}},\ }\href@noop {} {\  (\bibinfo {year}
  {2021})},\ \Eprint {https://arxiv.org/abs/2108.03063} {arXiv:2108.03063
  [hep-ph]} \BibitemShut {NoStop}%
\bibitem [{\citenamefont {Arzoumanian}\ \emph {et~al.}(2021)\citenamefont
  {Arzoumanian} \emph {et~al.}}]{NANOGrav:2021flc}%
  \BibitemOpen
  \bibfield  {author} {\bibinfo {author} {\bibfnamefont {Z.}~\bibnamefont
  {Arzoumanian}} \emph {et~al.} (\bibinfo {collaboration} {NANOGrav}),\
  }\bibfield  {title} {\bibinfo {title} {{Searching for Gravitational Waves
  from Cosmological Phase Transitions with the NANOGrav 12.5-Year Dataset}},\
  }\href {https://doi.org/10.1103/PhysRevLett.127.251302} {\bibfield  {journal}
  {\bibinfo  {journal} {Phys. Rev. Lett.}\ }\textbf {\bibinfo {volume} {127}},\
  \bibinfo {pages} {251302} (\bibinfo {year} {2021})},\ \Eprint
  {https://arxiv.org/abs/2104.13930} {arXiv:2104.13930 [astro-ph.CO]}
  \BibitemShut {NoStop}%
\bibitem [{\citenamefont {Agazie}\ \emph {et~al.}(2023)\citenamefont {Agazie}
  \emph {et~al.}}]{NANOGrav:2023gor}%
  \BibitemOpen
  \bibfield  {author} {\bibinfo {author} {\bibfnamefont {G.}~\bibnamefont
  {Agazie}} \emph {et~al.} (\bibinfo {collaboration} {NANOGrav}),\ }\bibfield
  {title} {\bibinfo {title} {{The NANOGrav 15 yr Data Set: Evidence for a
  Gravitational-wave Background}},\ }\href
  {https://doi.org/10.3847/2041-8213/acdac6} {\bibfield  {journal} {\bibinfo
  {journal} {Astrophys. J. Lett.}\ }\textbf {\bibinfo {volume} {951}},\
  \bibinfo {pages} {L8} (\bibinfo {year} {2023})},\ \Eprint
  {https://arxiv.org/abs/2306.16213} {arXiv:2306.16213 [astro-ph.HE]}
  \BibitemShut {NoStop}%
\bibitem [{\citenamefont {Chen}\ \emph {et~al.}(2021)\citenamefont {Chen} \emph
  {et~al.}}]{Chen:2021rqp}%
  \BibitemOpen
  \bibfield  {author} {\bibinfo {author} {\bibfnamefont {S.}~\bibnamefont
  {Chen}} \emph {et~al.},\ }\bibfield  {title} {\bibinfo {title}
  {{Common-red-signal analysis with 24-yr high-precision timing of the European
  Pulsar Timing Array: Inferences in the stochastic gravitational-wave
  background search}}\ }\href {https://doi.org/10.1093/mnras/stab2833}
  {10.1093/mnras/stab2833} (\bibinfo {year} {2021}),\ \Eprint
  {https://arxiv.org/abs/2110.13184} {arXiv:2110.13184 [astro-ph.HE]}
  \BibitemShut {NoStop}%
\bibitem [{\citenamefont {Antoniadis}\ \emph {et~al.}(2023)\citenamefont
  {Antoniadis} \emph {et~al.}}]{EPTA:2023fyk}%
  \BibitemOpen
  \bibfield  {author} {\bibinfo {author} {\bibfnamefont {J.}~\bibnamefont
  {Antoniadis}} \emph {et~al.} (\bibinfo {collaboration} {EPTA, InPTA:}),\
  }\bibfield  {title} {\bibinfo {title} {{The second data release from the
  European Pulsar Timing Array - III. Search for gravitational wave signals}},\
  }\href {https://doi.org/10.1051/0004-6361/202346844} {\bibfield  {journal}
  {\bibinfo  {journal} {Astron. Astrophys.}\ }\textbf {\bibinfo {volume}
  {678}},\ \bibinfo {pages} {A50} (\bibinfo {year} {2023})},\ \Eprint
  {https://arxiv.org/abs/2306.16214} {arXiv:2306.16214 [astro-ph.HE]}
  \BibitemShut {NoStop}%
\bibitem [{\citenamefont {Goncharov}\ \emph {et~al.}(2021)\citenamefont
  {Goncharov} \emph {et~al.}}]{Goncharov:2021oub}%
  \BibitemOpen
  \bibfield  {author} {\bibinfo {author} {\bibfnamefont {B.}~\bibnamefont
  {Goncharov}} \emph {et~al.},\ }\bibfield  {title} {\bibinfo {title} {{On the
  evidence for a common-spectrum process in the search for the nanohertz
  gravitational-wave background with the Parkes Pulsar Timing Array}}\ }\href
  {https://doi.org/10.3847/2041-8213/ac17f4} {10.3847/2041-8213/ac17f4}
  (\bibinfo {year} {2021}),\ \Eprint {https://arxiv.org/abs/2107.12112}
  {arXiv:2107.12112 [astro-ph.HE]} \BibitemShut {NoStop}%
\bibitem [{\citenamefont {Reardon}\ \emph {et~al.}(2023)\citenamefont {Reardon}
  \emph {et~al.}}]{Reardon:2023gzh}%
  \BibitemOpen
  \bibfield  {author} {\bibinfo {author} {\bibfnamefont {D.~J.}\ \bibnamefont
  {Reardon}} \emph {et~al.},\ }\bibfield  {title} {\bibinfo {title} {{Search
  for an Isotropic Gravitational-wave Background with the Parkes Pulsar Timing
  Array}},\ }\href {https://doi.org/10.3847/2041-8213/acdd02} {\bibfield
  {journal} {\bibinfo  {journal} {Astrophys. J. Lett.}\ }\textbf {\bibinfo
  {volume} {951}},\ \bibinfo {pages} {L6} (\bibinfo {year} {2023})},\ \Eprint
  {https://arxiv.org/abs/2306.16215} {arXiv:2306.16215 [astro-ph.HE]}
  \BibitemShut {NoStop}%
\bibitem [{\citenamefont {Xu}\ \emph {et~al.}(2023)\citenamefont {Xu} \emph
  {et~al.}}]{Xu:2023wog}%
  \BibitemOpen
  \bibfield  {author} {\bibinfo {author} {\bibfnamefont {H.}~\bibnamefont {Xu}}
  \emph {et~al.},\ }\bibfield  {title} {\bibinfo {title} {{Searching for the
  Nano-Hertz Stochastic Gravitational Wave Background with the Chinese Pulsar
  Timing Array Data Release I}},\ }\href
  {https://doi.org/10.1088/1674-4527/acdfa5} {\bibfield  {journal} {\bibinfo
  {journal} {Res. Astron. Astrophys.}\ }\textbf {\bibinfo {volume} {23}},\
  \bibinfo {pages} {075024} (\bibinfo {year} {2023})},\ \Eprint
  {https://arxiv.org/abs/2306.16216} {arXiv:2306.16216 [astro-ph.HE]}
  \BibitemShut {NoStop}%
\bibitem [{\citenamefont {Antoniadis}\ \emph {et~al.}(2022)\citenamefont
  {Antoniadis} \emph {et~al.}}]{Antoniadis:2022pcn}%
  \BibitemOpen
  \bibfield  {author} {\bibinfo {author} {\bibfnamefont {J.}~\bibnamefont
  {Antoniadis}} \emph {et~al.},\ }\bibfield  {title} {\bibinfo {title} {{The
  International Pulsar Timing Array second data release: Search for an
  isotropic gravitational wave background}},\ }\href
  {https://doi.org/10.1093/mnras/stab3418} {\bibfield  {journal} {\bibinfo
  {journal} {Mon. Not. Roy. Astron. Soc.}\ }\textbf {\bibinfo {volume} {510}},\
  \bibinfo {pages} {4873} (\bibinfo {year} {2022})},\ \Eprint
  {https://arxiv.org/abs/2201.03980} {arXiv:2201.03980 [astro-ph.HE]}
  \BibitemShut {NoStop}%
\bibitem [{\citenamefont {Agazie}\ \emph {et~al.}(2024)\citenamefont {Agazie}
  \emph {et~al.}}]{InternationalPulsarTimingArray:2023mzf}%
  \BibitemOpen
  \bibfield  {author} {\bibinfo {author} {\bibfnamefont {G.}~\bibnamefont
  {Agazie}} \emph {et~al.} (\bibinfo {collaboration} {International Pulsar
  Timing Array}),\ }\bibfield  {title} {\bibinfo {title} {{Comparing Recent
  Pulsar Timing Array Results on the Nanohertz Stochastic Gravitational-wave
  Background}},\ }\href {https://doi.org/10.3847/1538-4357/ad36be} {\bibfield
  {journal} {\bibinfo  {journal} {Astrophys. J.}\ }\textbf {\bibinfo {volume}
  {966}},\ \bibinfo {pages} {105} (\bibinfo {year} {2024})},\ \Eprint
  {https://arxiv.org/abs/2309.00693} {arXiv:2309.00693 [astro-ph.HE]}
  \BibitemShut {NoStop}%
\bibitem [{\citenamefont {Romero}\ \emph {et~al.}(2021)\citenamefont {Romero},
  \citenamefont {Martinovic}, \citenamefont {Callister}, \citenamefont {Guo},
  \citenamefont {Mart\'\i{}nez}, \citenamefont {Sakellariadou}, \citenamefont
  {Yang},\ and\ \citenamefont {Zhao}}]{Romero:2021kby}%
  \BibitemOpen
  \bibfield  {author} {\bibinfo {author} {\bibfnamefont {A.}~\bibnamefont
  {Romero}}, \bibinfo {author} {\bibfnamefont {K.}~\bibnamefont {Martinovic}},
  \bibinfo {author} {\bibfnamefont {T.~A.}\ \bibnamefont {Callister}}, \bibinfo
  {author} {\bibfnamefont {H.-K.}\ \bibnamefont {Guo}}, \bibinfo {author}
  {\bibfnamefont {M.}~\bibnamefont {Mart\'\i{}nez}}, \bibinfo {author}
  {\bibfnamefont {M.}~\bibnamefont {Sakellariadou}}, \bibinfo {author}
  {\bibfnamefont {F.-W.}\ \bibnamefont {Yang}},\ and\ \bibinfo {author}
  {\bibfnamefont {Y.}~\bibnamefont {Zhao}},\ }\bibfield  {title} {\bibinfo
  {title} {{Implications for First-Order Cosmological Phase Transitions from
  the Third LIGO-Virgo Observing Run}},\ }\href
  {https://doi.org/10.1103/PhysRevLett.126.151301} {\bibfield  {journal}
  {\bibinfo  {journal} {Phys. Rev. Lett.}\ }\textbf {\bibinfo {volume} {126}},\
  \bibinfo {pages} {151301} (\bibinfo {year} {2021})},\ \Eprint
  {https://arxiv.org/abs/2102.01714} {arXiv:2102.01714 [hep-ph]} \BibitemShut
  {NoStop}%
\bibitem [{\citenamefont {Rubakov}\ and\ \citenamefont
  {Gorbunov}(2017)}]{Rubakov:2017xzr}%
  \BibitemOpen
  \bibfield  {author} {\bibinfo {author} {\bibfnamefont {V.~A.}\ \bibnamefont
  {Rubakov}}\ and\ \bibinfo {author} {\bibfnamefont {D.~S.}\ \bibnamefont
  {Gorbunov}},\ }\href {https://doi.org/10.1142/10447} {\emph {\bibinfo {title}
  {{Introduction to the Theory of the Early Universe}: {Hot big bang
  theory}}}}\ (\bibinfo  {publisher} {World Scientific},\ \bibinfo {address}
  {Singapore},\ \bibinfo {year} {2017})\BibitemShut {NoStop}%
\bibitem [{\citenamefont {Escriv\`a}\ and\ \citenamefont
  {Romano}(2021)}]{Escriva:2021pmf}%
  \BibitemOpen
  \bibfield  {author} {\bibinfo {author} {\bibfnamefont {A.}~\bibnamefont
  {Escriv\`a}}\ and\ \bibinfo {author} {\bibfnamefont {A.~E.}\ \bibnamefont
  {Romano}},\ }\bibfield  {title} {\bibinfo {title} {{Effects of the shape of
  curvature peaks on the size of primordial black holes}},\ }\href
  {https://doi.org/10.1088/1475-7516/2021/05/066} {\bibfield  {journal}
  {\bibinfo  {journal} {JCAP}\ }\textbf {\bibinfo {volume} {05}},\ \bibinfo
  {pages} {066}},\ \Eprint {https://arxiv.org/abs/2103.03867} {arXiv:2103.03867
  [gr-qc]} \BibitemShut {NoStop}%
\bibitem [{\citenamefont {Caprini}\ \emph {et~al.}(2016)\citenamefont {Caprini}
  \emph {et~al.}}]{Caprini:2015zlo}%
  \BibitemOpen
  \bibfield  {author} {\bibinfo {author} {\bibfnamefont {C.}~\bibnamefont
  {Caprini}} \emph {et~al.},\ }\bibfield  {title} {\bibinfo {title} {{Science
  with the space-based interferometer eLISA. II: Gravitational waves from
  cosmological phase transitions}},\ }\href
  {https://doi.org/10.1088/1475-7516/2016/04/001} {\bibfield  {journal}
  {\bibinfo  {journal} {JCAP}\ }\textbf {\bibinfo {volume} {04}},\ \bibinfo
  {pages} {001}},\ \Eprint {https://arxiv.org/abs/1512.06239} {arXiv:1512.06239
  [astro-ph.CO]} \BibitemShut {NoStop}%
\bibitem [{\citenamefont {Caprini}\ \emph {et~al.}(2020)\citenamefont {Caprini}
  \emph {et~al.}}]{Caprini:2019egz}%
  \BibitemOpen
  \bibfield  {author} {\bibinfo {author} {\bibfnamefont {C.}~\bibnamefont
  {Caprini}} \emph {et~al.},\ }\bibfield  {title} {\bibinfo {title} {{Detecting
  gravitational waves from cosmological phase transitions with LISA: an
  update}},\ }\href {https://doi.org/10.1088/1475-7516/2020/03/024} {\bibfield
  {journal} {\bibinfo  {journal} {JCAP}\ }\textbf {\bibinfo {volume} {03}},\
  \bibinfo {pages} {024}},\ \Eprint {https://arxiv.org/abs/1910.13125}
  {arXiv:1910.13125 [astro-ph.CO]} \BibitemShut {NoStop}%
\bibitem [{\citenamefont {Gouttenoire}\ \emph {et~al.}(2022)\citenamefont
  {Gouttenoire}, \citenamefont {Jinno},\ and\ \citenamefont
  {Sala}}]{Gouttenoire:2021kjv}%
  \BibitemOpen
  \bibfield  {author} {\bibinfo {author} {\bibfnamefont {Y.}~\bibnamefont
  {Gouttenoire}}, \bibinfo {author} {\bibfnamefont {R.}~\bibnamefont {Jinno}},\
  and\ \bibinfo {author} {\bibfnamefont {F.}~\bibnamefont {Sala}},\ }\bibfield
  {title} {\bibinfo {title} {{Friction pressure on relativistic bubble
  walls}},\ }\href {https://doi.org/10.1007/JHEP05(2022)004} {\bibfield
  {journal} {\bibinfo  {journal} {JHEP}\ }\textbf {\bibinfo {volume} {05}},\
  \bibinfo {pages} {004}},\ \Eprint {https://arxiv.org/abs/2112.07686}
  {arXiv:2112.07686 [hep-ph]} \BibitemShut {NoStop}%
\bibitem [{\citenamefont {Gouttenoire}(2022)}]{Gouttenoire:2022gwi}%
  \BibitemOpen
  \bibfield  {author} {\bibinfo {author} {\bibfnamefont {Y.}~\bibnamefont
  {Gouttenoire}},\ }\href {https://doi.org/10.1007/978-3-031-11862-3} {\emph
  {\bibinfo {title} {{Beyond the Standard Model Cocktail}}}},\ Springer Theses\
  (\bibinfo  {publisher} {Springer},\ \bibinfo {address} {Cham},\ \bibinfo
  {year} {2022})\ \Eprint {https://arxiv.org/abs/2207.01633} {arXiv:2207.01633
  [hep-ph]} \BibitemShut {NoStop}%
\bibitem [{\citenamefont {Carr}\ and\ \citenamefont
  {Kuhnel}(2020)}]{Carr:2020xqk}%
  \BibitemOpen
  \bibfield  {author} {\bibinfo {author} {\bibfnamefont {B.}~\bibnamefont
  {Carr}}\ and\ \bibinfo {author} {\bibfnamefont {F.}~\bibnamefont {Kuhnel}},\
  }\bibfield  {title} {\bibinfo {title} {{Primordial Black Holes as Dark
  Matter: Recent Developments}},\ }\href
  {https://doi.org/10.1146/annurev-nucl-050520-125911} {\bibfield  {journal}
  {\bibinfo  {journal} {Ann. Rev. Nucl. Part. Sci.}\ }\textbf {\bibinfo
  {volume} {70}},\ \bibinfo {pages} {355} (\bibinfo {year} {2020})},\ \Eprint
  {https://arxiv.org/abs/2006.02838} {arXiv:2006.02838 [astro-ph.CO]}
  \BibitemShut {NoStop}%
\bibitem [{\citenamefont {Gouttenoire}(2023)}]{Gouttenoire:2023bqy}%
  \BibitemOpen
  \bibfield  {author} {\bibinfo {author} {\bibfnamefont {Y.}~\bibnamefont
  {Gouttenoire}},\ }\bibfield  {title} {\bibinfo {title} {{First-Order Phase
  Transition Interpretation of Pulsar Timing Array Signal Is Consistent with
  Solar-Mass Black Holes}},\ }\href
  {https://doi.org/10.1103/PhysRevLett.131.171404} {\bibfield  {journal}
  {\bibinfo  {journal} {Phys. Rev. Lett.}\ }\textbf {\bibinfo {volume} {131}},\
  \bibinfo {pages} {171404} (\bibinfo {year} {2023})},\ \Eprint
  {https://arxiv.org/abs/2307.04239} {arXiv:2307.04239 [hep-ph]} \BibitemShut
  {NoStop}%
\bibitem [{\citenamefont {Ellis}\ \emph {et~al.}(2024)\citenamefont {Ellis},
  \citenamefont {Fairbairn}, \citenamefont {Franciolini}, \citenamefont
  {H\"utsi}, \citenamefont {Iovino}, \citenamefont {Lewicki}, \citenamefont
  {Raidal}, \citenamefont {Urrutia}, \citenamefont {Vaskonen},\ and\
  \citenamefont {Veerm\"ae}}]{Ellis:2023oxs}%
  \BibitemOpen
  \bibfield  {author} {\bibinfo {author} {\bibfnamefont {J.}~\bibnamefont
  {Ellis}}, \bibinfo {author} {\bibfnamefont {M.}~\bibnamefont {Fairbairn}},
  \bibinfo {author} {\bibfnamefont {G.}~\bibnamefont {Franciolini}}, \bibinfo
  {author} {\bibfnamefont {G.}~\bibnamefont {H\"utsi}}, \bibinfo {author}
  {\bibfnamefont {A.}~\bibnamefont {Iovino}}, \bibinfo {author} {\bibfnamefont
  {M.}~\bibnamefont {Lewicki}}, \bibinfo {author} {\bibfnamefont
  {M.}~\bibnamefont {Raidal}}, \bibinfo {author} {\bibfnamefont
  {J.}~\bibnamefont {Urrutia}}, \bibinfo {author} {\bibfnamefont
  {V.}~\bibnamefont {Vaskonen}},\ and\ \bibinfo {author} {\bibfnamefont
  {H.}~\bibnamefont {Veerm\"ae}},\ }\bibfield  {title} {\bibinfo {title} {{What
  is the source of the PTA GW signal?}},\ }\href
  {https://doi.org/10.1103/PhysRevD.109.023522} {\bibfield  {journal} {\bibinfo
   {journal} {Phys. Rev. D}\ }\textbf {\bibinfo {volume} {109}},\ \bibinfo
  {pages} {023522} (\bibinfo {year} {2024})},\ \Eprint
  {https://arxiv.org/abs/2308.08546} {arXiv:2308.08546 [astro-ph.CO]}
  \BibitemShut {NoStop}%
\bibitem [{\citenamefont {Lewicki}\ \emph {et~al.}(2024)\citenamefont
  {Lewicki}, \citenamefont {Toczek},\ and\ \citenamefont
  {Vaskonen}}]{Lewicki:2024ghw}%
  \BibitemOpen
  \bibfield  {author} {\bibinfo {author} {\bibfnamefont {M.}~\bibnamefont
  {Lewicki}}, \bibinfo {author} {\bibfnamefont {P.}~\bibnamefont {Toczek}},\
  and\ \bibinfo {author} {\bibfnamefont {V.}~\bibnamefont {Vaskonen}},\
  }\bibfield  {title} {\bibinfo {title} {{Black holes and gravitational waves
  from slow phase transitions}},\ }\href@noop {} {\  (\bibinfo {year}
  {2024})},\ \Eprint {https://arxiv.org/abs/2402.04158} {arXiv:2402.04158
  [astro-ph.CO]} \BibitemShut {NoStop}%
\bibitem [{\citenamefont {Jinno}\ and\ \citenamefont
  {Takimoto}(2019)}]{Jinno:2017fby}%
  \BibitemOpen
  \bibfield  {author} {\bibinfo {author} {\bibfnamefont {R.}~\bibnamefont
  {Jinno}}\ and\ \bibinfo {author} {\bibfnamefont {M.}~\bibnamefont
  {Takimoto}},\ }\bibfield  {title} {\bibinfo {title} {{Gravitational Waves
  from Bubble Dynamics: Beyond the Envelope}},\ }\href
  {https://doi.org/10.1088/1475-7516/2019/01/060} {\bibfield  {journal}
  {\bibinfo  {journal} {JCAP}\ }\textbf {\bibinfo {volume} {01}},\ \bibinfo
  {pages} {060}},\ \Eprint {https://arxiv.org/abs/1707.03111} {arXiv:1707.03111
  [hep-ph]} \BibitemShut {NoStop}%
\bibitem [{\citenamefont {Konstandin}(2018)}]{Konstandin:2017sat}%
  \BibitemOpen
  \bibfield  {author} {\bibinfo {author} {\bibfnamefont {T.}~\bibnamefont
  {Konstandin}},\ }\bibfield  {title} {\bibinfo {title} {{Gravitational
  Radiation from a Bulk Flow Model}},\ }\href
  {https://doi.org/10.1088/1475-7516/2018/03/047} {\bibfield  {journal}
  {\bibinfo  {journal} {JCAP}\ }\textbf {\bibinfo {volume} {03}},\ \bibinfo
  {pages} {047}},\ \Eprint {https://arxiv.org/abs/1712.06869} {arXiv:1712.06869
  [astro-ph.CO]} \BibitemShut {NoStop}%
\bibitem [{\citenamefont {Hawking}(1974)}]{Hawking:1974rv}%
  \BibitemOpen
  \bibfield  {author} {\bibinfo {author} {\bibfnamefont {S.~W.}\ \bibnamefont
  {Hawking}},\ }\bibfield  {title} {\bibinfo {title} {{Black Hole
  Explosions}},\ }\href {https://doi.org/10.1038/248030a0} {\bibfield
  {journal} {\bibinfo  {journal} {Nature}\ }\textbf {\bibinfo {volume} {248}},\
  \bibinfo {pages} {30} (\bibinfo {year} {1974})}\BibitemShut {NoStop}%
\bibitem [{\citenamefont {Hawking}(1975)}]{Hawking:1975vcx}%
  \BibitemOpen
  \bibfield  {author} {\bibinfo {author} {\bibfnamefont {S.~W.}\ \bibnamefont
  {Hawking}},\ }\bibfield  {title} {\bibinfo {title} {{Particle Creation by
  Black Holes}},\ }\href {https://doi.org/10.1007/BF02345020} {\bibfield
  {journal} {\bibinfo  {journal} {Commun. Math. Phys.}\ }\textbf {\bibinfo
  {volume} {43}},\ \bibinfo {pages} {199} (\bibinfo {year} {1975})},\ \bibinfo
  {note} {[Erratum: Commun.Math.Phys. 46, 206 (1976)]}\BibitemShut {NoStop}%
\bibitem [{\citenamefont {Acharya}\ and\ \citenamefont
  {Khatri}(2020)}]{Acharya:2020jbv}%
  \BibitemOpen
  \bibfield  {author} {\bibinfo {author} {\bibfnamefont {S.~K.}\ \bibnamefont
  {Acharya}}\ and\ \bibinfo {author} {\bibfnamefont {R.}~\bibnamefont
  {Khatri}},\ }\bibfield  {title} {\bibinfo {title} {{Cmb and Bbn Constraints
  on Evaporating Primordial Black Holes Revisited}},\ }\href
  {https://doi.org/10.1088/1475-7516/2020/06/018} {\bibfield  {journal}
  {\bibinfo  {journal} {JCAP}\ }\textbf {\bibinfo {volume} {06}},\ \bibinfo
  {pages} {018}},\ \Eprint {https://arxiv.org/abs/2002.00898} {arXiv:2002.00898
  [astro-ph.CO]} \BibitemShut {NoStop}%
\bibitem [{\citenamefont {Keith}\ \emph {et~al.}(2020)\citenamefont {Keith},
  \citenamefont {Hooper}, \citenamefont {Blinov},\ and\ \citenamefont
  {McDermott}}]{Keith:2020jww}%
  \BibitemOpen
  \bibfield  {author} {\bibinfo {author} {\bibfnamefont {C.}~\bibnamefont
  {Keith}}, \bibinfo {author} {\bibfnamefont {D.}~\bibnamefont {Hooper}},
  \bibinfo {author} {\bibfnamefont {N.}~\bibnamefont {Blinov}},\ and\ \bibinfo
  {author} {\bibfnamefont {S.~D.}\ \bibnamefont {McDermott}},\ }\bibfield
  {title} {\bibinfo {title} {{Constraints on Primordial Black Holes from Big
  Bang Nucleosynthesis Revisited}},\ }\href
  {https://doi.org/10.1103/PhysRevD.102.103512} {\bibfield  {journal} {\bibinfo
   {journal} {Phys. Rev. D}\ }\textbf {\bibinfo {volume} {102}},\ \bibinfo
  {pages} {103512} (\bibinfo {year} {2020})},\ \Eprint
  {https://arxiv.org/abs/2006.03608} {arXiv:2006.03608 [astro-ph.CO]}
  \BibitemShut {NoStop}%
\bibitem [{\citenamefont {Gouttenoire}(2024)}]{Gouttenoire:2023pxh}%
  \BibitemOpen
  \bibfield  {author} {\bibinfo {author} {\bibfnamefont {Y.}~\bibnamefont
  {Gouttenoire}},\ }\bibfield  {title} {\bibinfo {title} {{Primordial Black
  Holes from Conformal Higgs}},\ }\bibfield  {journal} {\bibinfo  {journal}
  {Phys. Lett. B}\ }\href {https://doi.org/10.1016/j.physletb.2024.138800}
  {10.1016/j.physletb.2024.138800} (\bibinfo {year} {2024}),\ \Eprint
  {https://arxiv.org/abs/2311.13640} {arXiv:2311.13640 [hep-ph]} \BibitemShut
  {NoStop}%
\bibitem [{\citenamefont {Abbott}\ \emph
  {et~al.}(2019{\natexlab{b}})\citenamefont {Abbott} \emph
  {et~al.}}]{LIGOScientific:2019kan}%
  \BibitemOpen
  \bibfield  {author} {\bibinfo {author} {\bibfnamefont {B.~P.}\ \bibnamefont
  {Abbott}} \emph {et~al.} (\bibinfo {collaboration} {LIGO Scientific,
  Virgo}),\ }\bibfield  {title} {\bibinfo {title} {{Search for Subsolar Mass
  Ultracompact Binaries in Advanced LIGO\textquoteright{}s Second Observing
  Run}},\ }\href {https://doi.org/10.1103/PhysRevLett.123.161102} {\bibfield
  {journal} {\bibinfo  {journal} {Phys. Rev. Lett.}\ }\textbf {\bibinfo
  {volume} {123}},\ \bibinfo {pages} {161102} (\bibinfo {year}
  {2019}{\natexlab{b}})},\ \Eprint {https://arxiv.org/abs/1904.08976}
  {arXiv:1904.08976 [astro-ph.CO]} \BibitemShut {NoStop}%
\bibitem [{\citenamefont {Coleman}\ and\ \citenamefont
  {Weinberg}(1973)}]{Coleman:1973jx}%
  \BibitemOpen
  \bibfield  {author} {\bibinfo {author} {\bibfnamefont {S.~R.}\ \bibnamefont
  {Coleman}}\ and\ \bibinfo {author} {\bibfnamefont {E.~J.}\ \bibnamefont
  {Weinberg}},\ }\bibfield  {title} {\bibinfo {title} {{Radiative Corrections
  as the Origin of Spontaneous Symmetry Breaking}},\ }\href
  {https://doi.org/10.1103/PhysRevD.7.1888} {\bibfield  {journal} {\bibinfo
  {journal} {Phys. Rev. D}\ }\textbf {\bibinfo {volume} {7}},\ \bibinfo {pages}
  {1888} (\bibinfo {year} {1973})}\BibitemShut {NoStop}%
\bibitem [{\citenamefont {Witten}(1981)}]{Witten:1980ez}%
  \BibitemOpen
  \bibfield  {author} {\bibinfo {author} {\bibfnamefont {E.}~\bibnamefont
  {Witten}},\ }\bibfield  {title} {\bibinfo {title} {{Cosmological Consequences
  of a Light Higgs Boson}},\ }\href
  {https://doi.org/10.1016/0550-3213(81)90182-6} {\bibfield  {journal}
  {\bibinfo  {journal} {Nucl. Phys. B}\ }\textbf {\bibinfo {volume} {177}},\
  \bibinfo {pages} {477} (\bibinfo {year} {1981})}\BibitemShut {NoStop}%
\bibitem [{\citenamefont {Hempfling}(1996)}]{Hempfling:1996ht}%
  \BibitemOpen
  \bibfield  {author} {\bibinfo {author} {\bibfnamefont {R.}~\bibnamefont
  {Hempfling}},\ }\bibfield  {title} {\bibinfo {title} {{The Next-To-Minimal
  Coleman-Weinberg Model}},\ }\href
  {https://doi.org/10.1016/0370-2693(96)00446-7} {\bibfield  {journal}
  {\bibinfo  {journal} {Phys. Lett. B}\ }\textbf {\bibinfo {volume} {379}},\
  \bibinfo {pages} {153} (\bibinfo {year} {1996})},\ \Eprint
  {https://arxiv.org/abs/hep-ph/9604278} {arXiv:hep-ph/9604278} \BibitemShut
  {NoStop}%
\bibitem [{\citenamefont {Bardeen}(1995)}]{Bardeen:1995}%
  \BibitemOpen
  \bibfield  {author} {\bibinfo {author} {\bibfnamefont {W.~A.}\ \bibnamefont
  {Bardeen}},\ }\bibfield  {title} {\bibinfo {title} {{On naturalness in the
  standard model}},\ }in\ \href
  {http://lss.fnal.gov/cgi-bin/find_paper.pl?conf-95-391} {\emph {\bibinfo
  {booktitle} {{Ontake Summer Institute on Particle Physics Ontake Mountain,
  Japan, August 27-September 2, 1995}}}}\ (\bibinfo {year} {1995})\BibitemShut
  {NoStop}%
\bibitem [{\citenamefont {Iso}\ \emph {et~al.}(2017)\citenamefont {Iso},
  \citenamefont {Serpico},\ and\ \citenamefont {Shimada}}]{Iso:2017uuu}%
  \BibitemOpen
  \bibfield  {author} {\bibinfo {author} {\bibfnamefont {S.}~\bibnamefont
  {Iso}}, \bibinfo {author} {\bibfnamefont {P.~D.}\ \bibnamefont {Serpico}},\
  and\ \bibinfo {author} {\bibfnamefont {K.}~\bibnamefont {Shimada}},\
  }\bibfield  {title} {\bibinfo {title} {{QCD-Electroweak First-Order Phase
  Transition in a Supercooled Universe}},\ }\href
  {https://doi.org/10.1103/PhysRevLett.119.141301} {\bibfield  {journal}
  {\bibinfo  {journal} {Phys. Rev. Lett.}\ }\textbf {\bibinfo {volume} {119}},\
  \bibinfo {pages} {141301} (\bibinfo {year} {2017})},\ \Eprint
  {https://arxiv.org/abs/1704.04955} {arXiv:1704.04955 [hep-ph]} \BibitemShut
  {NoStop}%
\bibitem [{\citenamefont {von Harling}\ and\ \citenamefont
  {Servant}(2018)}]{vonHarling:2017yew}%
  \BibitemOpen
  \bibfield  {author} {\bibinfo {author} {\bibfnamefont {B.}~\bibnamefont {von
  Harling}}\ and\ \bibinfo {author} {\bibfnamefont {G.}~\bibnamefont
  {Servant}},\ }\bibfield  {title} {\bibinfo {title} {{QCD-induced Electroweak
  Phase Transition}},\ }\href {https://doi.org/10.1007/JHEP01(2018)159}
  {\bibfield  {journal} {\bibinfo  {journal} {JHEP}\ }\textbf {\bibinfo
  {volume} {01}},\ \bibinfo {pages} {159}},\ \Eprint
  {https://arxiv.org/abs/1711.11554} {arXiv:1711.11554 [hep-ph]} \BibitemShut
  {NoStop}%
\bibitem [{\citenamefont {Randall}\ and\ \citenamefont
  {Servant}(2007)}]{Randall:2006py}%
  \BibitemOpen
  \bibfield  {author} {\bibinfo {author} {\bibfnamefont {L.}~\bibnamefont
  {Randall}}\ and\ \bibinfo {author} {\bibfnamefont {G.}~\bibnamefont
  {Servant}},\ }\bibfield  {title} {\bibinfo {title} {{Gravitational Waves from
  Warped Spacetime}},\ }\href {https://doi.org/10.1088/1126-6708/2007/05/054}
  {\bibfield  {journal} {\bibinfo  {journal} {JHEP}\ }\textbf {\bibinfo
  {volume} {05}},\ \bibinfo {pages} {054}},\ \Eprint
  {https://arxiv.org/abs/hep-ph/0607158} {arXiv:hep-ph/0607158} \BibitemShut
  {NoStop}%
\bibitem [{\citenamefont {Jinno}\ and\ \citenamefont
  {Takimoto}(2017)}]{Jinno:2016knw}%
  \BibitemOpen
  \bibfield  {author} {\bibinfo {author} {\bibfnamefont {R.}~\bibnamefont
  {Jinno}}\ and\ \bibinfo {author} {\bibfnamefont {M.}~\bibnamefont
  {Takimoto}},\ }\bibfield  {title} {\bibinfo {title} {{Probing a classically
  conformal B-L model with gravitational waves}},\ }\href
  {https://doi.org/10.1103/PhysRevD.95.015020} {\bibfield  {journal} {\bibinfo
  {journal} {Phys. Rev.}\ }\textbf {\bibinfo {volume} {D95}},\ \bibinfo {pages}
  {015020} (\bibinfo {year} {2017})},\ \Eprint
  {https://arxiv.org/abs/1604.05035} {arXiv:1604.05035 [hep-ph]} \BibitemShut
  {NoStop}%
\bibitem [{\citenamefont {Brdar}\ \emph
  {et~al.}(2019{\natexlab{a}})\citenamefont {Brdar}, \citenamefont
  {Helmboldt},\ and\ \citenamefont {Kubo}}]{Brdar:2018num}%
  \BibitemOpen
  \bibfield  {author} {\bibinfo {author} {\bibfnamefont {V.}~\bibnamefont
  {Brdar}}, \bibinfo {author} {\bibfnamefont {A.~J.}\ \bibnamefont
  {Helmboldt}},\ and\ \bibinfo {author} {\bibfnamefont {J.}~\bibnamefont
  {Kubo}},\ }\bibfield  {title} {\bibinfo {title} {{Gravitational Waves from
  First-Order Phase Transitions: LIGO as a Window to Unexplored Seesaw
  Scales}},\ }\href {https://doi.org/10.1088/1475-7516/2019/02/021} {\bibfield
  {journal} {\bibinfo  {journal} {JCAP}\ }\textbf {\bibinfo {volume} {02}},\
  \bibinfo {pages} {021}},\ \Eprint {https://arxiv.org/abs/1810.12306}
  {arXiv:1810.12306 [hep-ph]} \BibitemShut {NoStop}%
\bibitem [{\citenamefont {Brdar}\ \emph
  {et~al.}(2019{\natexlab{b}})\citenamefont {Brdar}, \citenamefont
  {Helmboldt},\ and\ \citenamefont {Lindner}}]{Brdar:2019qut}%
  \BibitemOpen
  \bibfield  {author} {\bibinfo {author} {\bibfnamefont {V.}~\bibnamefont
  {Brdar}}, \bibinfo {author} {\bibfnamefont {A.~J.}\ \bibnamefont
  {Helmboldt}},\ and\ \bibinfo {author} {\bibfnamefont {M.}~\bibnamefont
  {Lindner}},\ }\bibfield  {title} {\bibinfo {title} {{Strong Supercooling as a
  Consequence of Renormalization Group Consistency}},\ }\href
  {https://doi.org/10.1007/JHEP12(2019)158} {\bibfield  {journal} {\bibinfo
  {journal} {JHEP}\ }\textbf {\bibinfo {volume} {12}},\ \bibinfo {pages}
  {158}},\ \Eprint {https://arxiv.org/abs/1910.13460} {arXiv:1910.13460
  [hep-ph]} \BibitemShut {NoStop}%
\bibitem [{\citenamefont {Marzo}\ \emph {et~al.}(2019)\citenamefont {Marzo},
  \citenamefont {Marzola},\ and\ \citenamefont {Vaskonen}}]{Marzo:2018}%
  \BibitemOpen
  \bibfield  {author} {\bibinfo {author} {\bibfnamefont {C.}~\bibnamefont
  {Marzo}}, \bibinfo {author} {\bibfnamefont {L.}~\bibnamefont {Marzola}},\
  and\ \bibinfo {author} {\bibfnamefont {V.}~\bibnamefont {Vaskonen}},\
  }\bibfield  {title} {\bibinfo {title} {{Phase transition and vacuum stability
  in the classically conformal B\textendash{}L model}},\ }\href
  {https://doi.org/10.1140/epjc/s10052-019-7076-x} {\bibfield  {journal}
  {\bibinfo  {journal} {Eur. Phys. J. C}\ }\textbf {\bibinfo {volume} {79}},\
  \bibinfo {pages} {601} (\bibinfo {year} {2019})},\ \Eprint
  {https://arxiv.org/abs/1811.11169} {arXiv:1811.11169 [hep-ph]} \BibitemShut
  {NoStop}%
\bibitem [{\citenamefont {Ellis}\ \emph {et~al.}(2019)\citenamefont {Ellis},
  \citenamefont {Lewicki}, \citenamefont {No},\ and\ \citenamefont
  {Vaskonen}}]{Ellis:2019oqb}%
  \BibitemOpen
  \bibfield  {author} {\bibinfo {author} {\bibfnamefont {J.}~\bibnamefont
  {Ellis}}, \bibinfo {author} {\bibfnamefont {M.}~\bibnamefont {Lewicki}},
  \bibinfo {author} {\bibfnamefont {J.~M.}\ \bibnamefont {No}},\ and\ \bibinfo
  {author} {\bibfnamefont {V.}~\bibnamefont {Vaskonen}},\ }\bibfield  {title}
  {\bibinfo {title} {{Gravitational wave energy budget in strongly supercooled
  phase transitions}},\ }\href {https://doi.org/10.1088/1475-7516/2019/06/024}
  {\bibfield  {journal} {\bibinfo  {journal} {JCAP}\ }\textbf {\bibinfo
  {volume} {06}},\ \bibinfo {pages} {024}},\ \Eprint
  {https://arxiv.org/abs/1903.09642} {arXiv:1903.09642 [hep-ph]} \BibitemShut
  {NoStop}%
\bibitem [{\citenamefont {Ellis}\ \emph {et~al.}(2020)\citenamefont {Ellis},
  \citenamefont {Lewicki},\ and\ \citenamefont {Vaskonen}}]{Ellis:2020nnr}%
  \BibitemOpen
  \bibfield  {author} {\bibinfo {author} {\bibfnamefont {J.}~\bibnamefont
  {Ellis}}, \bibinfo {author} {\bibfnamefont {M.}~\bibnamefont {Lewicki}},\
  and\ \bibinfo {author} {\bibfnamefont {V.}~\bibnamefont {Vaskonen}},\
  }\bibfield  {title} {\bibinfo {title} {{Updated predictions for gravitational
  waves produced in a strongly supercooled phase transition}},\ }\href
  {https://doi.org/10.1088/1475-7516/2020/11/020} {\bibfield  {journal}
  {\bibinfo  {journal} {JCAP}\ }\textbf {\bibinfo {volume} {11}},\ \bibinfo
  {pages} {020}},\ \Eprint {https://arxiv.org/abs/2007.15586} {arXiv:2007.15586
  [astro-ph.CO]} \BibitemShut {NoStop}%
\bibitem [{\citenamefont {Baldes}\ and\ \citenamefont
  {Garcia-Cely}(2019)}]{Baldes:2018emh}%
  \BibitemOpen
  \bibfield  {author} {\bibinfo {author} {\bibfnamefont {I.}~\bibnamefont
  {Baldes}}\ and\ \bibinfo {author} {\bibfnamefont {C.}~\bibnamefont
  {Garcia-Cely}},\ }\bibfield  {title} {\bibinfo {title} {{Strong gravitational
  radiation from a simple dark matter model}},\ }\href
  {https://doi.org/10.1007/JHEP05(2019)190} {\bibfield  {journal} {\bibinfo
  {journal} {JHEP}\ }\textbf {\bibinfo {volume} {05}},\ \bibinfo {pages}
  {190}},\ \Eprint {https://arxiv.org/abs/1809.01198} {arXiv:1809.01198
  [hep-ph]} \BibitemShut {NoStop}%
\bibitem [{\citenamefont {Prokopec}\ \emph {et~al.}(2019)\citenamefont
  {Prokopec}, \citenamefont {Rezacek},\ and\ \citenamefont
  {\'Swie\.zewska}}]{Prokopec:2018tnq}%
  \BibitemOpen
  \bibfield  {author} {\bibinfo {author} {\bibfnamefont {T.}~\bibnamefont
  {Prokopec}}, \bibinfo {author} {\bibfnamefont {J.}~\bibnamefont {Rezacek}},\
  and\ \bibinfo {author} {\bibfnamefont {B.}~\bibnamefont {\'Swie\.zewska}},\
  }\bibfield  {title} {\bibinfo {title} {{Gravitational waves from conformal
  symmetry breaking}},\ }\href {https://doi.org/10.1088/1475-7516/2019/02/009}
  {\bibfield  {journal} {\bibinfo  {journal} {JCAP}\ }\textbf {\bibinfo
  {volume} {02}},\ \bibinfo {pages} {009}},\ \Eprint
  {https://arxiv.org/abs/1809.11129} {arXiv:1809.11129 [hep-ph]} \BibitemShut
  {NoStop}%
\bibitem [{\citenamefont {Delle~Rose}\ \emph {et~al.}(2020)\citenamefont
  {Delle~Rose}, \citenamefont {Panico}, \citenamefont {Redi},\ and\
  \citenamefont {Tesi}}]{DelleRose:2019pgi}%
  \BibitemOpen
  \bibfield  {author} {\bibinfo {author} {\bibfnamefont {L.}~\bibnamefont
  {Delle~Rose}}, \bibinfo {author} {\bibfnamefont {G.}~\bibnamefont {Panico}},
  \bibinfo {author} {\bibfnamefont {M.}~\bibnamefont {Redi}},\ and\ \bibinfo
  {author} {\bibfnamefont {A.}~\bibnamefont {Tesi}},\ }\bibfield  {title}
  {\bibinfo {title} {{Gravitational Waves from Supercool Axions}},\ }\href
  {https://doi.org/10.1007/JHEP04(2020)025} {\bibfield  {journal} {\bibinfo
  {journal} {JHEP}\ }\textbf {\bibinfo {volume} {04}},\ \bibinfo {pages}
  {025}},\ \Eprint {https://arxiv.org/abs/1912.06139} {arXiv:1912.06139
  [hep-ph]} \BibitemShut {NoStop}%
\bibitem [{\citenamefont {Kierkla}\ \emph {et~al.}(2022)\citenamefont
  {Kierkla}, \citenamefont {Karam},\ and\ \citenamefont
  {Swiezewska}}]{Kierkla:2022odc}%
  \BibitemOpen
  \bibfield  {author} {\bibinfo {author} {\bibfnamefont {M.}~\bibnamefont
  {Kierkla}}, \bibinfo {author} {\bibfnamefont {A.}~\bibnamefont {Karam}},\
  and\ \bibinfo {author} {\bibfnamefont {B.}~\bibnamefont {Swiezewska}},\
  }\bibfield  {title} {\bibinfo {title} {{Conformal Model for Gravitational
  Waves and Dark Matter: a Status Update}},\ }\href@noop {} {\  (\bibinfo
  {year} {2022})},\ \Eprint {https://arxiv.org/abs/2210.07075}
  {arXiv:2210.07075 [astro-ph.CO]} \BibitemShut {NoStop}%
\bibitem [{\citenamefont {Konstandin}\ and\ \citenamefont
  {Servant}(2011{\natexlab{a}})}]{Konstandin:2011dr}%
  \BibitemOpen
  \bibfield  {author} {\bibinfo {author} {\bibfnamefont {T.}~\bibnamefont
  {Konstandin}}\ and\ \bibinfo {author} {\bibfnamefont {G.}~\bibnamefont
  {Servant}},\ }\bibfield  {title} {\bibinfo {title} {{Cosmological
  Consequences of Nearly Conformal Dynamics at the TeV scale}},\ }\href
  {https://doi.org/10.1088/1475-7516/2011/12/009} {\bibfield  {journal}
  {\bibinfo  {journal} {JCAP}\ }\textbf {\bibinfo {volume} {12}},\ \bibinfo
  {pages} {009}},\ \Eprint {https://arxiv.org/abs/1104.4791} {arXiv:1104.4791
  [hep-ph]} \BibitemShut {NoStop}%
\bibitem [{\citenamefont {Hambye}\ \emph {et~al.}(2018)\citenamefont {Hambye},
  \citenamefont {Strumia},\ and\ \citenamefont {Teresi}}]{Hambye:2018qjv}%
  \BibitemOpen
  \bibfield  {author} {\bibinfo {author} {\bibfnamefont {T.}~\bibnamefont
  {Hambye}}, \bibinfo {author} {\bibfnamefont {A.}~\bibnamefont {Strumia}},\
  and\ \bibinfo {author} {\bibfnamefont {D.}~\bibnamefont {Teresi}},\
  }\bibfield  {title} {\bibinfo {title} {{Super-Cool Dark Matter}},\ }\href
  {https://doi.org/10.1007/JHEP08(2018)188} {\bibfield  {journal} {\bibinfo
  {journal} {JHEP}\ }\textbf {\bibinfo {volume} {08}},\ \bibinfo {pages}
  {188}},\ \Eprint {https://arxiv.org/abs/1805.01473} {arXiv:1805.01473
  [hep-ph]} \BibitemShut {NoStop}%
\bibitem [{\citenamefont {Baldes}\ \emph
  {et~al.}(2021{\natexlab{a}})\citenamefont {Baldes}, \citenamefont
  {Gouttenoire},\ and\ \citenamefont {Sala}}]{Baldes:2020kam}%
  \BibitemOpen
  \bibfield  {author} {\bibinfo {author} {\bibfnamefont {I.}~\bibnamefont
  {Baldes}}, \bibinfo {author} {\bibfnamefont {Y.}~\bibnamefont
  {Gouttenoire}},\ and\ \bibinfo {author} {\bibfnamefont {F.}~\bibnamefont
  {Sala}},\ }\bibfield  {title} {\bibinfo {title} {{String Fragmentation in
  Supercooled Confinement and Implications for Dark Matter}},\ }\href
  {https://doi.org/10.1007/JHEP04(2021)278} {\bibfield  {journal} {\bibinfo
  {journal} {JHEP}\ }\textbf {\bibinfo {volume} {04}},\ \bibinfo {pages}
  {278}},\ \Eprint {https://arxiv.org/abs/2007.08440} {arXiv:2007.08440
  [hep-ph]} \BibitemShut {NoStop}%
\bibitem [{\citenamefont {Azatov}\ \emph
  {et~al.}(2021{\natexlab{a}})\citenamefont {Azatov}, \citenamefont
  {Vanvlasselaer},\ and\ \citenamefont {Yin}}]{Azatov:2021ifm}%
  \BibitemOpen
  \bibfield  {author} {\bibinfo {author} {\bibfnamefont {A.}~\bibnamefont
  {Azatov}}, \bibinfo {author} {\bibfnamefont {M.}~\bibnamefont
  {Vanvlasselaer}},\ and\ \bibinfo {author} {\bibfnamefont {W.}~\bibnamefont
  {Yin}},\ }\bibfield  {title} {\bibinfo {title} {{Dark Matter production from
  relativistic bubble walls}},\ }\href
  {https://doi.org/10.1007/JHEP03(2021)288} {\bibfield  {journal} {\bibinfo
  {journal} {JHEP}\ }\textbf {\bibinfo {volume} {03}},\ \bibinfo {pages}
  {288}},\ \Eprint {https://arxiv.org/abs/2101.05721} {arXiv:2101.05721
  [hep-ph]} \BibitemShut {NoStop}%
\bibitem [{\citenamefont {Baldes}\ \emph {et~al.}(2023)\citenamefont {Baldes},
  \citenamefont {Dichtl}, \citenamefont {Gouttenoire},\ and\ \citenamefont
  {Sala}}]{Baldes:2023fsp}%
  \BibitemOpen
  \bibfield  {author} {\bibinfo {author} {\bibfnamefont {I.}~\bibnamefont
  {Baldes}}, \bibinfo {author} {\bibfnamefont {M.}~\bibnamefont {Dichtl}},
  \bibinfo {author} {\bibfnamefont {Y.}~\bibnamefont {Gouttenoire}},\ and\
  \bibinfo {author} {\bibfnamefont {F.}~\bibnamefont {Sala}},\ }\bibfield
  {title} {\bibinfo {title} {{Bubbletrons}},\ }\href@noop {} {\  (\bibinfo
  {year} {2023})},\ \Eprint {https://arxiv.org/abs/2306.15555}
  {arXiv:2306.15555 [hep-ph]} \BibitemShut {NoStop}%
\bibitem [{\citenamefont {Baldes}\ \emph
  {et~al.}(2022{\natexlab{a}})\citenamefont {Baldes}, \citenamefont
  {Gouttenoire}, \citenamefont {Sala},\ and\ \citenamefont
  {Servant}}]{Baldes:2021aph}%
  \BibitemOpen
  \bibfield  {author} {\bibinfo {author} {\bibfnamefont {I.}~\bibnamefont
  {Baldes}}, \bibinfo {author} {\bibfnamefont {Y.}~\bibnamefont {Gouttenoire}},
  \bibinfo {author} {\bibfnamefont {F.}~\bibnamefont {Sala}},\ and\ \bibinfo
  {author} {\bibfnamefont {G.}~\bibnamefont {Servant}},\ }\bibfield  {title}
  {\bibinfo {title} {{Supercool composite Dark Matter beyond 100 TeV}},\ }\href
  {https://doi.org/10.1007/JHEP07(2022)084} {\bibfield  {journal} {\bibinfo
  {journal} {JHEP}\ }\textbf {\bibinfo {volume} {07}},\ \bibinfo {pages}
  {084}},\ \Eprint {https://arxiv.org/abs/2110.13926} {arXiv:2110.13926
  [hep-ph]} \BibitemShut {NoStop}%
\bibitem [{\citenamefont {Baldes}\ \emph
  {et~al.}(2022{\natexlab{b}})\citenamefont {Baldes}, \citenamefont
  {Gouttenoire},\ and\ \citenamefont {Sala}}]{Baldes:2022oev}%
  \BibitemOpen
  \bibfield  {author} {\bibinfo {author} {\bibfnamefont {I.}~\bibnamefont
  {Baldes}}, \bibinfo {author} {\bibfnamefont {Y.}~\bibnamefont
  {Gouttenoire}},\ and\ \bibinfo {author} {\bibfnamefont {F.}~\bibnamefont
  {Sala}},\ }\bibfield  {title} {\bibinfo {title} {{Hot and Heavy Dark Matter
  from Supercooling}},\ }\href@noop {} {\  (\bibinfo {year}
  {2022}{\natexlab{b}})},\ \Eprint {https://arxiv.org/abs/2207.05096}
  {arXiv:2207.05096 [hep-ph]} \BibitemShut {NoStop}%
\bibitem [{\citenamefont {Wong}\ and\ \citenamefont
  {Xie}(2023)}]{Wong:2023qon}%
  \BibitemOpen
  \bibfield  {author} {\bibinfo {author} {\bibfnamefont {X.}~\bibnamefont
  {Wong}}\ and\ \bibinfo {author} {\bibfnamefont {K.-P.}\ \bibnamefont {Xie}},\
  }\bibfield  {title} {\bibinfo {title} {{Freeze-in of WIMP dark matter}},\
  }\href@noop {} {\  (\bibinfo {year} {2023})},\ \Eprint
  {https://arxiv.org/abs/2304.00908} {arXiv:2304.00908 [hep-ph]} \BibitemShut
  {NoStop}%
\bibitem [{\citenamefont {Konstandin}\ and\ \citenamefont
  {Servant}(2011{\natexlab{b}})}]{Konstandin:2011ds}%
  \BibitemOpen
  \bibfield  {author} {\bibinfo {author} {\bibfnamefont {T.}~\bibnamefont
  {Konstandin}}\ and\ \bibinfo {author} {\bibfnamefont {G.}~\bibnamefont
  {Servant}},\ }\bibfield  {title} {\bibinfo {title} {{Natural Cold
  Baryogenesis from Strongly Interacting Electroweak Symmetry Breaking}},\
  }\href {https://doi.org/10.1088/1475-7516/2011/07/024} {\bibfield  {journal}
  {\bibinfo  {journal} {JCAP}\ }\textbf {\bibinfo {volume} {07}},\ \bibinfo
  {pages} {024}},\ \Eprint {https://arxiv.org/abs/1104.4793} {arXiv:1104.4793
  [hep-ph]} \BibitemShut {NoStop}%
\bibitem [{\citenamefont {Servant}(2014)}]{Servant:2014bla}%
  \BibitemOpen
  \bibfield  {author} {\bibinfo {author} {\bibfnamefont {G.}~\bibnamefont
  {Servant}},\ }\bibfield  {title} {\bibinfo {title} {{Baryogenesis from Strong
  $CP$ Violation and the QCD Axion}},\ }\href
  {https://doi.org/10.1103/PhysRevLett.113.171803} {\bibfield  {journal}
  {\bibinfo  {journal} {Phys. Rev. Lett.}\ }\textbf {\bibinfo {volume} {113}},\
  \bibinfo {pages} {171803} (\bibinfo {year} {2014})},\ \Eprint
  {https://arxiv.org/abs/1407.0030} {arXiv:1407.0030 [hep-ph]} \BibitemShut
  {NoStop}%
\bibitem [{\citenamefont {Azatov}\ \emph
  {et~al.}(2021{\natexlab{b}})\citenamefont {Azatov}, \citenamefont
  {Vanvlasselaer},\ and\ \citenamefont {Yin}}]{Azatov:2021irb}%
  \BibitemOpen
  \bibfield  {author} {\bibinfo {author} {\bibfnamefont {A.}~\bibnamefont
  {Azatov}}, \bibinfo {author} {\bibfnamefont {M.}~\bibnamefont
  {Vanvlasselaer}},\ and\ \bibinfo {author} {\bibfnamefont {W.}~\bibnamefont
  {Yin}},\ }\bibfield  {title} {\bibinfo {title} {{Baryogenesis via
  relativistic bubble walls}},\ }\href
  {https://doi.org/10.1007/JHEP10(2021)043} {\bibfield  {journal} {\bibinfo
  {journal} {JHEP}\ }\textbf {\bibinfo {volume} {10}},\ \bibinfo {pages}
  {043}},\ \Eprint {https://arxiv.org/abs/2106.14913} {arXiv:2106.14913
  [hep-ph]} \BibitemShut {NoStop}%
\bibitem [{\citenamefont {Baldes}\ \emph
  {et~al.}(2021{\natexlab{b}})\citenamefont {Baldes}, \citenamefont {Blasi},
  \citenamefont {Mariotti}, \citenamefont {Sevrin},\ and\ \citenamefont
  {Turbang}}]{Baldes:2021vyz}%
  \BibitemOpen
  \bibfield  {author} {\bibinfo {author} {\bibfnamefont {I.}~\bibnamefont
  {Baldes}}, \bibinfo {author} {\bibfnamefont {S.}~\bibnamefont {Blasi}},
  \bibinfo {author} {\bibfnamefont {A.}~\bibnamefont {Mariotti}}, \bibinfo
  {author} {\bibfnamefont {A.}~\bibnamefont {Sevrin}},\ and\ \bibinfo {author}
  {\bibfnamefont {K.}~\bibnamefont {Turbang}},\ }\bibfield  {title} {\bibinfo
  {title} {{Baryogenesis via relativistic bubble expansion}},\ }\href
  {https://doi.org/10.1103/PhysRevD.104.115029} {\bibfield  {journal} {\bibinfo
   {journal} {Phys. Rev. D}\ }\textbf {\bibinfo {volume} {104}},\ \bibinfo
  {pages} {115029} (\bibinfo {year} {2021}{\natexlab{b}})},\ \Eprint
  {https://arxiv.org/abs/2106.15602} {arXiv:2106.15602 [hep-ph]} \BibitemShut
  {NoStop}%
\bibitem [{\citenamefont {Baldes}\ and\ \citenamefont
  {Olea-Romacho}(2024)}]{Baldes:2023rqv}%
  \BibitemOpen
  \bibfield  {author} {\bibinfo {author} {\bibfnamefont {I.}~\bibnamefont
  {Baldes}}\ and\ \bibinfo {author} {\bibfnamefont {M.~O.}\ \bibnamefont
  {Olea-Romacho}},\ }\bibfield  {title} {\bibinfo {title} {{Primordial black
  holes as dark matter: interferometric tests of phase transition origin}},\
  }\href {https://doi.org/10.1007/JHEP01(2024)133} {\bibfield  {journal}
  {\bibinfo  {journal} {JHEP}\ }\textbf {\bibinfo {volume} {01}},\ \bibinfo
  {pages} {133}},\ \Eprint {https://arxiv.org/abs/2307.11639} {arXiv:2307.11639
  [hep-ph]} \BibitemShut {NoStop}%
\bibitem [{\citenamefont {Flores}\ \emph {et~al.}(2024)\citenamefont {Flores},
  \citenamefont {Kusenko},\ and\ \citenamefont {Sasaki}}]{Flores:2024lng}%
  \BibitemOpen
  \bibfield  {author} {\bibinfo {author} {\bibfnamefont {M.~M.}\ \bibnamefont
  {Flores}}, \bibinfo {author} {\bibfnamefont {A.}~\bibnamefont {Kusenko}},\
  and\ \bibinfo {author} {\bibfnamefont {M.}~\bibnamefont {Sasaki}},\
  }\bibfield  {title} {\bibinfo {title} {{Revisiting formation of primordial
  black holes in a supercooled first-order phase transition}},\ }\href@noop {}
  {\  (\bibinfo {year} {2024})},\ \Eprint {https://arxiv.org/abs/2402.13341}
  {arXiv:2402.13341 [hep-ph]} \BibitemShut {NoStop}%
\bibitem [{\citenamefont {Quiros}(1999)}]{Quiros:1999jp}%
  \BibitemOpen
  \bibfield  {author} {\bibinfo {author} {\bibfnamefont {M.}~\bibnamefont
  {Quiros}},\ }\bibfield  {title} {\bibinfo {title} {{Finite Temperature Field
  Theory and Phase Transitions}},\ }in\ \href@noop {} {\emph {\bibinfo
  {booktitle} {{Ictp Summer School in High-Energy Physics and Cosmology}}}}\
  (\bibinfo {year} {1999})\ pp.\ \bibinfo {pages} {187--259},\ \Eprint
  {https://arxiv.org/abs/hep-ph/9901312} {arXiv:hep-ph/9901312} \BibitemShut
  {NoStop}%
\bibitem [{\citenamefont {Hindmarsh}\ and\ \citenamefont
  {Hijazi}(2019)}]{Hindmarsh:2019phv}%
  \BibitemOpen
  \bibfield  {author} {\bibinfo {author} {\bibfnamefont {M.}~\bibnamefont
  {Hindmarsh}}\ and\ \bibinfo {author} {\bibfnamefont {M.}~\bibnamefont
  {Hijazi}},\ }\bibfield  {title} {\bibinfo {title} {{Gravitational waves from
  first order cosmological phase transitions in the Sound Shell Model}},\
  }\href {https://doi.org/10.1088/1475-7516/2019/12/062} {\bibfield  {journal}
  {\bibinfo  {journal} {JCAP}\ }\textbf {\bibinfo {volume} {12}},\ \bibinfo
  {pages} {062}},\ \Eprint {https://arxiv.org/abs/1909.10040} {arXiv:1909.10040
  [astro-ph.CO]} \BibitemShut {NoStop}%
\bibitem [{\citenamefont {Turner}\ \emph {et~al.}(1992)\citenamefont {Turner},
  \citenamefont {Weinberg},\ and\ \citenamefont {Widrow}}]{Turner:1992tz}%
  \BibitemOpen
  \bibfield  {author} {\bibinfo {author} {\bibfnamefont {M.~S.}\ \bibnamefont
  {Turner}}, \bibinfo {author} {\bibfnamefont {E.~J.}\ \bibnamefont
  {Weinberg}},\ and\ \bibinfo {author} {\bibfnamefont {L.~M.}\ \bibnamefont
  {Widrow}},\ }\bibfield  {title} {\bibinfo {title} {{Bubble Nucleation in
  First Order Inflation and Other Cosmological Phase Transitions}},\ }\href
  {https://doi.org/10.1103/PhysRevD.46.2384} {\bibfield  {journal} {\bibinfo
  {journal} {Phys. Rev. D}\ }\textbf {\bibinfo {volume} {46}},\ \bibinfo
  {pages} {2384} (\bibinfo {year} {1992})}\BibitemShut {NoStop}%
\bibitem [{\citenamefont {Deng}(2020)}]{Deng:2020mds}%
  \BibitemOpen
  \bibfield  {author} {\bibinfo {author} {\bibfnamefont {H.}~\bibnamefont
  {Deng}},\ }\bibfield  {title} {\bibinfo {title} {{Primordial black hole
  formation by vacuum bubbles. Part II}},\ }\href
  {https://doi.org/10.1088/1475-7516/2020/09/023} {\bibfield  {journal}
  {\bibinfo  {journal} {JCAP}\ }\textbf {\bibinfo {volume} {09}},\ \bibinfo
  {pages} {023}},\ \Eprint {https://arxiv.org/abs/2006.11907} {arXiv:2006.11907
  [astro-ph.CO]} \BibitemShut {NoStop}%
\bibitem [{\citenamefont {Vilenkin}\ and\ \citenamefont
  {Shellard}(2000)}]{Vilenkin:2000jqa}%
  \BibitemOpen
  \bibfield  {author} {\bibinfo {author} {\bibfnamefont {A.}~\bibnamefont
  {Vilenkin}}\ and\ \bibinfo {author} {\bibfnamefont {E.~P.~S.}\ \bibnamefont
  {Shellard}},\ }\href@noop {} {\emph {\bibinfo {title} {{Cosmic Strings and
  Other Topological Defects}}}}\ (\bibinfo  {publisher} {Cambridge University
  Press},\ \bibinfo {year} {2000})\BibitemShut {NoStop}%
\bibitem [{\citenamefont {Watkins}\ and\ \citenamefont
  {Widrow}(1992)}]{Watkins:1991zt}%
  \BibitemOpen
  \bibfield  {author} {\bibinfo {author} {\bibfnamefont {R.}~\bibnamefont
  {Watkins}}\ and\ \bibinfo {author} {\bibfnamefont {L.~M.}\ \bibnamefont
  {Widrow}},\ }\bibfield  {title} {\bibinfo {title} {{Aspects of reheating in
  first order inflation}},\ }\href
  {https://doi.org/10.1016/0550-3213(92)90362-F} {\bibfield  {journal}
  {\bibinfo  {journal} {Nucl. Phys. B}\ }\textbf {\bibinfo {volume} {374}},\
  \bibinfo {pages} {446} (\bibinfo {year} {1992})}\BibitemShut {NoStop}%
\bibitem [{\citenamefont {Kolb}\ and\ \citenamefont
  {Riotto}(1997)}]{Kolb:1996jr}%
  \BibitemOpen
  \bibfield  {author} {\bibinfo {author} {\bibfnamefont {E.~W.}\ \bibnamefont
  {Kolb}}\ and\ \bibinfo {author} {\bibfnamefont {A.}~\bibnamefont {Riotto}},\
  }\bibfield  {title} {\bibinfo {title} {{Preheating and symmetry restoration
  in collisions of vacuum bubbles}},\ }\href
  {https://doi.org/10.1103/PhysRevD.55.3313} {\bibfield  {journal} {\bibinfo
  {journal} {Phys. Rev. D}\ }\textbf {\bibinfo {volume} {55}},\ \bibinfo
  {pages} {3313} (\bibinfo {year} {1997})},\ \Eprint
  {https://arxiv.org/abs/astro-ph/9602095} {arXiv:astro-ph/9602095}
  \BibitemShut {NoStop}%
\bibitem [{\citenamefont {Falkowski}\ and\ \citenamefont
  {No}(2013)}]{Falkowski:2012fb}%
  \BibitemOpen
  \bibfield  {author} {\bibinfo {author} {\bibfnamefont {A.}~\bibnamefont
  {Falkowski}}\ and\ \bibinfo {author} {\bibfnamefont {J.~M.}\ \bibnamefont
  {No}},\ }\bibfield  {title} {\bibinfo {title} {{Non-thermal Dark Matter
  Production from the Electroweak Phase Transition: Multi-TeV WIMPs and
  'Baby-Zillas'}},\ }\href {https://doi.org/10.1007/JHEP02(2013)034} {\bibfield
   {journal} {\bibinfo  {journal} {JHEP}\ }\textbf {\bibinfo {volume} {02}},\
  \bibinfo {pages} {034}},\ \Eprint {https://arxiv.org/abs/1211.5615}
  {arXiv:1211.5615 [hep-ph]} \BibitemShut {NoStop}%
\bibitem [{\citenamefont {Cutting}\ \emph {et~al.}(2018)\citenamefont
  {Cutting}, \citenamefont {Hindmarsh},\ and\ \citenamefont
  {Weir}}]{Cutting:2018tjt}%
  \BibitemOpen
  \bibfield  {author} {\bibinfo {author} {\bibfnamefont {D.}~\bibnamefont
  {Cutting}}, \bibinfo {author} {\bibfnamefont {M.}~\bibnamefont {Hindmarsh}},\
  and\ \bibinfo {author} {\bibfnamefont {D.~J.}\ \bibnamefont {Weir}},\
  }\bibfield  {title} {\bibinfo {title} {{Gravitational waves from vacuum
  first-order phase transitions: from the envelope to the lattice}},\ }\href
  {https://doi.org/10.1103/PhysRevD.97.123513} {\bibfield  {journal} {\bibinfo
  {journal} {Phys. Rev. D}\ }\textbf {\bibinfo {volume} {97}},\ \bibinfo
  {pages} {123513} (\bibinfo {year} {2018})},\ \Eprint
  {https://arxiv.org/abs/1802.05712} {arXiv:1802.05712 [astro-ph.CO]}
  \BibitemShut {NoStop}%
\bibitem [{\citenamefont {Cutting}\ \emph {et~al.}(2021)\citenamefont
  {Cutting}, \citenamefont {Escartin}, \citenamefont {Hindmarsh},\ and\
  \citenamefont {Weir}}]{Cutting:2020nla}%
  \BibitemOpen
  \bibfield  {author} {\bibinfo {author} {\bibfnamefont {D.}~\bibnamefont
  {Cutting}}, \bibinfo {author} {\bibfnamefont {E.~G.}\ \bibnamefont
  {Escartin}}, \bibinfo {author} {\bibfnamefont {M.}~\bibnamefont
  {Hindmarsh}},\ and\ \bibinfo {author} {\bibfnamefont {D.~J.}\ \bibnamefont
  {Weir}},\ }\bibfield  {title} {\bibinfo {title} {{Gravitational waves from
  vacuum first order phase transitions II: from thin to thick walls}},\ }\href
  {https://doi.org/10.1103/PhysRevD.103.023531} {\bibfield  {journal} {\bibinfo
   {journal} {Phys. Rev. D}\ }\textbf {\bibinfo {volume} {103}},\ \bibinfo
  {pages} {023531} (\bibinfo {year} {2021})},\ \Eprint
  {https://arxiv.org/abs/2005.13537} {arXiv:2005.13537 [astro-ph.CO]}
  \BibitemShut {NoStop}%
\bibitem [{\citenamefont {Aarts}\ \emph {et~al.}(2000)\citenamefont {Aarts},
  \citenamefont {Bonini},\ and\ \citenamefont {Wetterich}}]{Aarts:2000mg}%
  \BibitemOpen
  \bibfield  {author} {\bibinfo {author} {\bibfnamefont {G.}~\bibnamefont
  {Aarts}}, \bibinfo {author} {\bibfnamefont {G.~F.}\ \bibnamefont {Bonini}},\
  and\ \bibinfo {author} {\bibfnamefont {C.}~\bibnamefont {Wetterich}},\
  }\bibfield  {title} {\bibinfo {title} {{On Thermalization in classical scalar
  field theory}},\ }\href {https://doi.org/10.1016/S0550-3213(00)00447-8}
  {\bibfield  {journal} {\bibinfo  {journal} {Nucl. Phys. B}\ }\textbf
  {\bibinfo {volume} {587}},\ \bibinfo {pages} {403} (\bibinfo {year}
  {2000})},\ \Eprint {https://arxiv.org/abs/hep-ph/0003262}
  {arXiv:hep-ph/0003262} \BibitemShut {NoStop}%
\bibitem [{\citenamefont {Micha}\ and\ \citenamefont
  {Tkachev}(2003)}]{Micha:2002ey}%
  \BibitemOpen
  \bibfield  {author} {\bibinfo {author} {\bibfnamefont {R.}~\bibnamefont
  {Micha}}\ and\ \bibinfo {author} {\bibfnamefont {I.~I.}\ \bibnamefont
  {Tkachev}},\ }\bibfield  {title} {\bibinfo {title} {{Relativistic turbulence:
  A Long way from preheating to equilibrium}},\ }\href
  {https://doi.org/10.1103/PhysRevLett.90.121301} {\bibfield  {journal}
  {\bibinfo  {journal} {Phys. Rev. Lett.}\ }\textbf {\bibinfo {volume} {90}},\
  \bibinfo {pages} {121301} (\bibinfo {year} {2003})},\ \Eprint
  {https://arxiv.org/abs/hep-ph/0210202} {arXiv:hep-ph/0210202} \BibitemShut
  {NoStop}%
\bibitem [{\citenamefont {Arrizabalaga}\ \emph {et~al.}(2005)\citenamefont
  {Arrizabalaga}, \citenamefont {Smit},\ and\ \citenamefont
  {Tranberg}}]{Arrizabalaga:2005tf}%
  \BibitemOpen
  \bibfield  {author} {\bibinfo {author} {\bibfnamefont {A.}~\bibnamefont
  {Arrizabalaga}}, \bibinfo {author} {\bibfnamefont {J.}~\bibnamefont {Smit}},\
  and\ \bibinfo {author} {\bibfnamefont {A.}~\bibnamefont {Tranberg}},\
  }\bibfield  {title} {\bibinfo {title} {{Equilibration in phi**4 theory in 3+1
  dimensions}},\ }\href {https://doi.org/10.1103/PhysRevD.72.025014} {\bibfield
   {journal} {\bibinfo  {journal} {Phys. Rev. D}\ }\textbf {\bibinfo {volume}
  {72}},\ \bibinfo {pages} {025014} (\bibinfo {year} {2005})},\ \Eprint
  {https://arxiv.org/abs/hep-ph/0503287} {arXiv:hep-ph/0503287} \BibitemShut
  {NoStop}%
\bibitem [{\citenamefont {Niedermann}\ and\ \citenamefont
  {Sloth}(2020)}]{Niedermann:2020dwg}%
  \BibitemOpen
  \bibfield  {author} {\bibinfo {author} {\bibfnamefont {F.}~\bibnamefont
  {Niedermann}}\ and\ \bibinfo {author} {\bibfnamefont {M.~S.}\ \bibnamefont
  {Sloth}},\ }\bibfield  {title} {\bibinfo {title} {{Resolving the Hubble
  tension with new early dark energy}},\ }\href
  {https://doi.org/10.1103/PhysRevD.102.063527} {\bibfield  {journal} {\bibinfo
   {journal} {Phys. Rev. D}\ }\textbf {\bibinfo {volume} {102}},\ \bibinfo
  {pages} {063527} (\bibinfo {year} {2020})},\ \Eprint
  {https://arxiv.org/abs/2006.06686} {arXiv:2006.06686 [astro-ph.CO]}
  \BibitemShut {NoStop}%
\bibitem [{\citenamefont {Harada}\ \emph {et~al.}(2016)\citenamefont {Harada},
  \citenamefont {Yoo}, \citenamefont {Kohri}, \citenamefont {Nakao},\ and\
  \citenamefont {Jhingan}}]{Harada:2016mhb}%
  \BibitemOpen
  \bibfield  {author} {\bibinfo {author} {\bibfnamefont {T.}~\bibnamefont
  {Harada}}, \bibinfo {author} {\bibfnamefont {C.-M.}\ \bibnamefont {Yoo}},
  \bibinfo {author} {\bibfnamefont {K.}~\bibnamefont {Kohri}}, \bibinfo
  {author} {\bibfnamefont {K.-i.}\ \bibnamefont {Nakao}},\ and\ \bibinfo
  {author} {\bibfnamefont {S.}~\bibnamefont {Jhingan}},\ }\bibfield  {title}
  {\bibinfo {title} {{Primordial black hole formation in the matter-dominated
  phase of the Universe}},\ }\href {https://doi.org/10.3847/1538-4357/833/1/61}
  {\bibfield  {journal} {\bibinfo  {journal} {Astrophys. J.}\ }\textbf
  {\bibinfo {volume} {833}},\ \bibinfo {pages} {61} (\bibinfo {year} {2016})},\
  \Eprint {https://arxiv.org/abs/1609.01588} {arXiv:1609.01588 [astro-ph.CO]}
  \BibitemShut {NoStop}%
\bibitem [{\citenamefont {Dasgupta}\ \emph {et~al.}(2020)\citenamefont
  {Dasgupta}, \citenamefont {Laha},\ and\ \citenamefont
  {Ray}}]{Dasgupta:2019cae}%
  \BibitemOpen
  \bibfield  {author} {\bibinfo {author} {\bibfnamefont {B.}~\bibnamefont
  {Dasgupta}}, \bibinfo {author} {\bibfnamefont {R.}~\bibnamefont {Laha}},\
  and\ \bibinfo {author} {\bibfnamefont {A.}~\bibnamefont {Ray}},\ }\bibfield
  {title} {\bibinfo {title} {{Neutrino and positron constraints on spinning
  primordial black hole dark matter}},\ }\href
  {https://doi.org/10.1103/PhysRevLett.125.101101} {\bibfield  {journal}
  {\bibinfo  {journal} {Phys. Rev. Lett.}\ }\textbf {\bibinfo {volume} {125}},\
  \bibinfo {pages} {101101} (\bibinfo {year} {2020})},\ \Eprint
  {https://arxiv.org/abs/1912.01014} {arXiv:1912.01014 [hep-ph]} \BibitemShut
  {NoStop}%
\bibitem [{\citenamefont {Laha}\ \emph {et~al.}(2020)\citenamefont {Laha},
  \citenamefont {Mu\~noz},\ and\ \citenamefont {Slatyer}}]{Laha:2020ivk}%
  \BibitemOpen
  \bibfield  {author} {\bibinfo {author} {\bibfnamefont {R.}~\bibnamefont
  {Laha}}, \bibinfo {author} {\bibfnamefont {J.~B.}\ \bibnamefont {Mu\~noz}},\
  and\ \bibinfo {author} {\bibfnamefont {T.~R.}\ \bibnamefont {Slatyer}},\
  }\bibfield  {title} {\bibinfo {title} {{INTEGRAL constraints on primordial
  black holes and particle dark matter}},\ }\href
  {https://doi.org/10.1103/PhysRevD.101.123514} {\bibfield  {journal} {\bibinfo
   {journal} {Phys. Rev. D}\ }\textbf {\bibinfo {volume} {101}},\ \bibinfo
  {pages} {123514} (\bibinfo {year} {2020})},\ \Eprint
  {https://arxiv.org/abs/2004.00627} {arXiv:2004.00627 [astro-ph.CO]}
  \BibitemShut {NoStop}%
\bibitem [{\citenamefont {Ray}\ \emph {et~al.}(2021)\citenamefont {Ray},
  \citenamefont {Laha}, \citenamefont {Mu\~noz},\ and\ \citenamefont
  {Caputo}}]{Ray:2021mxu}%
  \BibitemOpen
  \bibfield  {author} {\bibinfo {author} {\bibfnamefont {A.}~\bibnamefont
  {Ray}}, \bibinfo {author} {\bibfnamefont {R.}~\bibnamefont {Laha}}, \bibinfo
  {author} {\bibfnamefont {J.~B.}\ \bibnamefont {Mu\~noz}},\ and\ \bibinfo
  {author} {\bibfnamefont {R.}~\bibnamefont {Caputo}},\ }\bibfield  {title}
  {\bibinfo {title} {{Near future MeV telescopes can discover asteroid-mass
  primordial black hole dark matter}},\ }\href
  {https://doi.org/10.1103/PhysRevD.104.023516} {\bibfield  {journal} {\bibinfo
   {journal} {Phys. Rev. D}\ }\textbf {\bibinfo {volume} {104}},\ \bibinfo
  {pages} {023516} (\bibinfo {year} {2021})},\ \Eprint
  {https://arxiv.org/abs/2102.06714} {arXiv:2102.06714 [astro-ph.CO]}
  \BibitemShut {NoStop}%
\end{thebibliography}%

\end{document}